\documentclass[aps,prf]{revtex4-2}
\usepackage{float}
\usepackage{graphicx}
\usepackage{epstopdf, epsfig}
\usepackage{caption}
\usepackage{subcaption}
\usepackage{amsmath}
\usepackage{amssymb}
\usepackage{color}
\usepackage{placeins}

\usepackage{dcolumn}%
\usepackage{bm}%

\begin{document}

\title{A Variational Formulation of Resolvent Analysis}

\author{Benedikt Barthel}
 \email{bbarthel@caltech.edu.}
 \affiliation{Graduate Aerospace Laboratories, California Institute of Technology,
Pasadena, CA 91125, USA}%

\author{Salvador Gomez}%
\affiliation{Graduate Aerospace Laboratories, California Institute of Technology,
Pasadena, CA 91125, USA}%

\author{Beverley J. McKeon}
\affiliation{Graduate Aerospace Laboratories, California Institute of Technology,
Pasadena, CA 91125, USA}%

\date{\today}

% \keywords{low dimensional models, computational methods, transition to turbulence}

\begin{abstract}
The conceptual picture underlying resolvent analysis(RA) is that the nonlinear term in the Navier-Stokes(NS) equations acts as an intrinsic forcing to the linear dynamics, a description inspired by control theory. The inverse of the linear operator, defined as the resolvent, is interpreted as a transfer function between the forcing and the velocity response. From a theoretical point of view this is an attractive approach since it allows for the vast mathematical machinery of control theory to be brought to bear on the problem.  However, from a practical point of view, this is not always advantageous. The inversion of the linear operator inherent in the control theoretic definition obscures the physical interpretation of the governing equations and is prohibitive to analytical manipulation, and for large systems leads to significant computational cost and memory requirements. In this work we suggest an alternative, inverse free, definition of the resolvent basis based on an extension of the Courant–Fischer–Weyl min-max principle in which resolvent modes are defined as stationary points of a constrained variational problem. This definition leads to a straightforward approach to approximate the resolvent (response) modes of complex flows as expansions in any arbitrary basis.  The proposed method avoids matrix inversions and requires only the spectral decomposition of a matrix of significantly reduced size as compared to the original system. To illustrate this method and the advantages of the variational formulation we present three examples. First, we consider streamwise constant fluctuations in turbulent channel flow where an asymptotic analysis allows us to derive closed form expressions for the optimal resolvent modes. Second, to illustrate the cost saving potential, and investigate the limits, of the proposed method we apply our method to both a 2-dimensional, 3-component equilibrium solution in Couette flow and, finally, to a streamwise developing turbulent boundary layer. For these larger systems we achieve a model reduction of up to two orders of magnitude. Such savings have the potential to open up RA to the investigation of larger domains and more complex flow configurations.
\end{abstract}

\maketitle

\section{Introduction}
Resolvent analysis (RA) can be used to give insight into the forced response of a linearized dynamical system. This concept was introduced by \citet{trefethen_hydrodynamic_1993} and \citet{jovanovic_componentwise_2005} who considered the stability and amplification of linearly stable flows to external forcing. These ideas were later applied to turbulent flows by \citet{mckeon_critical-layer_2010} who interpreted the nonlinear term in the Navier-Stokes equations (NSE) as a forcing to the linearized system. The conceptual framework of RA is inspired by control theory (CT), such that the resolvent operator, the inverse of the linearized operator, is interpreted as a transfer function from the forcing to the response. A singular value decomposition (SVD) of the discretized resolvent operator provides two distinct orthonormal bases (left and right singular modes) for both the response and the forcing, ordered by a set of gains (singular values) which quantify the linear amplification of the system. This CT-inspired framework has proven theoretically useful since it is conceptually straightforward and benefits from years of established mathematical machinery. However, from a practical point of view the reliance on the inversion of the linear operator poses some difficulties. It obscures the analytical tractability of the equations and is computationally costly for all but the simplest systems.

Early studies using RA were largely focused on wall bounded shear flows with only a single non-homogeneous spatial dimension for which cost of the inversion and SVD of the operator is trivial \citep{jovanovic_componentwise_2005,mckeon_critical-layer_2010,hwang_linear_2010,moarref_model-based_2013,sharma_scaling_2017}. In these cases, linearly amplified length scales identified by RA were found to correlate with the energetically active scales observed in experiments and simulations, and the corresponding resolvent modes capture the qualitative features of the coherent structures observed in wall turbulence \citep{mckeon_engine_2017}. In particular, resolvent modes have been found to exhibit self similar behaviour characteristic of the attached eddy hypothesis proposed by Townsend~\citep{townsend_structure_1951,moarref_model-based_2013,mckeon_self-similar_2019}. \citet{towne_spectral_2018} have elaborated the assumptions under which resolvent response modes correlate with spectral proper orthogonal decomposition (SPOD) modes computed from data, illustrating that RA can predict coherent structure in the full flow field. More recently RA has also been extended to 2D flows such as boundary layers \citep{sipp_characterization_2013,rigas_nonlinear_2021}, the flow behind bluff bodies \citep{symon_non-normality_2018,symon_mean_2020}, exact coherent states (ECS) \citep{rosenberg_efficient_2019}, and turbulent jets \citep{schmidt_spectral_2018,pickering_optimal_2021}. In particular, modal analysis techniques including RA have been used by a variety of authors to implement flow control strategies, for example, to suppress vortex shedding \citep{gomez_data-driven_2017}, and delay flow separation \citep{yeh_resolvent-analysis-based_2019}. For these 2D flows the computational cost and memory requirements becomes considerable and thus the further extension to 3D flows has generally remained limited.

The community has endeavoured to address these computational challenges through innovation in novel methods of estimating resolvent modes. One area of research has been in so called ``matrix free" methods such as the work of \citet{martini_efficient_2021} who use the transient and steady state responses of the periodically forced linearized system and its corresponding adjoint system to estimate the action of the resolvent operator. Another avenue of investigation inspired by the field of data analysis has been in ``equation free" methods such as \citet{herrmann_data-driven_2021} who use dynamic mode decomposition (DMD) modes to estimate the linear dynamics of a system from a time series of data. Others have made use of iterative Arnoldi Algorithms that replace the cost of calculating the SVD and a matrix inverse with the cost of an LU decomposition and a few matrix multiplications \citep{sipp_characterization_2013,schmidt_spectral_2018}. Furthermore, algorithms such as randomized SVD and others have made it possible to efficiently and accurately compute singular modes of data sets that would otherwise be prohibitively expensive \citep{schmid_stability_2001,halko_finding_2011,moarref_model-based_2013,tropp_streaming_2019,ribeiro_randomized_2020}.

The previously cited research has focused on the CT interpretation of RA and the SVD-based definition of resolvent modes. In this work we take an alternative approach and propose an equivalent definition based on an extension of the Courant–Fischer–Weyl min-max principle (CFL). The CFL principle itself has been used previously by \citet{dawson_shape_2019} who formulated a simplified variational problem to estimate the shape of the vorticity component of the optimal resolvent mode in wall bounded shear flows. We believe the explicit extension from the CFL principle, to what we coin ``variational resolvent analysis" (VRA), which constitutes an alternative definition of the resolvent basis that includes all modes, to be novel in the resolvent literature.

This new definition is based on the solutions of the Euler-Lagrange equations associated with the constrained variation of the operator norm of the linearized dynamics. Critically, this definition does not involve the inversion of any operator, which is useful from both a theoretical and practical sense. The inversion of large matrices is both costly and obscures the intuitive interpretation of the underlying linear differential operator. While in general the resulting Euler-Lagrange equations remain difficult to solve exactly, this variational formulation allows for the approximation of resolvent modes as an expansion in any convenient basis, for example the much cheaper one-dimensional resolvent basis in a two- or three-dimensional problem, an analytical basis such as that described by \cite{dawson_shape_2019} or a data-driven one. Further, it requires only the eigenvalue decomposition of a matrix of reduced size. In this paper we illustrate how this variational definition is useful in both gaining physical insights by allowing for analytical progress in simplified systems, and by reducing computational cost in complex systems. To illustrate the former we consider the case of streamwise constant fluctuations in wall bounded shear flows, and to investigate the latter we perform RA around a 2D/3C exact coherent solution. We find that we can accurately approximate the resolvent response modes and reduce the computational complexity by an order of magnitude. Finally, the VRA formulation is applied to a streamwise developing turbulent boundary layer, where the near wall modes can be predicted with a $97\%$ reduction in computational cost using resolvent modes calculated using a 1D mean flow.

The paper is organized as follows. In \S\ref{sec:math} we derive the proposed variational definition. In \S\ref{sec:kx0} we use the variational formulation to analyze streamwise constant structures in turbulent channel flow. In \S\ref{sec:ECS} and \S\ref{sec:BL} we consider RA applied to both streamwise periodic and streamwise developing two-dimensional, three velocity component (2D/3C) systems to illustrate the computational cost and memory saving potential of the proposed VRA formulation. In \S\ref{sec:direction} we analyze the uncertainty and potential sources of error in our method. We provide discussion of the results and the outlook for future applications in \S\ref{sec:discussion} and conclude in \S\ref{sec:conclusion}.

%% Mathematical Formulation Section
\section{Mathematical Formulation}\label{sec:math}
Let us consider a general forced linear system
\begin{equation}\label{system}
    \frac{\partial \mathbf{u}}{\partial t} - \mathbf{A}\mathbf{u} = \mathbf{f}
\end{equation}
where $\mathbf{A}$ represents a spatial-linear differential operator and $\mathbf{u}(\mathbf{x},t), \mathbf{f}(\mathbf{x},t) \in \mathrm{C}^{\infty}$. The state variables $\mathbf{u}$ and $\mathbf{f}$ are referred to as the `response' and `forcing' respectively.  We consider the temporal Fourier transfer of (\ref{system}) and define the spatio-temporal linear operator 
\begin{equation}\label{L}
    \mathbf{L}\left(\omega\right)\equiv i\omega\mathbf{I} - \mathbf{A}
\end{equation}
as well as the resolvent operator 
\begin{equation}\label{H}
    \mathbf{H}\left(\omega\right) \equiv \mathbf{L}\left(\omega\right)^{-1}
\end{equation}
which is classically interpreted as a transfer function from the forcing to the response.
\begin{equation}\label{RAsystem}
    \mathbf{u} = \mathbf{H}\mathbf{f}
\end{equation}
For readability we have dropped explicit reference to the dependence on $\omega$. An SVD of the resolvent 
\begin{equation}\label{RAsvd}
    \mathbf{H} = \sum_{j=1}^{\infty} \boldsymbol{\psi}_j \sigma_j \boldsymbol{\phi}_j^{H}
\end{equation}
results in a pair of distinct sets of basis functions for the response $(\boldsymbol{\psi}_j)$ and forcing $(\boldsymbol{\phi}_j)$ and are referred to as the resolvent `response modes' and `forcing modes' respectively. These are ordered by their gains $\sigma_{j}$ that are ordered in descending order, representing the $j$th largest linear gain possible. Here and throughout this work superscript $^H$ denotes the Hermitian adjoint, or for discrete matrices the conjugate transpose.

% variational formulation subsection
\subsection{A variational definition of resolvent modes}
A key contribution of this work is the observation that resolvent response modes may be equivalently defined as the stationary points, $\mathbf{q}^*$, of the operator norm of $\mathbf{L}$ under the condition that the argument $\mathbf{q}^*$ satisfies some norm constraint. More explicitly, the resolvent modes of the linear operator $\mathbf{H}$ are defined as the stationary points of the functional
\begin{equation}\label{J}
    J = \|\mathbf{L}\mathbf{q}\|_a^2 
\end{equation}
subject to the constraint
\begin{equation}\label{normcon1}
   \|\mathbf{q}\|_b^2 = 1.
\end{equation}
We note that in general the norms $\|\mathbf{x}||_a \equiv \mathbf{x}^H\mathbf{Q}_a\mathbf{x}$ and $\|\mathbf{x}||_b \equiv \mathbf{x}^H\mathbf{Q}_b\mathbf{x}$ need not be the same, such as for example in the Orr-Sommerfeld and Squire decomposition discussed in \S\ref{sec:kx0}. Following the notation of \citet{herrmann_data-driven_2021} the Cholesky factorization may be used to decompose the weight matrix \begin{equation}\label{Q}
    \mathbf{Q}_a = \mathbf{F}_a^H\mathbf{F}_a
\end{equation}
This allows a general norm to be related to the Euclidean 2 norm. In other words, we can express any arbitrary user defined norm as
\begin{equation}\label{norm}
\|\mathbf{x}||_\alpha = \|\mathbf{F}_\alpha\mathbf{x}||_2 
\end{equation}
where $\alpha$ is simply a label used to distinguish between different norms. 

The method of Lagrange multipliers allows us to combine(\ref{J}), (\ref{normcon1}) and the definition (\ref{norm}) to formulate a constrained variational problem and define a Lagrangian
\begin{equation}\label{lag1}
\mathcal{L}\left(\mathbf{q}\right) = \|\mathbf{F}_a \mathbf{L}\mathbf{q}||^2_2 - \sigma^{-2}\|\mathbf{F}_b\mathbf{q}\|_2^2 = \mathbf{q}^H\mathbf{L}^H \mathbf{Q}_a\mathbf{L} \mathbf{q} - \sigma^{-2}\mathbf{q}^H\mathbf{Q}_b \mathbf{q}.
\end{equation}
% \begin{equation}\label{lag2}
% \mathcal{L}\left(\mathbf{q}\right) = \mathbf{q}^H\mathbf{L}^H \mathbf{Q}_a\mathbf{L} \mathbf{q} - \sigma^{-2}\mathbf{q}^H\mathbf{Q}_b \mathbf{q}.
% \end{equation}
Here $\mathbf{L}$ and $\mathbf{F}$ may be either interpreted as continuous differential operators or discrete matrices. The vanishing of the variation with respect to the conjugate state $\mathbf{q}^*$ is a necessary and sufficient condition for the stationarity of (\ref{lag1}). The reader is referred to appendix \ref{app:CR} for a derivation of this property based on the work of \citep{wirtinger_zur_1927, brandwood_complex_1983}. The resolvent response modes of $\mathbf{H} = \mathbf{L}^{-1}$ are then defined as the solutions to the Euler-Lagrange equations given by
\begin{equation}\label{EL1}
    \frac{\delta \mathcal{L}}{\delta \mathbf{q}} = \mathbf{L}^H \mathbf{Q}_a\mathbf{L} \mathbf{q} - \sigma^{-2}\mathbf{Q}_b \mathbf{q} = 0.
\end{equation}
Equation \ref{EL1} constitutes an eigenvalue problem and thus has a countably infinite set of solutions which we index by the subscript $j$. 
\begin{equation}\label{ELj}
    \mathbf{L}^H \mathbf{Q}_a\mathbf{L} \boldsymbol{\psi}_j = \sigma_j^{-2}\mathbf{Q}_b \boldsymbol{\psi}_j
\end{equation}
We have denoted the eigenvalue $\sigma_j^{-2}$ and the eigenfunctions $\boldsymbol{\psi}_j$ such that the singular values and resolvent response modes of $\mathbf{H}$ are given by $\sigma_j$ and $\psi_j$ respectively. The resolvent forcing modes are recovered through
\begin{equation} \label{RA_definition_of_forcing}
    \boldsymbol{\phi}_j = \sigma_j\mathbf{L}\boldsymbol{\psi}_j.
\end{equation}
Note that the $\boldsymbol{\psi}_j$ are guaranteed to be orthogonal since the matrices in (\ref{ELj}) are Hermitian, and the $\boldsymbol{\phi}_j$ are orthogonal w.r.t $\mathbf{Q}_a$ since
\begin{equation}
    \boldsymbol{\phi}^H_i \mathbf{Q}_a \boldsymbol{\phi}_j = \sigma_i \sigma_j \boldsymbol{\psi}^H_i \mathbf{L}^H\mathbf{Q}_a\mathbf{L} \boldsymbol{\psi}_j = \sigma_i \sigma^{-1}_j \boldsymbol{\psi}^H_i \mathbf{Q}_b \boldsymbol{\psi}_j = \delta_{ij}.
\end{equation}
% Proof of equivalence subsection
\subsection{Proof of equivalence}
We will now illustrate the equivalence of (\ref{ELj}) to the standard SVD-based definition. For simplicity we consider the case where $\|\mathbf{x}||_a=\|\mathbf{x}||_b$. Again, following the notation of \citet{herrmann_data-driven_2021}, the SVD of the properly weighted resolvent operator is given by
\begin{equation}\label{Hf}
    \mathbf{H}_F \equiv \mathbf{F}\mathbf{H}\mathbf{F}^{-1} = \boldsymbol{\Psi}_F \boldsymbol{\Sigma} \boldsymbol{\Phi}^H_F.
\end{equation}
The physical resolvent forcing and response modes are then recovered by left multiplication by $\mathbf{F}^{-1}$, such that $\boldsymbol{\Phi} = \mathbf{F}^{-1}\boldsymbol{\Phi}_F$ and  $\boldsymbol{\Psi} = \mathbf{F}^{-1}\boldsymbol{\Psi}_F$, whose columns give the individual modes $\boldsymbol{\phi}_j$ and $\boldsymbol{\psi}_j$ respectively. We focus first on the resolvent response modes $\boldsymbol{\Psi}$. Beginning from the definition of the weighted resolvent we can write
\begin{equation}
    \mathbf{H}_F\mathbf{H}^H_F = \boldsymbol{\Psi}_F \boldsymbol{\Sigma}^2 \boldsymbol{\Psi}^H_F.
\end{equation}
Next we use (\ref{H}), (\ref{Q}), and (\ref{Hf})  to write the above expression in terms of the linear operator $\mathbf{L}$,
\begin{equation}
    \mathbf{F}\left(\mathbf{L}^{-1} \mathbf{Q}^{-1} \mathbf{L}^{-H}\right)\mathbf{F}^H  = \boldsymbol{\Psi}_F \boldsymbol{\Sigma}^2 \boldsymbol{\Psi}^H_F.
\end{equation}
Taking the inverse of both sides and noting the unitary nature of $\boldsymbol{\Psi}_F$ we find
\begin{equation}
    \mathbf{F}^{-H}\left(\mathbf{L}^H \mathbf{Q} \mathbf{L} \right)\mathbf{F}^{-1} = \boldsymbol{\Psi}_F \boldsymbol{\Sigma}^{-2} \boldsymbol{\Psi}^H_F.
\end{equation}
Finally, we right multiply by $\boldsymbol{\Psi}_F$ and left multiply by $\mathbf{F}^H$ to arrive at
\begin{equation}
    \left(\mathbf{L}^H \mathbf{Q} \mathbf{L} \right)\boldsymbol{\Psi} = \mathbf{Q} \boldsymbol{\Psi} \boldsymbol{\Sigma}^{-2}
\end{equation}
which is equivalent to (\ref{ELj}). Again, the resolvent forcing modes are then recovered through 
\begin{equation}
    \boldsymbol{\Phi} = \mathbf{L}\boldsymbol{\Psi}\boldsymbol{\Sigma}.
\end{equation}
This establishes the equivalence of the variational and SVD-based definitions of resolvent modes. We would like to emphasize that a consequence of this equivalence is that the completeness property of the SVD-based basis also applies to the variational computed basis.
The case where  $\|\mathbf{x}||_a \neq \|\mathbf{x}||_b$ follows similar arguments but for the sake of brevity is not included here. 
\subsection{Resolvent Mode Estimation}

In general, the Euler-Lagrange equations (\ref{ELj}) are both analytically intractable and computationally intensive for complex flows with multiple non-homogeneous spatial dimensions. However, the variational definition introduced above provides a convenient way to estimate resolvent modes as an expansion in any convenient basis. Suppose we wish to estimate the resolvent response modes of some system $\boldsymbol{\psi}(\mathbf{x})$, and let $\mathbf{q}_j(\mathbf{x})$ with $(j =1...r)$ be some known basis defined on the same domain. We can then write the resolvent response modes as an expansion in this basis.
\begin{equation}
    \boldsymbol{\psi} = a_j \mathbf{q}_j
\end{equation}
Inserting this expansion into (\ref{lag1}) transforms the continuous vector field $\mathbf{q} \in \mathrm{C}^{\infty}$ into a discrete field $\mathbf{a} \in \mathrm{C}^{r}$, where $\mathbf{a}$ is the vector of amplitudes $a_j$. The Euler-Lagrange equation (\ref{EL1}) takes the form
\begin{equation}\label{EVP_basic}
    \mathbf{M}\mathbf{a} - \sigma^{-2} \mathbf{Q}\mathbf{a} = 0,
\end{equation}
where $\mathbf{M},\mathbf{Q} \in \mathrm{C}^{r\times r}$, $M_{ij} \equiv   \mathbf{q}_i^H \mathbf{L}^H \mathbf{Q}_a \mathbf{L} \mathbf{q}_j $, and $Q_{ij} \equiv  \mathbf{q}_i^H \mathbf{Q}_b \mathbf{q}_j$. 
The eigenvectors $\mathbf{a}$ contain the amplitudes $a_j$ which optimally approximate the resolvent response modes and the $\sigma$ are the approximate singular values. For $r$ basis elements we will have $\mathbf{M}, \mathbf{Q} \in \mathbb{C}^{r \times r}$ and thus we will obtain $r$ eigenvalue/eigenvector pairs, representing $n$ singular mode/singular value pairs. The necessary $r$ depends on both the efficiency of the model basis and the desired level of accuracy. However, we show in the following examples that for large systems a reduction over the dimension of the original system by up to two orders of magnitudes is possible due to the lack of matrix inversion. Throughout the paper we use $r$ to refer to the size of the reduced system (\ref{EVP_basic}) and $n$ to refer to the size of the original system.

%% Subsection kx = 0 eigenfunction expansion
\section{1D resolvent analysis: turbulent channel flow}\label{sec:kx0}
\subsection{The Orr-Sommerfeld Squire System}
As a first example we consider the incompressible linearized NSE for streamwise constant fluctuations about a turbulent mean in a wall bounded shear flow. This example illustrates the fundamental theory and highlights the analytical manipulation enabled by the VRA framework. The equations are nondimensionalized using the channel half-height and friction velocity. A Fourier transform in the homogeneous spatial directions and time results in a system parametrized by the Reynolds number, $R$, and the wave number triplet, $\mathbf{k} = [k_x,k_z,\omega]$. Here $k_x$ and $k_z$ denote the wavenumbers in streamwise and spanwise directions respectively, and $\omega$ again represents the temporal frequency. We focus on streamwise constant fluctuations which are useful models of the streamwise elongated structures known to play a crucial role in the sustenance of turbulence \citep{jimenez_minimal_1991}. Therefore, for the remainder of \S\ref{sec:kx0} we assume $k_x = 0$. 

The forced linearized NSE can be written as 
\begin{equation}\label{OS_SQ_sys}
 \begin{bmatrix}
L_{OS} & 0\\
\bar{U}_y & L_{SQ}
\end{bmatrix}
 \begin{bmatrix} 
v(y)\\
u(y)
\end{bmatrix}
=
 \begin{bmatrix}
g_v(y)\\
g_u(y)
\end{bmatrix}.
\end{equation}
Here $y \in [-1,1]$ and $[v,u]$ are the wall-normal and streamwise velocity fluctuations about the streamwise, spanwise, and temporal averaged mean velocity $\bar{U}$. The spanwise velocity is recovered through the continuity equation as $w = i k_z^{-1} v_y$. The right hand side $[g_v,g_u]^T$ represents an unknown forcing. The relevant boundary conditions are thus $v(\pm1) = v_y(\pm1) = u(\pm1) = 0$. Note that we write (\ref{OS_SQ_sys}) in terms of $u$ instead of the classical formulation in terms of the wall normal vorticity $\eta = i k_z u - i k_x w$, since if $k_x = 0$ then $u \sim \eta$. Note that this implies that the off-diagonal term in (\ref{OS_SQ_sys}) does not include the $i k_z$ present in more classical formulations. The Orr-Sommerfeld and Squire operators in (\ref{OS_SQ_sys}) simplify to

\begin{equation}\label{Los}
L_{OS} = -i\omega\nabla^2 - \frac{1}{R}\nabla^4
\end{equation}
\begin{equation}\label{Lsq}
L_{SQ} = -i\omega - \frac{1}{R}\nabla^2
\end{equation}
% \begin{equation}\label{Los}
% L_{OS} =  L_{SQ}\nabla^2,
% \end{equation}
where $\nabla^2 \equiv \partial_{yy} - k_z^2$. The inner product defining the kinetic energy norm is
\begin{equation}
    \langle \mathbf{q}_i,\mathbf{q}_j \rangle_{KE} \equiv \langle v^*_i v_j + k_z^{-2} v^*_{i,y} v_{j,y} + u^*_i u_j\rangle
\end{equation}
% \begin{equation}
%     \|\cdot\|_{KE} \equiv \langle v^2 + k_z^{-2} v_{y}^2 +u^2 \rangle
% \end{equation}
where $\langle f(y) \rangle \equiv \int_{-1}^{1}f(y)dy$, $\mathbf{q} = [v,u]$, and $\|\mathbf{q}\|_{KE} = \sqrt{\langle \mathbf{q},\mathbf{q} \rangle}_{KE} $. It is convenient to also define the following norm associated with the OS operator induced by 
\begin{equation}
    \langle \cdot \rangle_{OS} \equiv \langle v^*_i v_j + k_z^{-2} v^*_{i,y} v_{j,y} \rangle
\end{equation}
% \begin{equation}
%     \|\cdot\|_{OS} \equiv \langle v^2 + k_z^{-2} v_{y}^2  \rangle
% \end{equation}
which represents the contribution of $v$ (and thus $w$) to the kinetic energy and where again the norm is defined as $\|v\|_{OS} = \sqrt{\langle v,v \rangle}_{OS} $.
Lastly, it is numerically convenient to implement (\ref{OS_SQ_sys}) as  
\begin{equation}\label{OS_SQ_sys_num}
 \begin{bmatrix}
\nabla^{-2}L_{OS} & 0\\
\bar{U}_y & L_{SQ}
\end{bmatrix}
 \begin{bmatrix} 
v\\
u
\end{bmatrix}
=
 \begin{bmatrix}
\nabla^{-2}g_v\\
g_u
\end{bmatrix}
=
 \begin{bmatrix}
\tilde{g}_v\\
g_u
\end{bmatrix}.
\end{equation}
In order to compare our variational results to the direct SVD we use the definition (\ref{OS_SQ_sys_num}) going forward.

\subsection{The Orr-Sommerfeld and Squire Families}
It is instructive to decompose the system into the Orr-Sommerfeld (OS) and Squire (SQ) families of modes as suggested by \citet{rosenberg_efficient_2019}. The OS family corresponds to the forced response due to $g_v$,
\begin{equation}\label{OS_sys}
 \begin{bmatrix}
\nabla^{-2}L_{OS} & 0\\
\bar{U}_y & L_{SQ}
\end{bmatrix}
 \begin{bmatrix} 
v^{OS}\\
u^{OS}
\end{bmatrix}
=
 \begin{bmatrix}
\tilde{g}_v\\
0
\end{bmatrix}
\end{equation}
which upon elimination of $\tilde{g}_v$ from the equation for $u^{OS}$ results in a decoupled system reminiscent of the classical OS/SQ decomposition of linear stability theory \citep{drazin_hydrodynamic_2004,schmid_stability_2001}.
\begin{equation}\label{Lvg}
   \nabla^{-2} L_{OS}v^{OS} = \tilde{g}_v
\end{equation}
\begin{equation}\label{uUv}
    L_{SQ}u^{OS} = - \bar{U}_y v^{OS}
\end{equation}
The SQ family of modes, on the other hand, is the forced response to $g_u$, where by construction $v^{SQ} = 0$.
\begin{equation}\label{SQ_sys}
L_{SQ} u^{SQ} = g_{u}
\end{equation}
Since (\ref{SQ_sys}) is a normal scalar operator, the resolvent forcing and response modes are proportional to the eigenmodes of $L_{SQ}$, and the singular values are equal to the inverse of the norm of the eigenvalues of $L_{SQ}$.
\begin{equation} \label{psiSQj}
\psi^{SQ}_{j}(y)  =  \left[0,\sin \left(\frac{j\pi}{2}(y+1)\right)\right]
\end{equation}
\begin{equation} \label{phiSQj}
\phi^{SQ}_{j}(y)  = \left[0,e^{i \arctan \left(\frac{-4 R \omega}{\pi^2j^2 + 4k_z^2} \right)} \sin \left(\frac{j\pi}{2}(y+1)\right)\right]
\end{equation}
\begin{equation} \label{sigmaSQj}
\sigma^{SQ}_j = \left(\frac{1}{16 R^2}\left(\pi^2j^2 + 4k_z^2\right)^2 + \omega^2\right)^{-1/2}
\end{equation}

The problem thus reduces to finding the OS family of modes associated with (\ref{OS_sys}), which in accordance with \S\ref{sec:math}, are defined as the stationary points of the associated Lagrangian
\begin{equation}\label{lagkx0_1}
\mathcal{L}\left(\mathbf{q}^{OS}\right) = \|\nabla^{-2}L_{OS}v^{OS}||^2_{OS} - \sigma^{-2}\|\mathbf{q}^{OS}||_{KE}^2 
\end{equation}
where $\mathbf{q}^{OS}\equiv [v^{OS},u^{OS}]$ and we have made use of the fact that $g_u=0$ to simplify the operator norm. In order to eliminate the streamwise velocity $u^{OS}$ we expand the solution to (\ref{uUv}) in eigenfunctions of $L_{SQ}$ given by (\ref{psiSQj}).
\begin{equation} \label{uos}
    u^{OS} = -\frac{1}{\lambda^{SQ}_n} \langle v^{OS} \bar{U}_y u^{SQ}_n \rangle u^{SQ}_n
\end{equation}
This allows us to write the kinetic energy constraint as
\begin{equation}\label{normnse}
\|\mathbf{q}\|^2_{KE} = \langle |v|^2 + k_z^{-2} |v_y|^2 + |u(v)|^2\rangle = \|v\|^2_{KE}
\end{equation}
where the third term $u(v)^2$ is given by the square of (\ref{uos}). This allows us to rewrite (\ref{lagkx0_1}) as 
\begin{equation}\label{lagkx0_2}
\mathcal{L}\left(v^{OS}\right) = \|\nabla^{-2}L_{OS}v^{OS}||^2_{OS} - \sigma^{-2}\|v^{OS}||_{KE}^2 
\end{equation}
with associated Euler-Lagrange equation
\begin{equation}\label{ELkx0_2}
\frac{\delta}{\delta v }\left( \|\nabla^{-2}L_{OS}v^{OS}||^2_{OS} - \sigma^{-2}\|v^{OS}||_{KE}^2 \right) = 0.
\end{equation}

\begin{figure}
\includegraphics[trim = 60 0 0 0, scale=0.38]{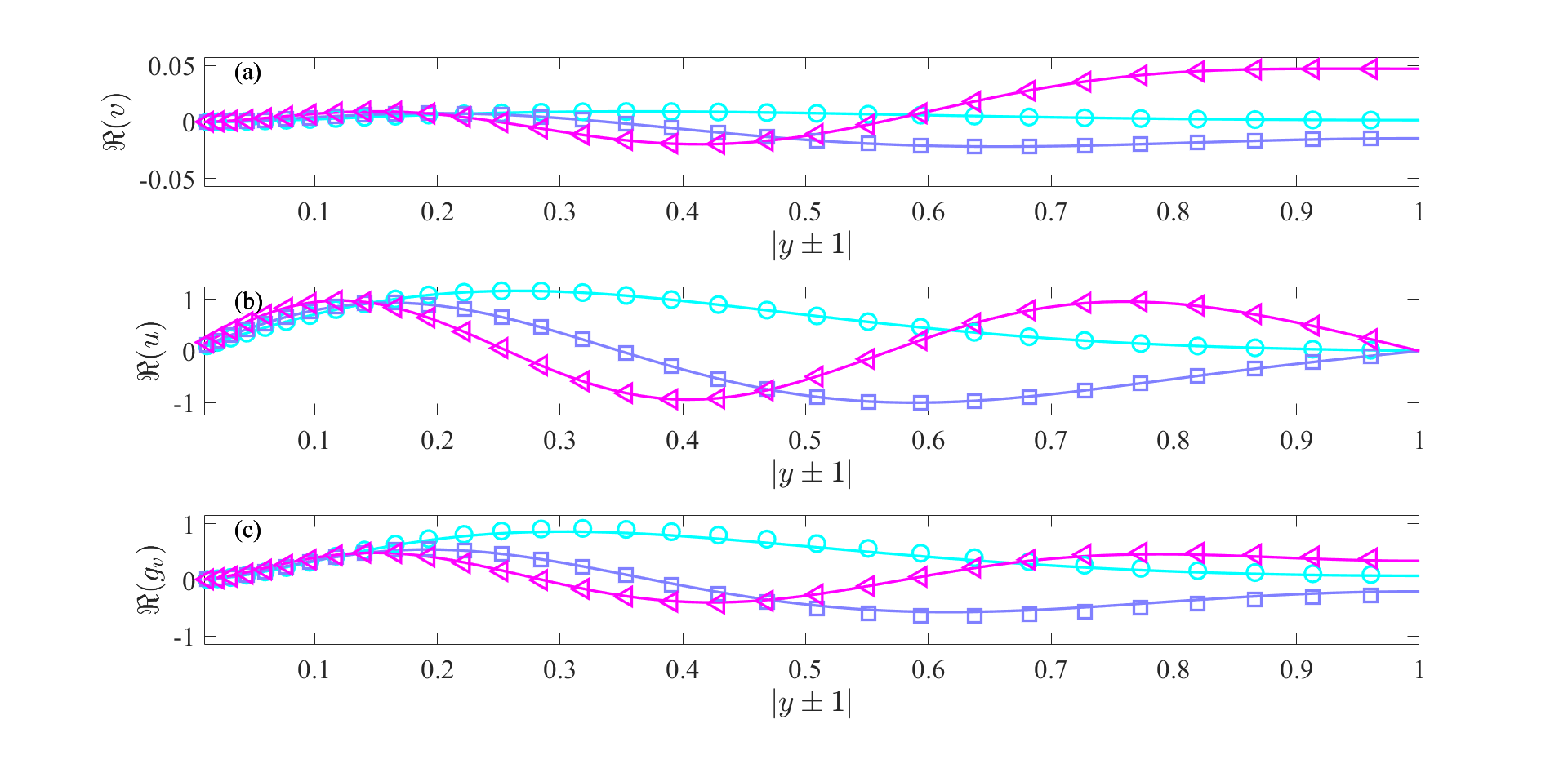}
\caption{Real part of the wall normal component $v$ (a), streamwise component $u$ (b), and forcing $g_v$ (c) of the of the $1^{st}$, $3^{rd}$, and $5^{th}$, Orr-Sommerfeld family of resolvent modes. Real part (left column), imaginary part (right column). Reference modes computed via direct SVD are shown in solid lines, VRA reconstruction using $20$ basis eigenmodes is shown in symbols. $k_z = 6$,  $\omega = 0.1$, and $R=1000$. } \label{fig:singularmodes_kx0}
\end{figure}

\begin{figure}
\includegraphics[trim = 40 0 0 0, scale=0.38]{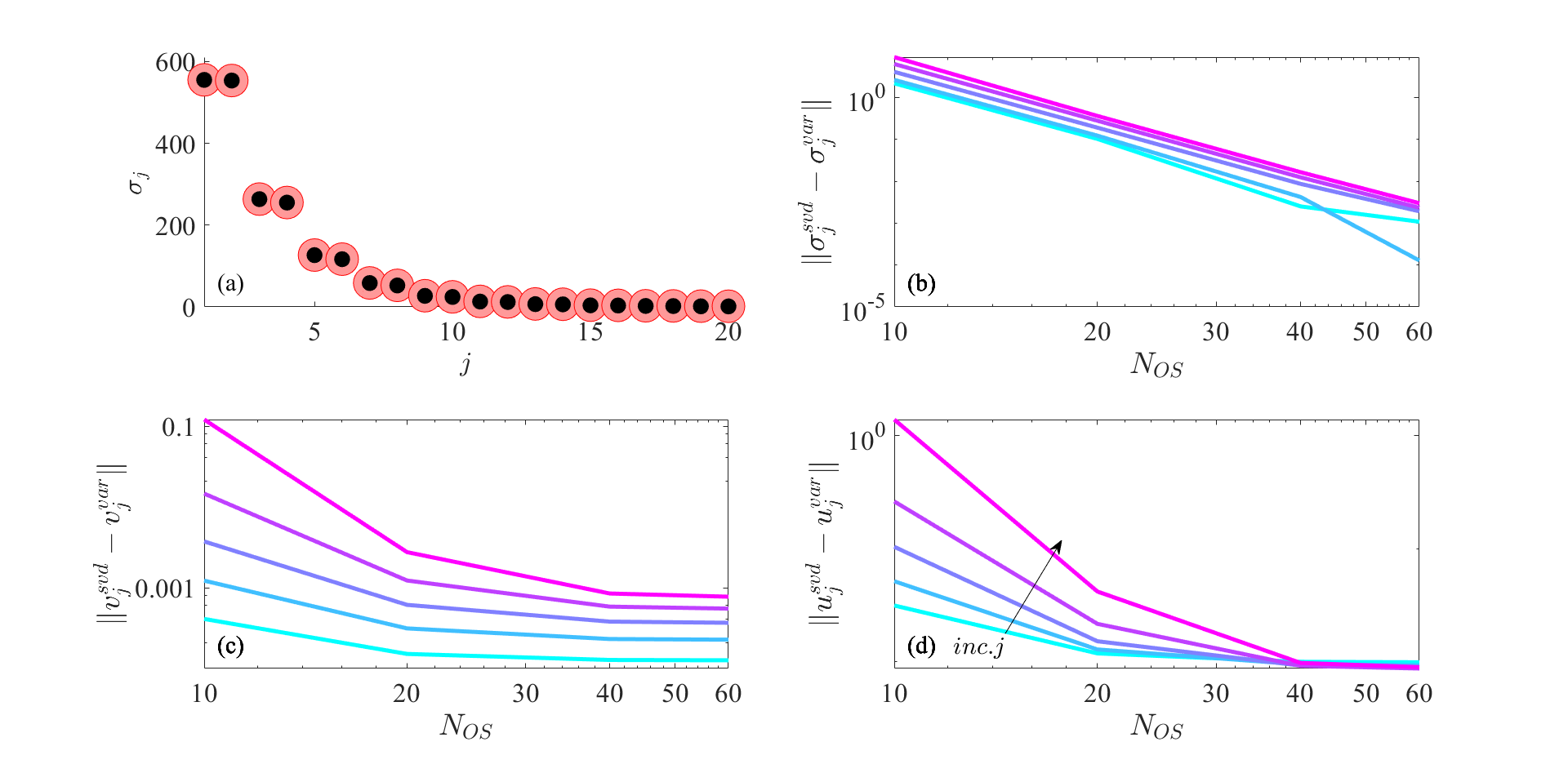}
\caption{Orr-Sommerfeld family of singular values (a) with reference values computed via direct SVD in red, and  variational reconstruction using $20$ basis eigenmodes in black. Error in variational reconstruction as a function of basis elements in $\sigma_j$  (b), $v_j$ (c), and $u_j$ (d) of $ \boldsymbol{\psi}_j$ for $j =1,3,5,7,9$.  $k_z = 6$, $\omega = 0.1$, and $R=1000$ as a function of the retained basis elements $ r = N_{OS}$.}
\label{fig:error_kx0}
\end{figure}

For $k_x = 0$, the eigenfunctions of $L_{OS}$ may also be derived analytically \citep{dolph_application_1958,jovanovic_componentwise_2005}. Using standard methods they are found to be
\begin{equation}\label{vj}
\begin{aligned}
v_j(y;k_z)  =  A_j\left[\cos\left(\gamma_j\left(y+1\right)\right)-\cosh\left(k_z\left(y+1\right)\right)\right] + \\      B_j\left[\sin\left(\gamma_j\left(y+1\right)\right)-\gamma_j k_z^{-1}\sinh\left(k_z\left(y+1\right)\right)\right]
\end{aligned}
\end{equation}
\begin{equation}
\lambda^{OS}_j =\frac{1}{R}\left(\gamma_j^2 + k_z^2\right) - i\omega 
\end{equation}
where $A_j, B_j$ and $\gamma_j$ are defined in appendix \ref{app:LOS} and satisfy $L_{OS} v_j = \lambda^{OS}_j \nabla^2 v_j$ and $\langle v_i, v_j \rangle_{OS} = \delta_{ij}$. Expanding the solution to (\ref{ELkx0_2}) in the basis of OS eigenfunctions (\ref{vj}) such that
\begin{equation}
    v^{OS}=a_m v_m  
\end{equation}
allows us to transform the variation into an optimization over the coefficients $a_j$. 
\begin{equation}
\frac{\partial}{\partial a} \left(\|\nabla^{-2}L_{OS} a_j v_j\|^2_{OS} - \sigma^{-2}( \|a_j v_j\|^2_{KE} - 1)\right) = \frac{\partial}{\partial a}\left(|\lambda^{OS}_j|^2a_j ^2 - \sigma^{-2} a_i a_j(\delta_{ij} + U_{in}U^H_{nj})\right) = 0
\end{equation}
Here the quantity $U_{in}$ represents the projection of the OS eigenfunctions onto the SQ eigenfunctions through (\ref{uos}).
\begin{equation}
    U_{in}  \equiv -\frac{1}{\lambda^{SQ}_n} \langle v_i \bar{U}_y u^{SQ}_n \rangle
\end{equation}
Upon carrying out the above differentiation with respect to $a$ we find the eigenvalue problem
\begin{equation}\label{eigenProblem}
 \|\boldsymbol{\Lambda}^{OS}\|^2\mathbf{a} =\sigma^{-2}  \left(\mathbf{I}+\mathbf{U}\mathbf{U}^H \right)\mathbf{a}
\end{equation}
where $\boldsymbol{\Lambda}_{ij}^{OS} = |\lambda^{OS}_i|^2\delta_{ij}$. The eigenvectors $\mathbf{a}$ correspond to the the coefficients which optimally represent the resolvent response modes of the system (\ref{OS_sys}) as a linear combination of the eigenbasis (\ref{vj}). 
\begin{equation}\label{psiOSj}
    \boldsymbol{\psi}^{OS}_j = [a^j_m v_m, u(a^j_m v_m)]
\end{equation}
The singular values $\sigma_j$ are given by the eigenvalues of (\ref{eigenProblem}) and the forcing modes are recovered through
\begin{equation}\label{phiOSj}
    \boldsymbol{\phi}^{OS}_j = [\sigma_j \nabla^{-2} L_{OS}v_j^{OS}, 0] = [\sigma_j \lambda^{OS}_m a^j_m v_m, 0].
\end{equation}
Together with Squire family of resolvent modes (\ref{psiSQj}-\ref{sigmaSQj}) the Orr-Sommerfeld family given by (\ref{psiOSj}) and (\ref{phiOSj}) fully describe the resolvent basis. In Figures \ref{fig:singularmodes_kx0} we plot the real part of the variationally reconstructed Orr-Sommerfeld response and forcing modes along side their numerically computed counterparts for the wave number triplet $[k_x,k_z,\omega] = [0,6,0.1]$ and $R = 1000$. The singular values plots are plotted in \ref{fig:error_kx0}a. For this example, the VRA model uses $r=N_{OS}=20$ basis elements, this value is chosen to show a balance between the accuracy and model reduction capabilities of the method. Although for this example the computational cost is trivial, the reduction in size of the relevant matrices and avoiding the need for matrix inversion reduces the computation time by two orders of magnitude. To quantify the convergence of our method we plot in Figure \ref{fig:error_kx0} the error in the VRA reconstruction of $v$, $u$, and $\sigma$ as a function of the number of retained OS eigenfunctions $(r=N_{OS})$ included in the variational reconstruction. The error is defined as
\begin{equation}
    e_q = \sqrt{\int_{-1}^{1}|q_{svd} - q_{var}|^2dy}
\end{equation}
where $q = u,v$ and the subscripts $_{var}$ and $_{svd}$ denote the quantities computed using the VRA model and direct SVD respectively. In all cases we observe monotonic convergence. In this this example the VRA model is extremely effective at reconstructing the results of the direct SVD since our model basis exactly spans the range of $L_{OS}$.

%% Subsection analytical approximation
%% ORIGINAL DERIVATION OF EULER LAGRANGE EQUATIONS
\subsection{Analytical approximation of $\mathbf{\psi}_1$}\label{sec:analytical approximations}
In this section we demonstrate how, under certain assumptions, the variational resolvent formulation allows for the analytical approximation of the leading OS resolvent mode $\boldsymbol{\psi}^{OS}_1$. Written explicitly, the Lagrangian associated with  (\ref{OS_SQ_sys}) is 
\begin{equation}
    \mathcal{L}(v) = \left(\omega^2\left(\nabla^2v \right)^2 + \frac{1}{R^2}\left(\nabla^4v \right)^2\right) + \frac{1}{k_z^2}\left(\omega^2\left(\nabla^2v_y \right)^2  + \frac{1}{R^2}\left(\nabla^4v_y \right)^2\right) -\frac{1}{\sigma_1^2}\left(  u(v)^2 + v^2 + k_z^{-2}v_y^2\right)
\end{equation}
where $u$ is the solution to
\begin{equation} \label{ELV1u}
    -i \omega u - \frac{1}{R}\nabla^2u = -\bar{U}_y v.
\end{equation}
Here $u$ and $v$ are the streamwise and wall normal components of $\boldsymbol{\psi}^{OS}_1$ and $\sigma_1$ is the leading OS singular value. The associated Euler-Lagrange equation written in terms of $v$ is then
\begin{equation}\label{ELV1}
    \frac{1}{k^2_z}\left(\frac{1}{R^2}\nabla^{10} +\omega^2\nabla^6\right) v + \frac{1}{\sigma_1^2}\left(  \left(L_{SQ}^{-1}\bar{U}_y\right)^H L_{SQ}^{-1} \bar{U}_y v-\frac{1}{k_z^2}\nabla^2 v  \right) = 0
\end{equation}
with boundary conditions $v(\pm1)=v_y(\pm1)=u(\pm1)=0$. Note that we use the the original definition (\ref{OS_SQ_sys}) not the numerical implementation (\ref{OS_SQ_sys_num}) to derive (\ref{ELV1}). This is done to avoid the analytically cumbersome treatment of the $\nabla^{-2}$ operator. The problem is now parameterized by $\omega$, $R$, and $k_z$. Our analysis will consider the appropriate limits of each in turn.

It has been shown that for $k_x=0$ the most linearly amplified frequency is $\omega = 0$, therefore we will consider the limit as $\omega \rightarrow 0$. Since, in this limit (\ref{ELV1}) is regularly perturbed problem, the leading order solution may be found by simply setting $\omega =0$. 
We may further simplify (\ref{ELV1}) by considering a high Reynolds number limit $R \rightarrow \infty$. Analysis of (\ref{OS_sys}) reveals that for $\omega = 0$, $\sigma_1 \sim R^2$ as $R\rightarrow \infty$ (see appendix \ref{app:sigma}). This allows us introduce the small parameter $\epsilon \equiv R^{-1}$ such that (\ref{ELV1}) and (\ref{ELV1u}) take the form
\begin{equation}\label{ELV3}
    \frac{1}{k^2_z}\nabla^{10}  v + \frac{\epsilon^2}{\tilde{\sigma}_1^2}\left(\frac{1}{\epsilon^2}\ \left( \nabla^{-2}\bar{U}_y\right)^H\nabla^{-2} \bar{U}_y v-\frac{1}{k_z^2}\nabla^2 v  \right) = 0
\end{equation}
\begin{equation} \label{ELV3u}
    \epsilon \nabla^2u_0 = \bar{U}_y v
\end{equation}
where $\tilde{\sigma} \neq f(R)$. We note that (\ref{ELV3u}) implies that $v \sim \epsilon u$ and expand the solution in an asymptotic series.
\begin{equation}
\begin{split}
    v & =  \epsilon v_1 + \epsilon^2 v_2 + \mathcal{O}(\epsilon^3) \\
    u & = u_0 + \epsilon u_1 + \epsilon^2 u_2 + \mathcal{O}(\epsilon^3)
\end{split}    
\end{equation}
The leading order solution to (\ref{ELV3}) and (\ref{ELV3u}) then satisfy
\begin{equation}\label{ELV4}
    \frac{1}{k^2_z}\nabla^{10}  v_1 + \frac{1}{\tilde{\sigma}_1^2}\left(\left( \nabla^{-2}\bar{U}_y\right)^H\nabla^{-2} \bar{U}_y v_1\right) = 0
\end{equation}
\begin{equation} \label{ELV4u}
    \nabla^2u_0 = \bar{U}_y v_1
\end{equation}
and the norm constraint takes the form
\begin{equation}\label{normcon4}
    \|u_0\|^2 = 1.
\end{equation}
In this work we focus on the leading order solution, and thus to avoid notational clutter we drop the subscripts $_0$ and $_1$ moving forward.

While we have managed to eliminate the nonlinearity, the second term in (\ref{ELV4}) remains prohibitive to analytical progress. In order to proceed we consider the $y\rightarrow-y$ symmetry of (\ref{OS_SQ_sys}) which dictates that the resolvent modes come in pairs, one of which is even about the center of the channel, and one of which is odd. If additionally, the modes have compact support, as is generally the case, we have $\boldsymbol{\psi}_1(y)=\boldsymbol{\psi}_1(-y)=\boldsymbol{\psi}_2(y)=-\boldsymbol{\psi}_2(-y)$, and therefore it is sufficient to solve for the mode shape in one half of the domain. 

We assume that $v$ is indeed locally supported and thus introduce the scaling $Y=k_z |y\pm1|$ under the assumption $k_z \gg 1$ and make the transformation $u(y),v(y)\rightarrow U(Y), V(Y)$. If we formally take the limit as $k_z \rightarrow \infty$ we may transform the domain from $y\in[-1,1]$, to the ``semi-infinite"  half channel: $Y\in[0,\infty]$ and recover the solution in the other half through $\boldsymbol{\psi}_1(y)=\boldsymbol{\psi}_1(-y)=\boldsymbol{\psi}_2(y)=-\boldsymbol{\psi}_2(-y)$.

Finally, in order to make progress we require some suitable approximation of the mean velocity profile. Since we are working within a high Reynolds number limit we choose to assume that the mean velocity obeys a logarithmic profile over the entirety of the semi-infinite domain.  This is a reasonable assumption since in high Reynolds number channel flow the log-law applies to a large fraction of the channel. Our approach thus implicitly assumes the support of the resolvent modes is localized within this region where the log-law approximation is valid. The mean shear is then given in our scaled variables by $\bar{U}_Y = k_z(\kappa Y)^{-1}$, where $\kappa$ is the Von Karman constant. We note that the mean shear diverges as like $Y^{-1}$ as $Y\rightarrow 0$, however, since $V(0) = V_Y(0) = 0$ we have $V(Y)\sim Y^2$ as $Y \rightarrow 0$, and thus the right hand side of (\ref{ELV4u}) remains bounded as $Y\rightarrow 0$.

Inspection of (\ref{ELV4u}) and (\ref{normcon4}) reveals that the appropriate scaling of the velocity components is given by $\tilde{U}(Y) = k_z^{1/2}U(Y)$ and $\tilde{V}(Y) = k_z^{3/2}V(Y)$. Additionally, we define the scaled Laplacian $\tilde{\nabla}^2\equiv \partial_{YY} -  1$ such that $\nabla^2 \rightarrow k_z^2\tilde{\nabla}^2$, and note that for $k_x = \omega =0$ and $k_z \rightarrow \infty$ the singular value scales as $\tilde{\sigma}_1 \sim k_z^{-3}$ (see appendix \ref{app:sigma}). Thus we can write (\ref{ELV4}) in our scaled variables as
\begin{equation}\label{ELV5}
    \nabla^{10}  \tilde{V} + \frac{1}{\kappa^2k_z^4\gamma_1^2}\left(\left( \tilde{\nabla}^{-2}Y^{-1}\right)^H\tilde{\nabla}^{-2} Y^{-1} \right)\tilde{V} = 0
\end{equation}
where $\gamma_1$ is a constant. We expand $\tilde{U}$ and $\tilde{V}$ in asymptotic series 
\begin{equation}
\begin{split}
    \tilde{V} & = \tilde{V}_0 + k_z^{-4} \tilde{V}_1 + \mathcal{O}(k_z^{-8}) \\
    \tilde{U} & = \tilde{U}_0 + k_z^{-4} \tilde{U}_1 + \mathcal{O}(k_z^{-8})
\end{split}    
\end{equation}
which upon substitution into (\ref{ELV5}) allows us to eliminate the norm constraint  at leading order and reduce the problem of deriving the leading OS resolvent mode to   
\begin{equation}\label{ELV6}
    \tilde{\nabla}^{10}  \tilde{V} = 0
\end{equation}
\begin{equation}\label{ELV6u}
    \tilde{\nabla}^{2}  \tilde{U} = \frac{1}{\kappa Y}\tilde{V}
\end{equation}
where we have again dropped the subscripts to avoid notational clutter. The relevant boundary conditions are $\tilde{V}(0)=\tilde{V}_Y(0)=\tilde{U}(0)=\tilde{V}(\infty) = \tilde{U}(\infty)=0$. The remaining constants of integration are then chosen such that $\|\nabla^{-2}L_{OS}V\|^2_{OS}$ is minimized and $\|U\|^2 = 1$. Here we choose to minimize $\|\nabla^{-2}L_{OS}V\|^2_{OS}$  instead of $\|L_{OS}V\|^2_{OS}$  in order to facilitate comparison with the numerically computed modes. However, we note that minimizing the latter functional leads to a very similar solution. Using standard methods the solutions satisfying the boundary conditions are found to be 
\begin{equation}\label{V1}
    V(Y) = \frac{k_z^{3/2}}{R}\left(a + b Y +c Y^2 \right)Y^2e^{-Y}
\end{equation}
\begin{equation}\label{U1}
    U(Y) = -\frac{k^{1/2}_z}{24 \kappa}\left( 3 c Y^3 + (4b+6c)Y^2 + (6a + 6b + 9c)\left(Y + 1\right)\right)Ye^{-Y}
\end{equation}
The three remaining constants of integration, $a,b,c$, are found by minimizing $\|\nabla^{-2}L_{OS}V\|^2_{OS}$ subject to the constraint $\|U\|^2= 1$. Straight forward integration results in 
\begin{equation}\label{normLv}
    \|\nabla^{-2}L_{OS}V\|^2_{OS} =\frac{1}{R^2}\|\nabla^{2}V\|^2_{OS} = 4\frac{k_z^6}{R^4}\left(6a^2 + 9b^2 + 36 bc + 72c^2\right)
\end{equation}
and
\begin{equation}\label{normU}
    \|U\|^2 = \frac{1}{\kappa^2}\left( \frac{7}{32}a^2 + \frac{1}{128}(112b +228c)a + \frac{31}{32}b^2 + \frac{351}{64}bc + \frac{1089}{128}c^2 \right) = 1.
\end{equation}
Minimizing (\ref{normLv}) subject to (\ref{normU}) results in the eigenvalue problem
\begin{equation}\label{analyticalEVP}
    \begin{bmatrix}
    12 & 0 & 0\\
    0 & 18 & 36\\
    0 & 36 & 144
    \end{bmatrix}
    \begin{bmatrix}
    a\\b\\c
    \end{bmatrix}
    = \frac{1}{\sigma_1^2}\left(\frac{R^4}{4\kappa^2k_z^{6}} \right)
    \begin{bmatrix}
        7/16 & 7/8 & 9/4\\
    7/8 & 31/16 & 351/64\\
    9/4 & 351/64 & 1089/64
    \end{bmatrix}
        \begin{bmatrix}
    a\\b\\c
    \end{bmatrix}.
\end{equation}
Assuming $\kappa = 0.4$ the minimizing solution that satisfies the norm constraint is found to be
\begin{equation}\label{coeffs}
    [a,b,c] = [0.1283, 0.1066, 0.0431]\frac{1}{\sqrt{2}} .
\end{equation}
The leading singular value is
\begin{equation}\label{sigma1}
    \sigma_1 =\|\nabla^{-2}L_{OS}\|_{OS}^{-1} = \frac{R^2}{2k_z^3}.
\end{equation}
The wall normal component ${g_v}$ of the optimal resolvent forcing mode $\boldsymbol{\phi}_1^{OS}$ is recovered through
\begin{equation}
    \nabla^{2} g_v = \sigma_1 L_{OS} v
\end{equation}
subject to the boundary conditions $g_v(\pm1)= 0$. Using (\ref{sigma1}) and letting $g_v(y) \rightarrow G_V(Y)$ this takes the form
\begin{equation}
    \tilde{\nabla}^2 G_v(Y) = \sigma_1 L_{OS} v = -\frac{R}{2 k_z} \tilde{\nabla}^4 V(Y).
\end{equation}
The solution satisfying the boundary condition $G_V(0) = 0$ is found to be
\begin{equation}\label{gv1}
    G_V(Y) = k_z^{1/2}\left(4c Y^3 +(3b - 6c)Y^2 + (2a -3 b)Y\right)e^{-Y}.
\end{equation}

\begin{figure}
    \centering
    \includegraphics[trim=80 0 0 0,  scale =  0.4]{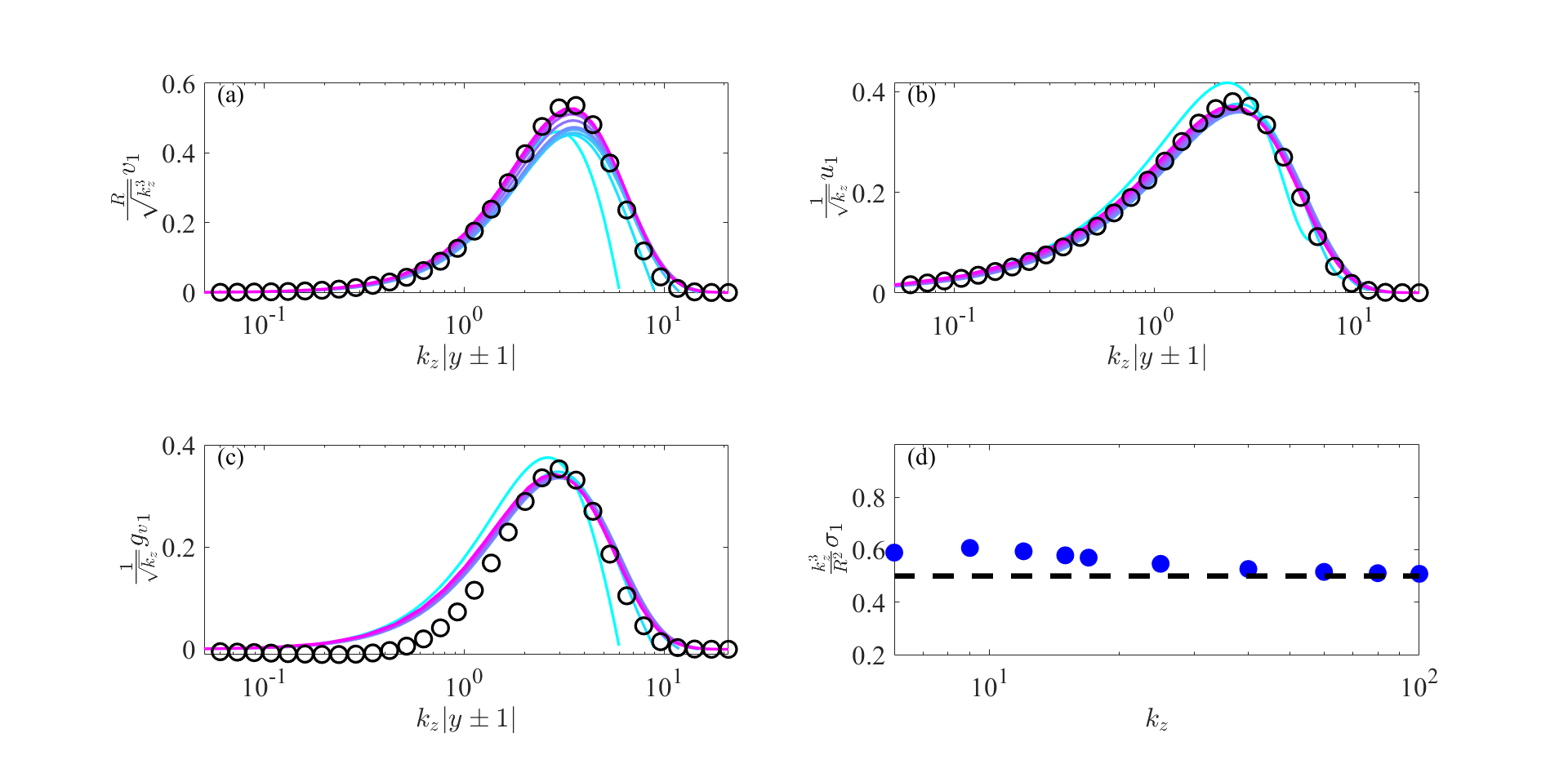}
    \caption{Optimal resolvent modes: $v$ (a), $u$ (b), $g_v$ (c) and singular value (d) for $k_x = 0$, $\omega = 0, R = 10,000$ and a range of $k_z$. Numerically calculated modes shown in colored lines, analytically derived modes are shown in the black open circles. From light to dark, colors indicate increasing $k_z$ from 6 to 100.}
    \label{fig:analytical_modes}
\end{figure}

The solutions (\ref{V1}), (\ref{U1}), and (\ref{gv1}) with optimal coefficients (\ref{coeffs}) are plotted in Figure \ref{fig:analytical_modes} alongside numerically computed resolvent modes for $R=10,000$ and $\omega = 0$ over a range of $k_z$. Note that for $k_x=0$, and $\omega\rightarrow 0$ the symmetries of (\ref{OS_SQ_sys_num}) result in numerical resolvent modes with constant arbitrary phase, which for ease of comparison we set to zero. With the exception of the $u$ component for the smallest wave number $(k_z = 6)$, the derived scaling laws lead to reasonable collapse in both the numerically computed resolvent response and forcing modes. As $k_z \rightarrow 1$ the assumption of local support in $y$ is no longer valid. In this limit $\boldsymbol{\psi}_1$ tends to have significant support at the channel center. 

%For $k_z > 6$ the collapse is best for the streamwise velocity $u$ and the wall-normal forcing $g_v$, with the wall-normal velocity $v$ displaying some scatter. 

For the response modes the analytically-derived mode accurately predicts the shape, amplitude, and localization of the numerically computed modes. The analytical prediction of the wall normal velocity is most accurate for the largest wave numbers, tending to slightly over predict the amplitude of the smaller wave number modes. This is most likely due to the fact that the amplitude of $v$ is smaller by a factor of $R=10,000$ and is thus susceptible to some numerical uncertainty since it does not meaningfully contribute to the norm. The streamwise velocity more closely obeys the derived scaling laws, and thus the analytical model accurately predicts the shape of the numerically computed modes for all $k_z >6$.

The prediction of the forcing mode is slightly less accurate. While we capture the location and amplitude of the peak, the model underpredicts the true mode closer to the wall. The discrepancy in the forcing despite accurate reconstruction of the response is due to the sensitivity of the action of linear operator $f = L_{OS}v$ to perturbations in the argument $v$. This is discussed in detail in \S\ref{sec:direction}. 

Finally, in Figure \ref{fig:analytical_modes} we also plot the numerically computed leading singular values along side the analytical prediction (\ref{sigma1}). While the analytically obtained value of $\sigma_1$ slightly under-predicts the true singular values for the smaller values of $k_z$, the numerical singular values do converge to the analytical prediction with increasing $k_z$, consistent with the assumption made in our model that $k_z\gg1$. This under prediction is consistent with the fact that the true singular value represents the global maximum gain.

% \FloatBarrier

%% ECS SECTION
\FloatBarrier
\section{2D resolvent analysis: periodic mean flow}\label{sec:ECS}
\begin{figure}
    \centering
    \includegraphics[trim = 40 0 0 0,  scale =  0.35]{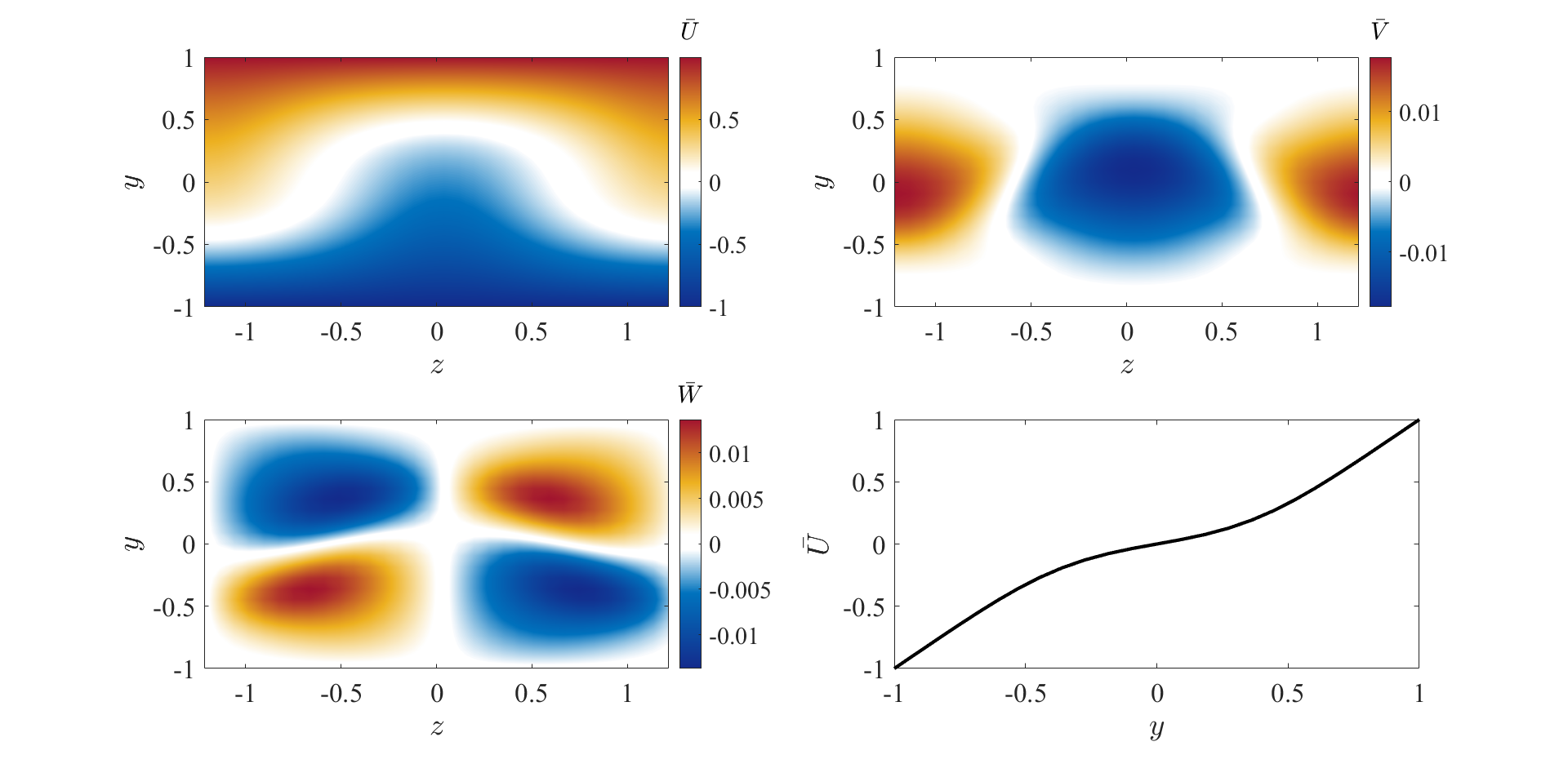}
    \caption{Exact coherent state EQ1 at $R=400$ used to compute 2D/3C resolvent modes. Clockwise from top left: $U(y,z)$, $V(y,z)$, $W(y,z)$, and spanwise average $\bar{U}(y)$ used to compute the 1D basis modes.}
    \label{fig:ECSmean}
\end{figure}

In this section we use VRA to efficiently and accurately compute resolvent modes about a 2D/3C mean flow. We consider the equilibrium solution EQ1 found in plane Couette flow by \citet{nagata_three-dimensional_1990}. The data was obtained from the open-source database \textit{channelflow.org} \citep{gibson_visualizing_2008,gibson_spectral_2014}. In this case the flow has two non-homogeneous spatial dimensions, the wall normal direction $y\in [-1,1]$ and the spanwise direction $z\in [-L_z/2,L_z/2]$ with $L_z = 0.8\pi$. The spanwise periodic EQ1 solution is shown in Figure \ref{fig:ECSmean}. The 2D/3C resolvent modes computed about this flow are then parameterized by the streamwise wavenumber and frequency pair, $[k_x,\omega]$. We choose as our modeling basis the local 1D resolvent modes about the mean flow $\bar{U}(y)$ given by the spanwise average of the EQ1 solution: $q_j(y,z) = \boldsymbol{\psi}^{1D}_j(y;k_z,k_x,c)e^{i k_z z}$. In other words, we seek to approximate the 2D/3C resolvent modes from the 1D resolvent basis as
\begin{equation}\label{ECSexpansion}
    \boldsymbol{\psi}^{2D}(y,z;k_x,c) = a_j q_j(y,z).
\end{equation}
The expansion coefficients $a_j$ are found by solving the eigenvalue problem
\begin{equation}\label{ECS_EVP}
    \mathbf{M}\mathbf{a} -  \sigma^{-2}\mathbf{Q}\mathbf{a} = 0
\end{equation}
where $M_{i,j} = \langle \mathbf{L}^{2D}q_i, \mathbf{L}^{2D}q_j\rangle$, and $Q_{i,j} = \langle q_i, q_j\rangle$. The operator $\mathbf{L}^{2D}$ is the NS operator, in velocity-vorticity form, linearized about the 2D/3C mean flow, the details of which are discussed in \citet{rosenberg_computing_2019}. The operator is discretized in $N_y = 33$ Chebychev points in the wall normal direction, and $N_z = 32$ linearly spaced points in the spanwise direction, for a total of $N_{2D} = 2\times N_y \times N_z = 2112$ degrees of freedom. 

The 1D resolvent modes are computed for the same $k_x$ as the 2D modes, and a range of $N_c = 3$ linearly spaced wavespeeds $0.8 c \leq c^{1D} \leq 1.2 c$ where $c = \omega/k_x$. We use a range of $c^{1D}$ since the 2D mode is expected to be localized near but not necessarily exactly at the critical layer where $c=\bar{U}(y)$. To account for the variation in $z$ we include a range of $N_{k_z} = 11$ spanwise wavenumbers $k_z = [-5...0...5]\times2\pi/L_z$. We found that increasing the number of retained harmonics beyond this range did not meaningfully change the results. At each wave number triplet $[k_x,k_z,c]$ we include $N_{SVD} = 8$ resolvent modes, resulting in a total of $r = N_c \times N_{k_z} \times N_{SVD} = 254$ degrees of freedom. These values were chosen to demonstrate a balance between accuracy and the cost saving potential of the proposed method (The reader is referred to Appendix~\ref{app:inputbasis} for an illustration of some representative basis elements). Once $\mathbf{L}^{2D}$ is known, the construction of the matrices $\mathbf{M}$ and $\mathbf{Q}$ takes approximately 0.5 seconds and the associated eigendecomposition takes approximately 0.01 seconds on a personal laptop. Meanwhile, the inversion and direct truncated SVD of the original system takes approximately 5 seconds using the built in Matlab functions $mldivide()$ and $svds()$.

\begin{figure}
    \centering

  \includegraphics[trim = 100 0 0 0,  scale =  0.37]{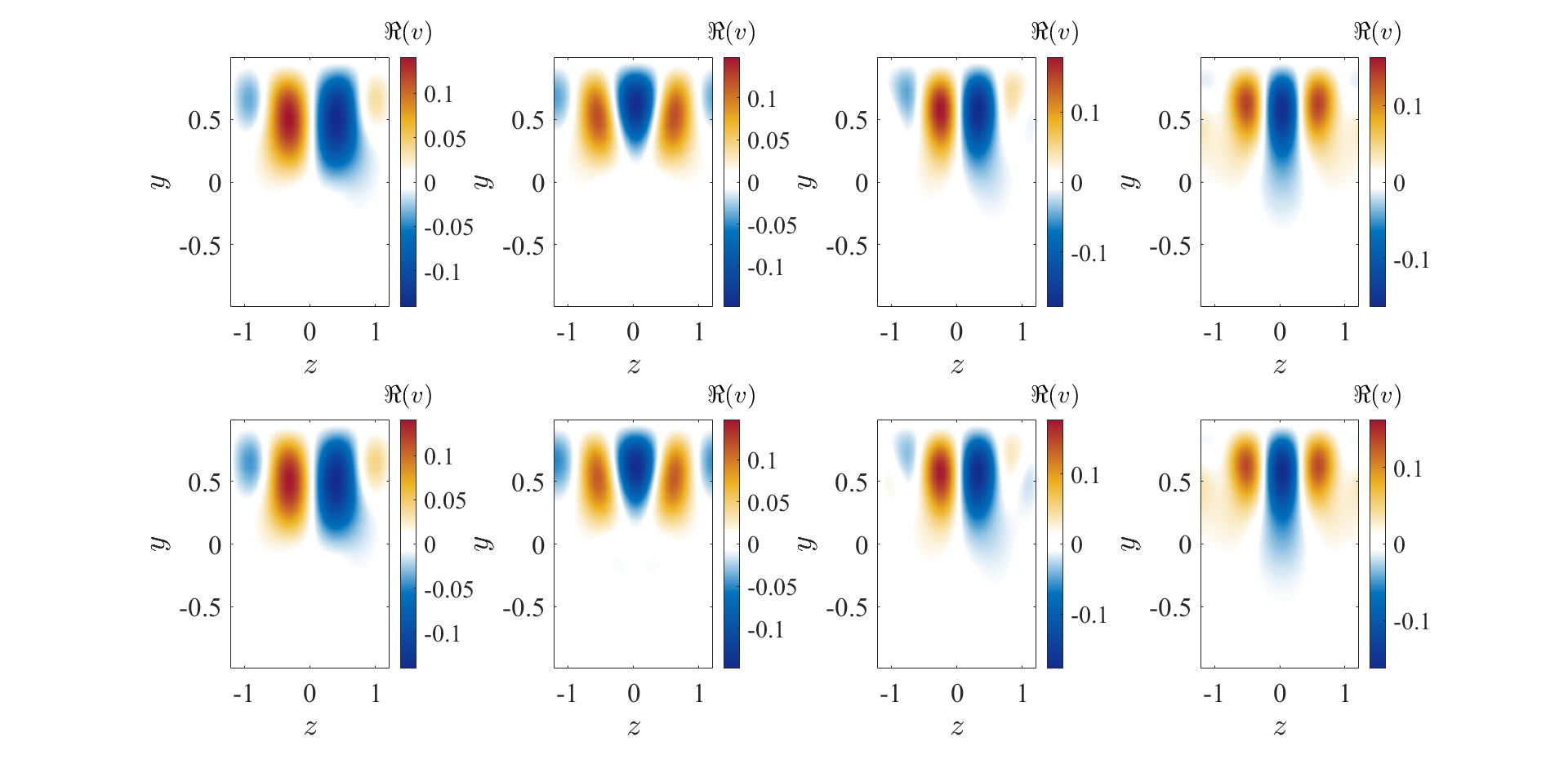}%

    \caption{Real part of the $v$ component of the first 4 resolvent response modes $(\boldsymbol{\psi}_j)$ for $k_x =  0.5$, $c = 0.75$, and $R = 400$. Top row: true modes, bottom row: VRA model with $N_{k_z} = 11$, $N_c = 3$, and $N_{SVD} = 8$. From left to right: $j=1,2,3,4$.}
    \label{fig:ECS_v}
\end{figure}

\begin{figure}
    \centering

  \includegraphics[trim = 100 0 0 0,  scale =  0.37]{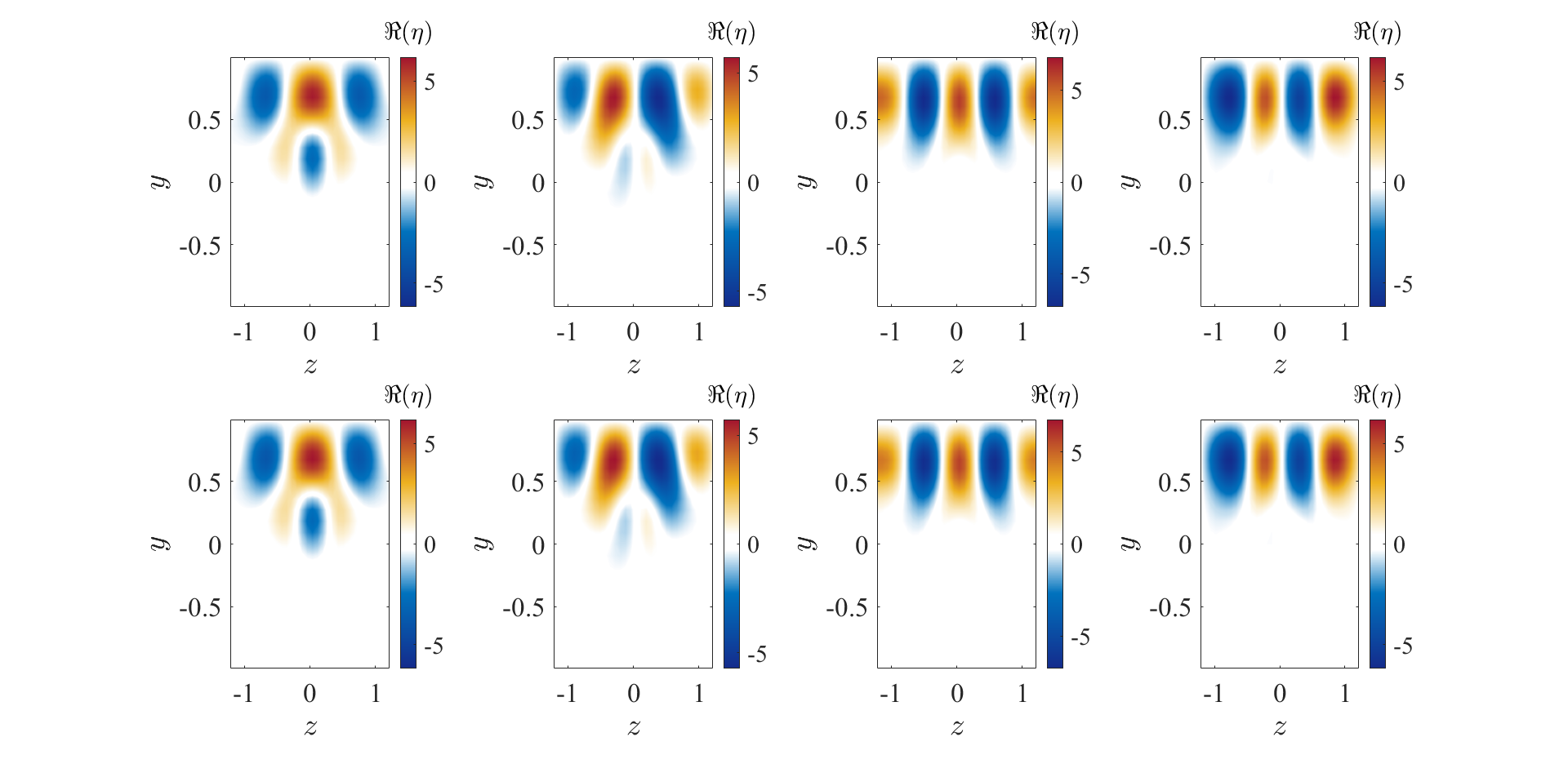}%

    \caption{Real part of the $\eta$ component of the first 4 resolvent response modes $(\boldsymbol{\psi}_j)$ for $k_x =  0.5$, $c = 0.75$, and $R = 400$.  Top row: true modes, bottom row: VRA model with $N_{k_z} = 11$, $N_c = 3$, and $N_{SVD} = 8$. From left to right: $j=1,2,3,4$. }
    \label{fig:ECS_eta}
\end{figure}

In Figures \ref{fig:ECS_v} and \ref{fig:ECS_eta} we compare the real part of the first four resolvent response modes of the variational reconstruction and the modes computed directly through the SVD of the 2D resolvent for $k_x = 0.5$ and $c = 0.75$ and $R = 400$. The variational approach very accurately reconstructs the true response modes considering the significant reduction in computational complexity. 

\begin{figure}
    \centering

  \includegraphics[trim = 100 0 0 0,  scale =  0.37]{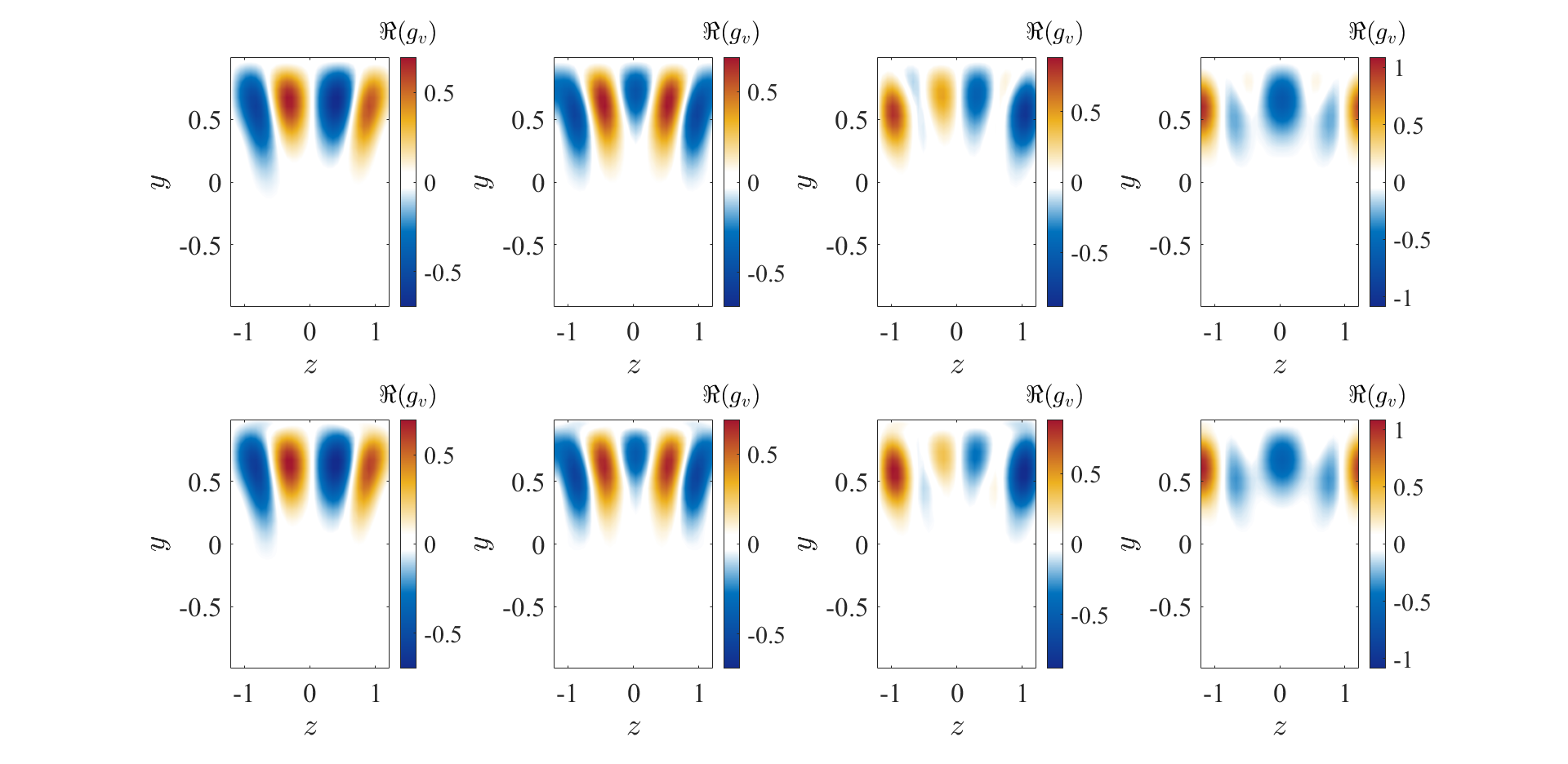}%

    \caption{Real part of the $v$ component of the first 4 resolvent forcing modes $(\boldsymbol{\phi}_j)$ for $k_x =  0.5$, $c = 0.75$, and $R = 400$. Top row: true modes, bottom row: VRA model with $N_{k_z} = 11$, $N_c = 3$, and $N_{SVD} = 8$. From left to right: $j=1,2,3,4$.}
    \label{fig:ECS_gv}
\end{figure}

\begin{figure}
    \centering

  \includegraphics[trim = 100 0 0 0,  scale =  0.37]{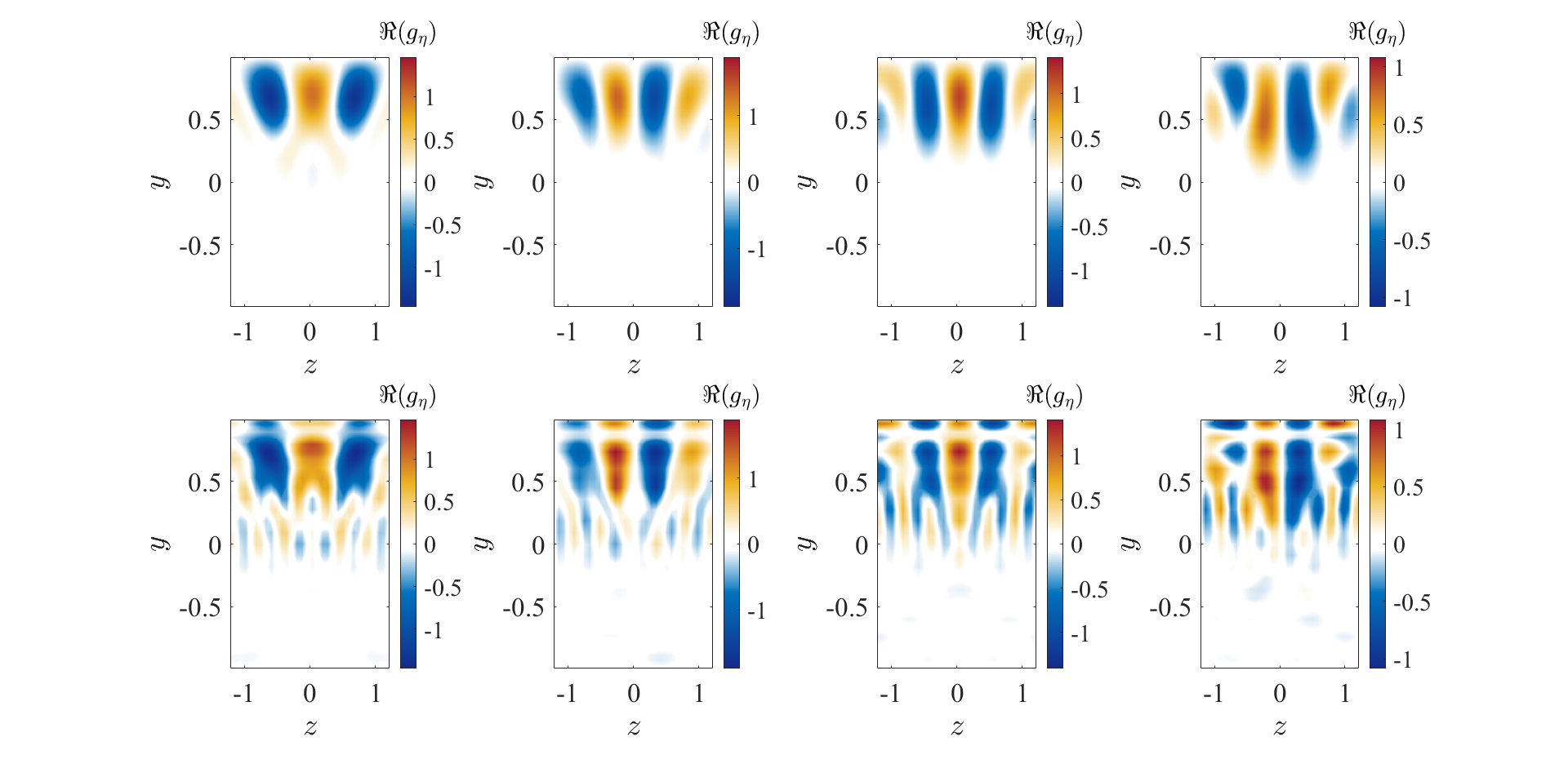}%

    \caption{Real part of the $\eta$ component of the first 4 resolvent forcing modes $(\boldsymbol{\phi}_j)$ for $k_x =  0.5$, $c = 0.75$, and $R = 400$.  Top row: true modes, bottom row: VRA model with $N_{k_z} = 11$, $N_c = 3$, and $N_{SVD} = 8$. From left to right: $j=1,2,3,4$. }
    \label{fig:ECS_geta}
\end{figure}

In Figures \ref{fig:ECS_gv} and \ref{fig:ECS_geta} we plot resolvent forcing modes computed from the response modes through $\boldsymbol{\phi}_j = \sigma_j\mathbf{L}^{2D}\boldsymbol{\psi}_j$. Interestingly we find that while the $g_v$ component is reproduced accurately the $g_{\eta}$ component shows significant discrepancy. While the qualitative shape of the $\eta$ component of the forcing mode is predicted by the VRA model, the mode is contaminated by higher harmonics. This contamination observed in the VRA reconstruction of the forcing modes, $\boldsymbol{\phi}_j$, despite the accurate reconstruction of the response modes, $\boldsymbol{\psi}_j$, is due to the directional amplification of the resolvent operator or equivalently, a sensitivity of the action of the linear operator $\mathbf{L}^{2D}\mathbf{q}$, to perturbations in the input $\mathbf{q}$. This phenomenon is discussed in detail in \S\ref{sec:direction}.

\begin{figure}
\centering
    \includegraphics[trim = 100 0 0 0,  scale =  0.37]{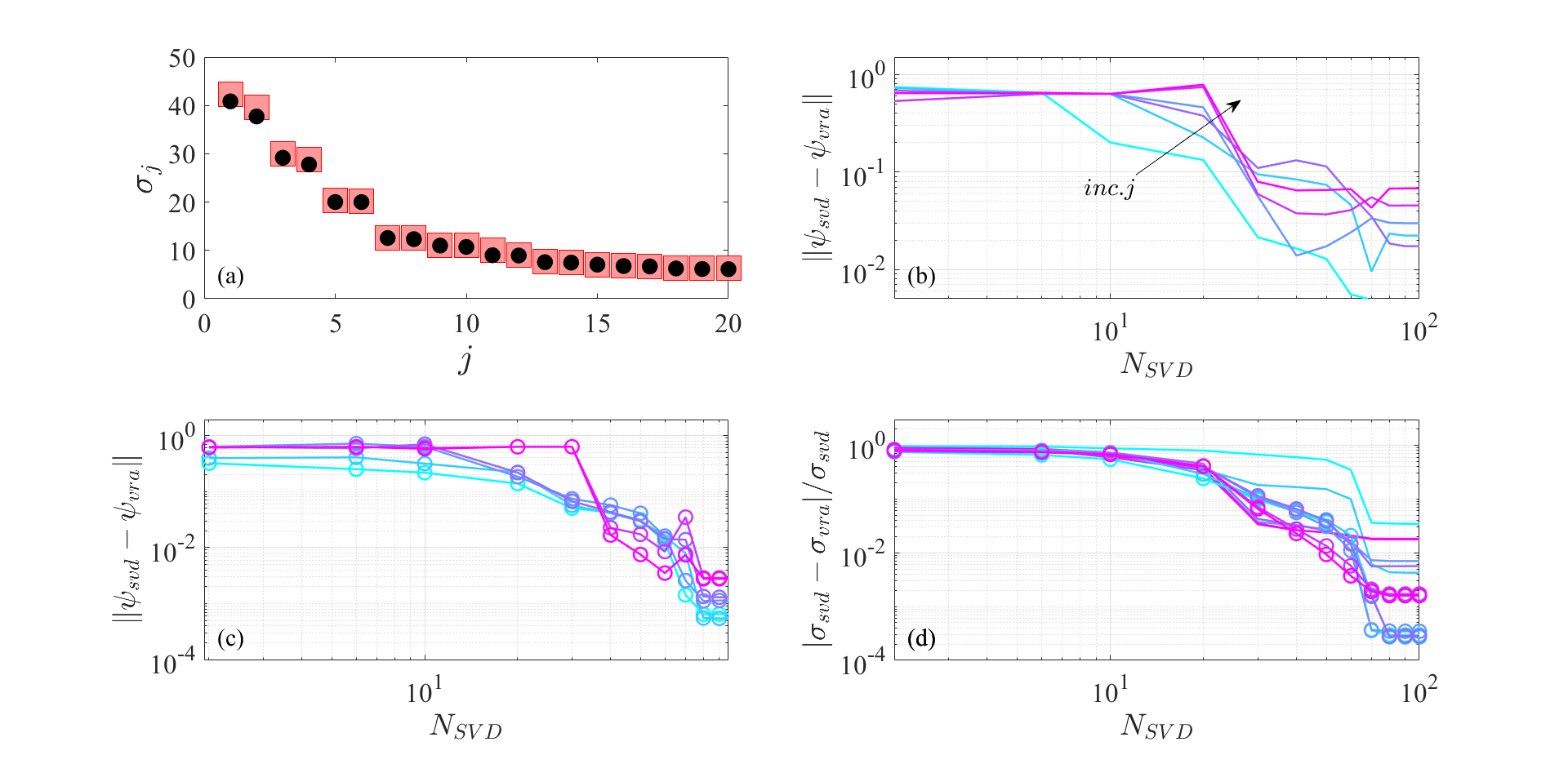}

\caption{Singular values for $k_x =  0.5$, $c = 0.75$, and $R = 400$ (a). SVD reference (red squares) and VRA model(black circles), same model parameters as in Figures \ref{fig:ECS_v}-\ref{fig:ECS_geta}. Integrated error of variational reconstruction of first six resolvent response modes (b) and (c), plotted separately for clarity, and first six singular values (d) as a function of retained singular basis elements $N_{SVD}$ for $N_c =1$ and $N_{k_z} = 11$. Results with $c=0.75$ are plotted in plain lines and those with $c=0$ are plotted with lines with open circles. From light to dark, colors indicate increasing $j$ from 1 to 6.}
    \label{fig:ECS_error}
\end{figure}

Additionally, in Figure \ref{fig:ECS_error}a we compare the variationally computed singular values with the true values computed via direct SVD. The singular values are estimated relatively accurately, with our model tending to slightly underestimate the leading singular values. As before, the true singular values represent the optimal gains and the predicted singular values are bounded above by the true values. For this example there is no significant separation of singular values, in other words the resolvent operator is not low rank, and yet our method still accurately predicts the singular values and resolvent response modes.

In order to quantify the convergence properties of the proposed method, for this example we fix $c^{1D} = c^{2D}$, include $k_z = [-5...0...5]\times2\pi/L_z$ such that $N_c = 1$ and $N_{k_z} = 11$ and compute the error as a function of the number of retained singular modes $N_{SVD}$.  The error is based on the kinetic energy norm and is defined as
\begin{equation}
    e \equiv \sqrt{ \frac{1}{2L_z}\int_{0}^{L_z} \int_{-1}^{1} |\boldsymbol{\psi}^{2D}_{svd}-\boldsymbol{\psi}^{2D}_{vra}|^2 dy dz}
\end{equation}
where $\boldsymbol{\psi} = [u,v,w]$. The error is plotted in Figure \ref{fig:ECS_error} alongside the relative error in singular values for two values of the wave speed, $c = 0.75$ and $ c = 0$. The former corresponds to the example plotted in Figures \ref{fig:ECS_v} through \ref{fig:ECS_error}a where there is no significant singular value separation. The latter case corresponds to a case where the 2D/3C resolvent is more low rank, $(\sigma_1/\sigma_2 \approx 6)$. In both cases our method is not only able to accurately approximate the leading singular mode and value but also a large range of suboptimal modes and singular values. Interestingly, we see that our method is more accurate in the case where there is less singular value separation. Furthermore, for the low rank case, $(c=0)$ the largest error in singular value is for $\sigma_1$. Again, these findings are a result of the directional nature of the resolvent operator and are discussed in detail in \S\ref{sec:direction}.

% plot mean
% \FloatBarrier
% plot PSIs

% Plot Phis

% Plot error and sigmas

\FloatBarrier
%% Boundary Layer Section
\section{2D Resolvent analysis: streamwise developing mean flow}\label{sec:BL}
In this section we use VRA to approximate the resolvent modes for a streamwise developing zero pressure gradient turbulent boundary layer (ZPGTBL). The streamwise developing nature of this flow necessitates large spatial domains and requires nonreflecting boundary conditions at the inlet and outlet of the domain. In this case the direct computation of the resolvent operator becomes impossible on a personal computer, and the size of the resulting matrices lead to memory requirements which become cumbersome even for high performance computers. Again we choose as our modeling basis 1D resolvent modes, in this case calculated using the mean flow $\bar{U}(y)$ at the inlet of the domain. Thus we have $q_j(x,y) = \boldsymbol{\psi}^{1D}_j(y;k_x,k_z,c)e^{i k_x x}$.

The reference 2D resolvent modes are computed using $\mathbf{L}^{2D}$, the LNSE about the mean flow, $\bar{\mathbf{U}}(x,y)$, under the assumption that the streamwise and wall normal directions are nonhomogenous. The mean flow is interpolated from mean profiles of a ZPGTBL DNS dataset described in \citet{schlatter_assessment_2010} with inlet $Re_{\tau} = u_{\tau}\delta_{99}/\nu \approx 700$. Variables without superscript are nondimensionalized with the velocity scale $U_{\infty}$, the free stream velocity, and $\delta_{99}$, the inlet boundary layer thickness, and variables with superscript $+$ denote rescaling with the local friction velocity, $u_{\tau}(x)$, and local friction lengthscale $\ell(x) = \nu/u_{\tau}$. The nonhomogeneous directions are discretized using a Chebyshev-Chebyshev grid, with $N_{y}$ points in $y \in [0,y_{max}]$ and $N_{x}$ points in $x \in [x_{i},x_{i} + L_x]$, where $L_x$ is the domain length in outer units. Our state $\mathbf{q}= [u,v,w,p]^{T}$ assumes the following wall normal boundary conditions: $\mathbf{u}(x,0) = \mathbf{0}$, $v_{y}(x,y) = 0$, and $\mathbf{u}_{y}(x,y_{max}) = \mathbf{0}$. At the inlet and outlet, we use Dirichlet boundary conditions and extrapolation boundary conditions with an artificial sponge layer applied to damp any artificial reflections due to the boundary conditions \cite{ran_effect_2017,ran_stochastic_2019}. The discretization was validated with the results from \citet{ran_effect_2017}. We note that using finite differences results in sparser operators that would reduce the computation times, but this was not explored in this paper. The modes are parametrized by the spanwise wavenumber $k_{z}$ and the temporal frequency, $\omega$. Here we consider three wavenumber - frequency combinations, two inner modes: $[k_z,\omega] = [43.9,1.8]$ and $[183,3.6]$, and an outer mode localized in the wake region: $[k_z,\omega] = [11.0,2.3]$. The latter is used to illustrate the current limitations of the proposed method.

Because the dimension of matrix $\mathbf{L}^{2D}$ is $4N_{x}N_{y} \times 4N_{x}N_{y}$, the matrix inversion and singular value decomposition are expensive, and scale with $\mathcal{O}((4N_{x}N_{y})^{3})$. To avoid such expensive calculations, an LU decomposition and Arnoldi Method is applied as in \citet{sipp_characterization_2013} and \citet{schmidt_spectral_2018} to compute the SVD of the resolvent by solving linear systems, as opposed to computing the matrix inverse. The most expensive computation, the LU decomposition, is handled with PARDISO, a sparse linear algebra solver which is part of the Intel math kernel library, as in \cite{jeun_input-output_2016}. Because of the low rank behavior that is often exhibited by the resolvent operator, the Arnoldi Method converges to the singular values and singular vectors in a few iterations. Although this strategy is considerably faster than computing the inverse and taking the SVD, the LU decomposition is still an expensive $\mathcal{O}((4N_{x}N_{y})^{3})$ operation.

The 1D resolvent modes used as the model basis are all calculated using the inlet mean velocity profile, the same $k_z$ as the 2D modes, a range of $N_{k_x}$ streamwise wavenumbers defined as integer multiples of $2\pi/L_x$, and $N_{c}$ wavespeeds, $c$. Although the model basis is computed using knowledge at one streamwise location, the coefficients of the basis are determined using $\mathbf{L}^{2D}$, which includes the streamwise variation of the mean. The multiple wavenumbers allow for constructive and destructive interference, creating the structure seen in the true response mode. Due to the critical layer mechanism, the 1D modes are localized at the critical layer, where $\bar{U}(y) = c$. To cover the wall-normal extent where we expect the 2D mode to be localized we then include a range of $N_c$ linearly spaced wavespeeds. At each wave number triplet $[k_x,k_z,c]$ we also include the leading $N_{SVD}$ resolvent modes, resulting in a total of $r  = N_{k_x} \times N_c \times N_{SVD}$ degrees of freedom. The modeling parameters, global mode spatial resolutions, and overall model reduction for the two examples considered here are summarized in Table \ref{tab:BL}. The reader is referred to Figure \ref{fig:inputbasis_BL} in Appendix~\ref{app:inputbasis} for an illustration of some representative basis elements. 

\subsection{Inner Modes}\label{sec:BLinner}
\begin{figure}
    \centering
    \subfloat[]{%
  \includegraphics[trim = 140 20 0 80,  scale =  0.17]{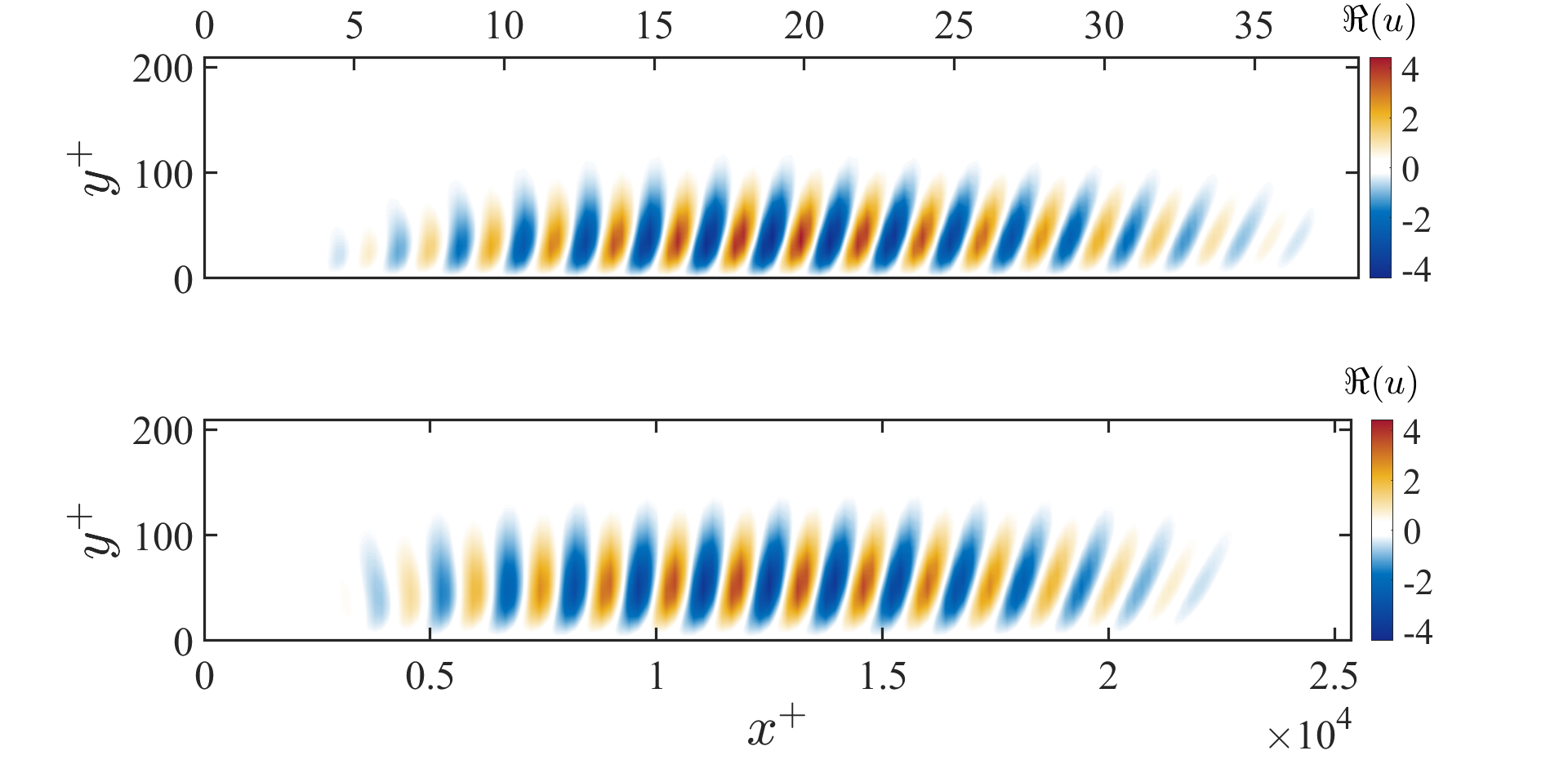}%
}
\subfloat[]{%
  \includegraphics[trim = 100 20 60 80,  scale =  0.17]{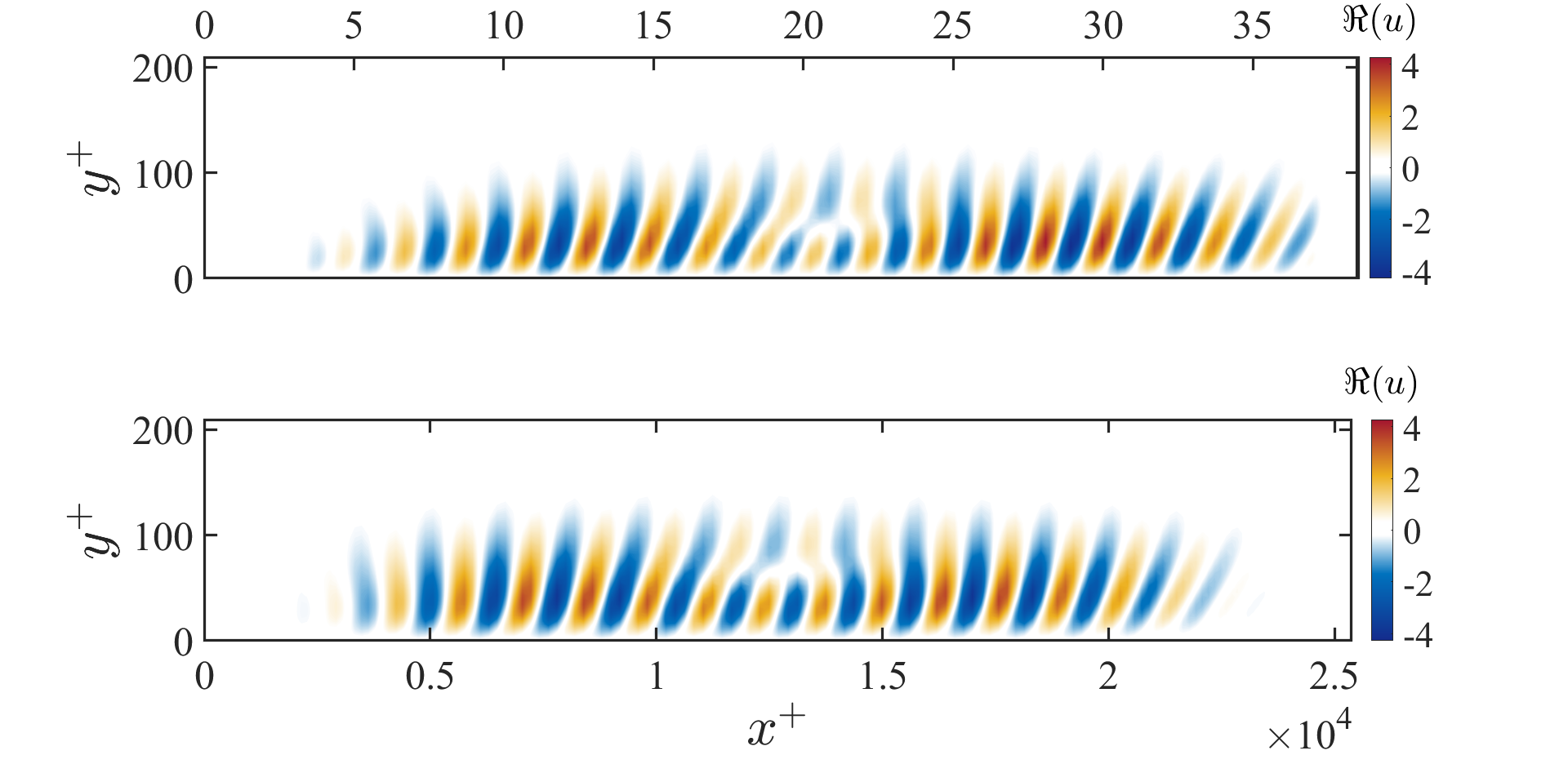}%
}

    \subfloat[]{%
  \includegraphics[trim = 140 20 0 20,  scale =  0.17]{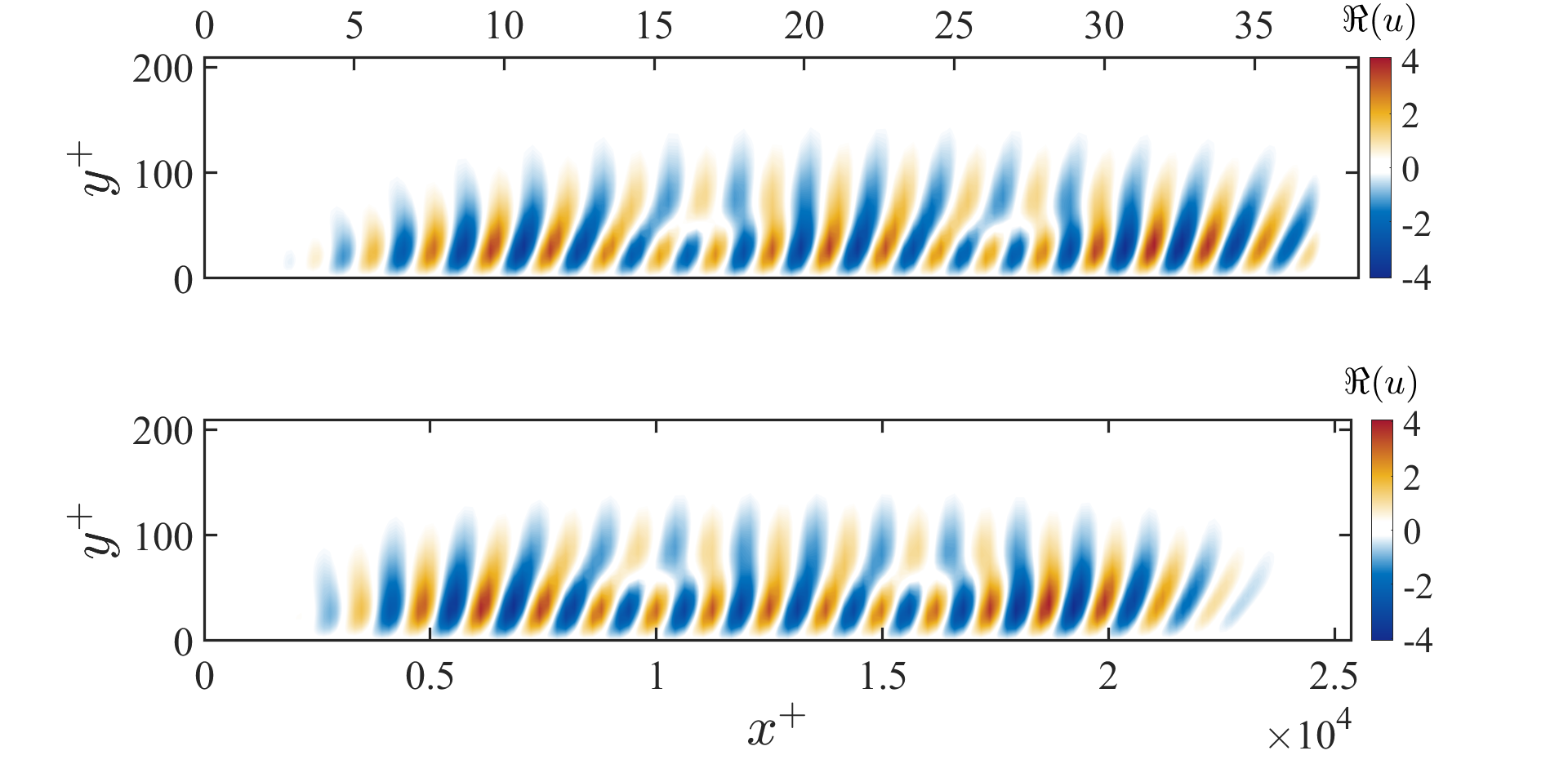}%
}
\subfloat[]{%
  \includegraphics[trim = 100 20 60 20,  scale =  0.17]{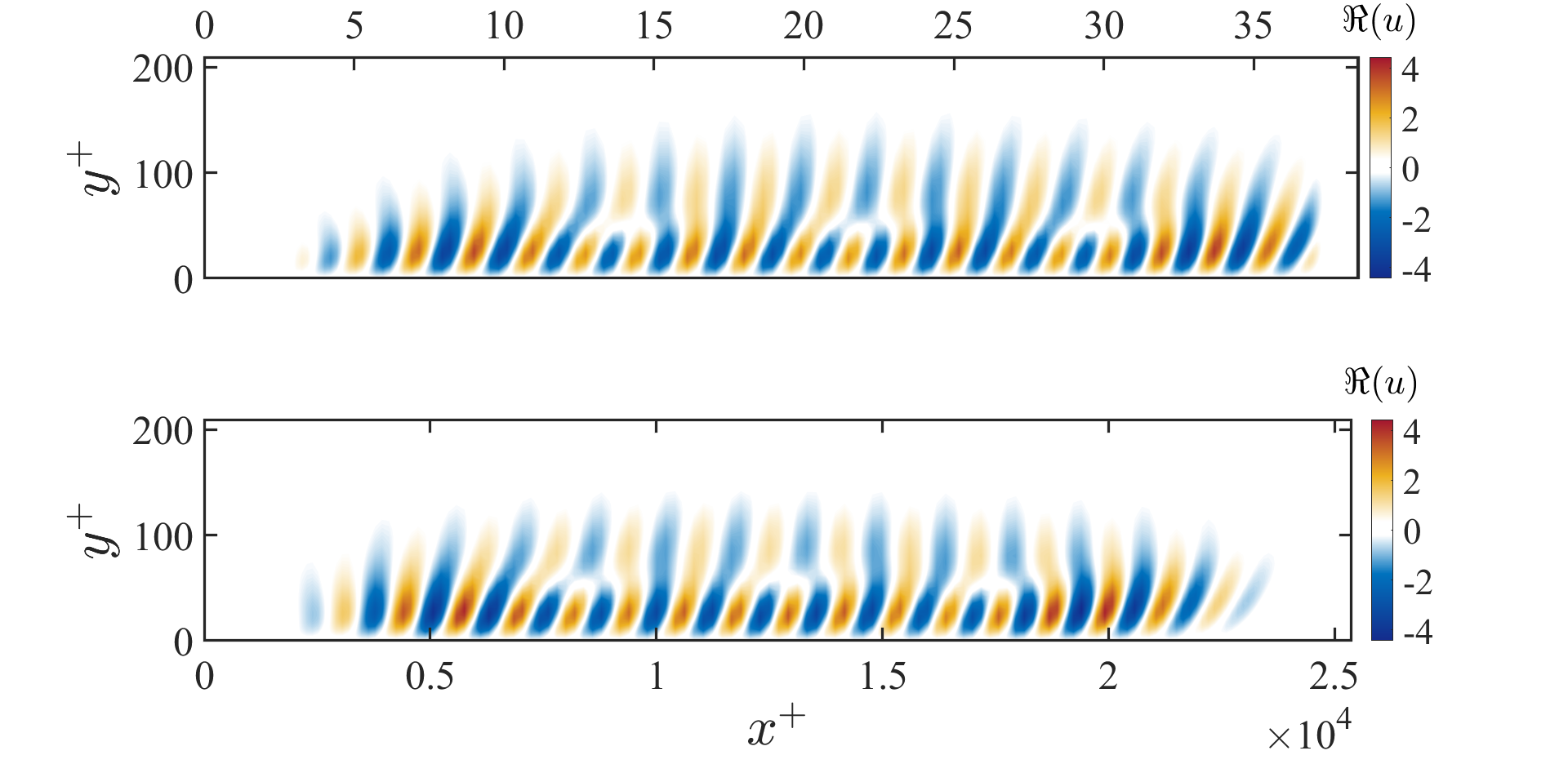}%
}
    \caption{First four resolvent response modes $(\boldsymbol{\psi}_j)$: real part of the streamwise component $u$. $j=1,2,3,4$ (a - d) for $Re_{\tau} \approx 700$ and $[k_z,\omega] =  [43.9, 1.8]$. Top panels: true global modes, bottom panels: VRA model. Upper x-axis: represents outer units $x$, lower x-axis represents inner units $x^+$. Model basis parameters are: $N_{k_x} = 26$, $N_c = 3$, $N_{SVD} = 6$. }
    \label{fig:BL_u_44}
\end{figure}

\begin{figure}
    \centering
    \subfloat[]{%
  \includegraphics[trim = 140 20 0 80,  scale =  0.17]{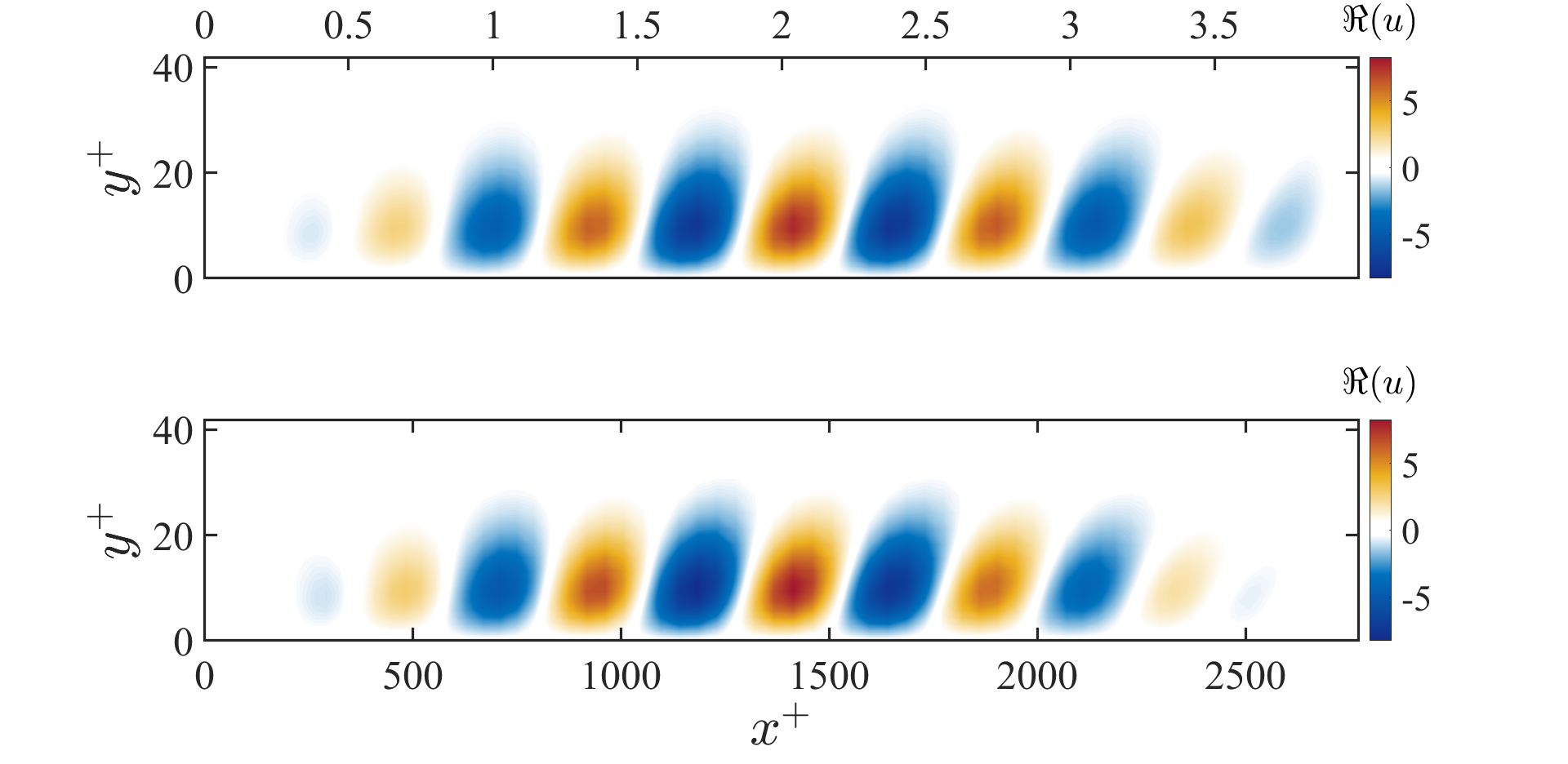}%
}
\subfloat[]{%
  \includegraphics[trim = 100 20 60 80,  scale =  0.17]{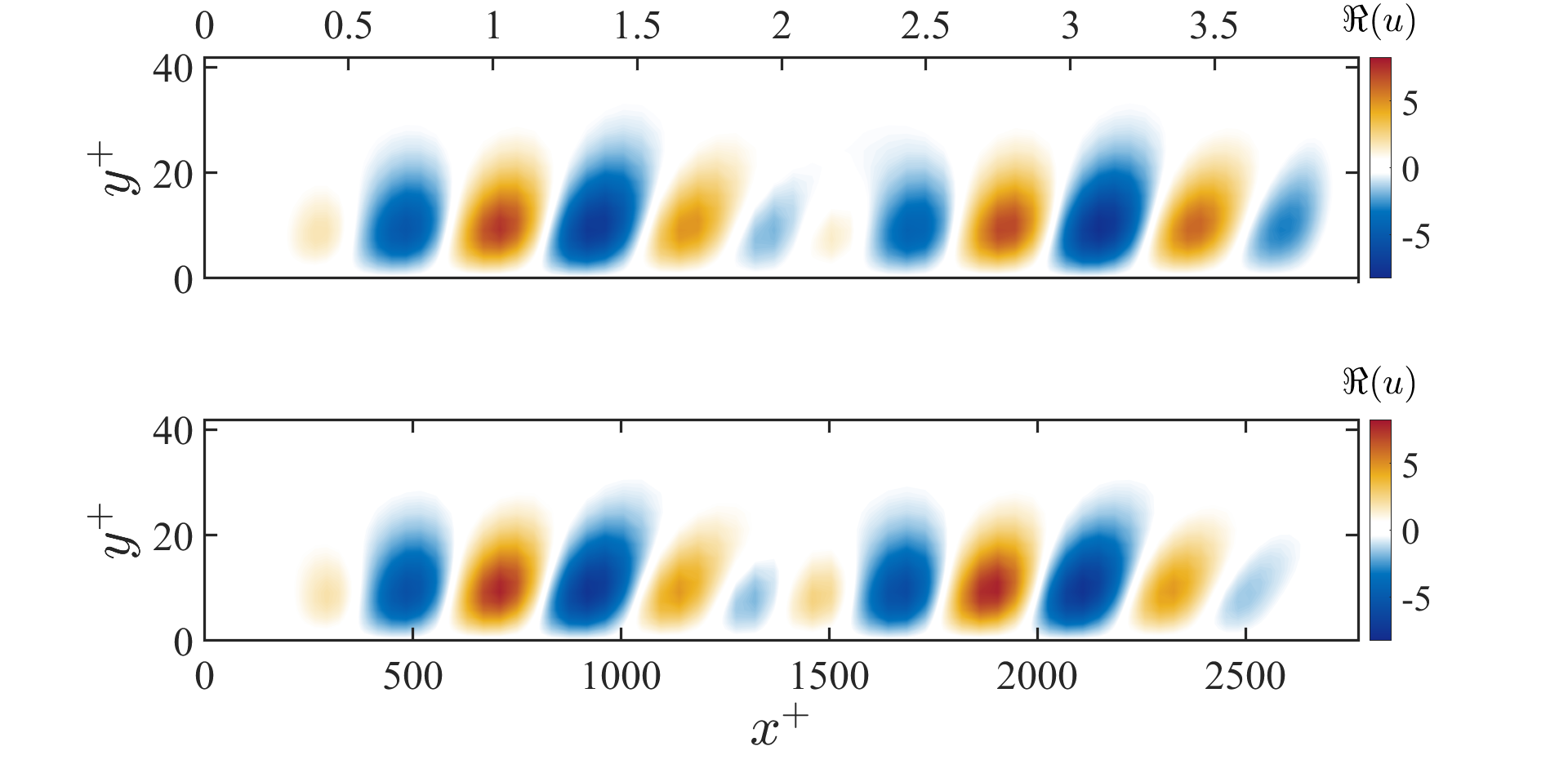}%
}

    \subfloat[]{%
  \includegraphics[trim = 140 20 0 20,  scale =  0.17]{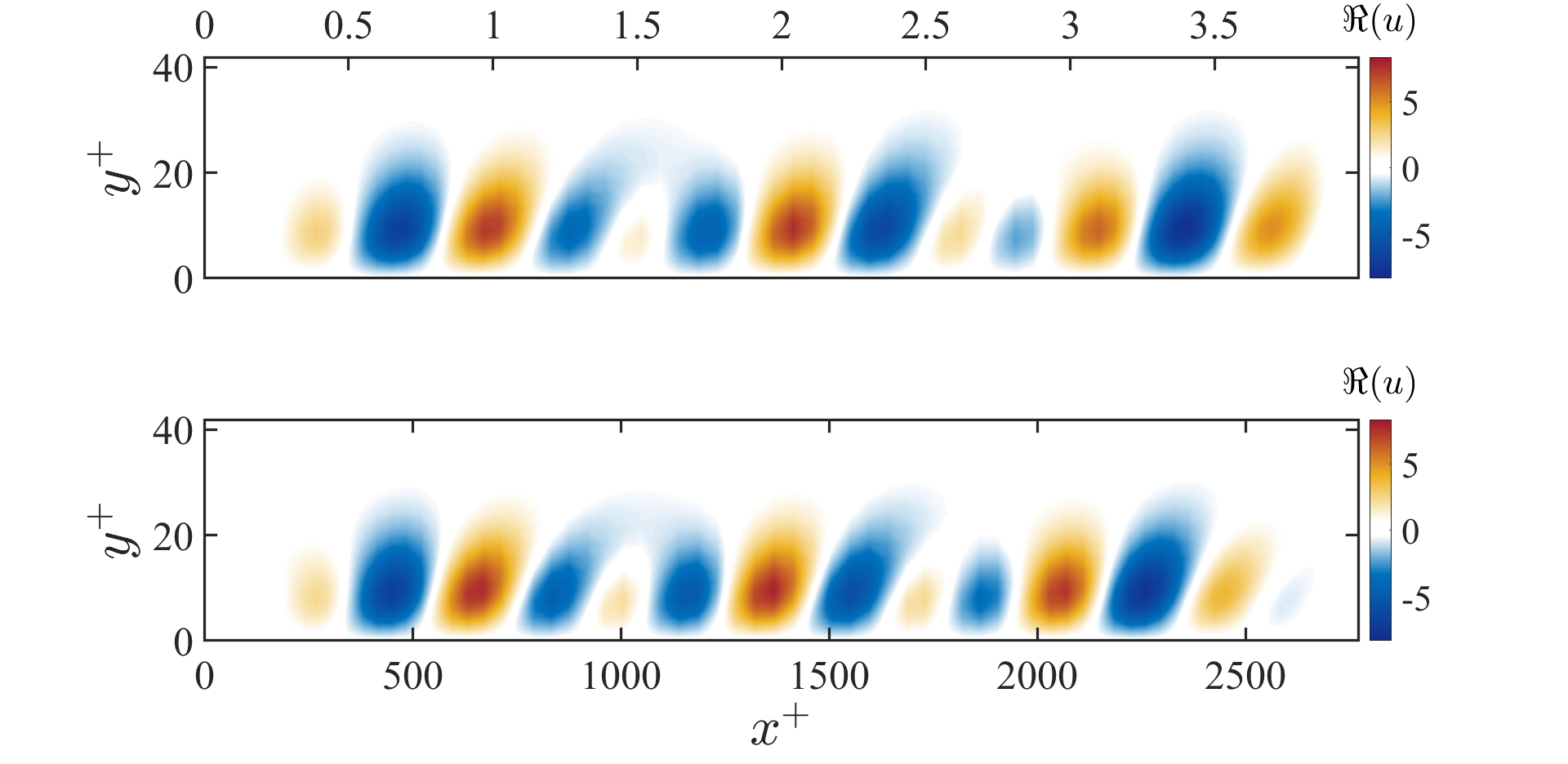}%
}
\subfloat[]{%
  \includegraphics[trim = 100 20 60 20,  scale =  0.17]{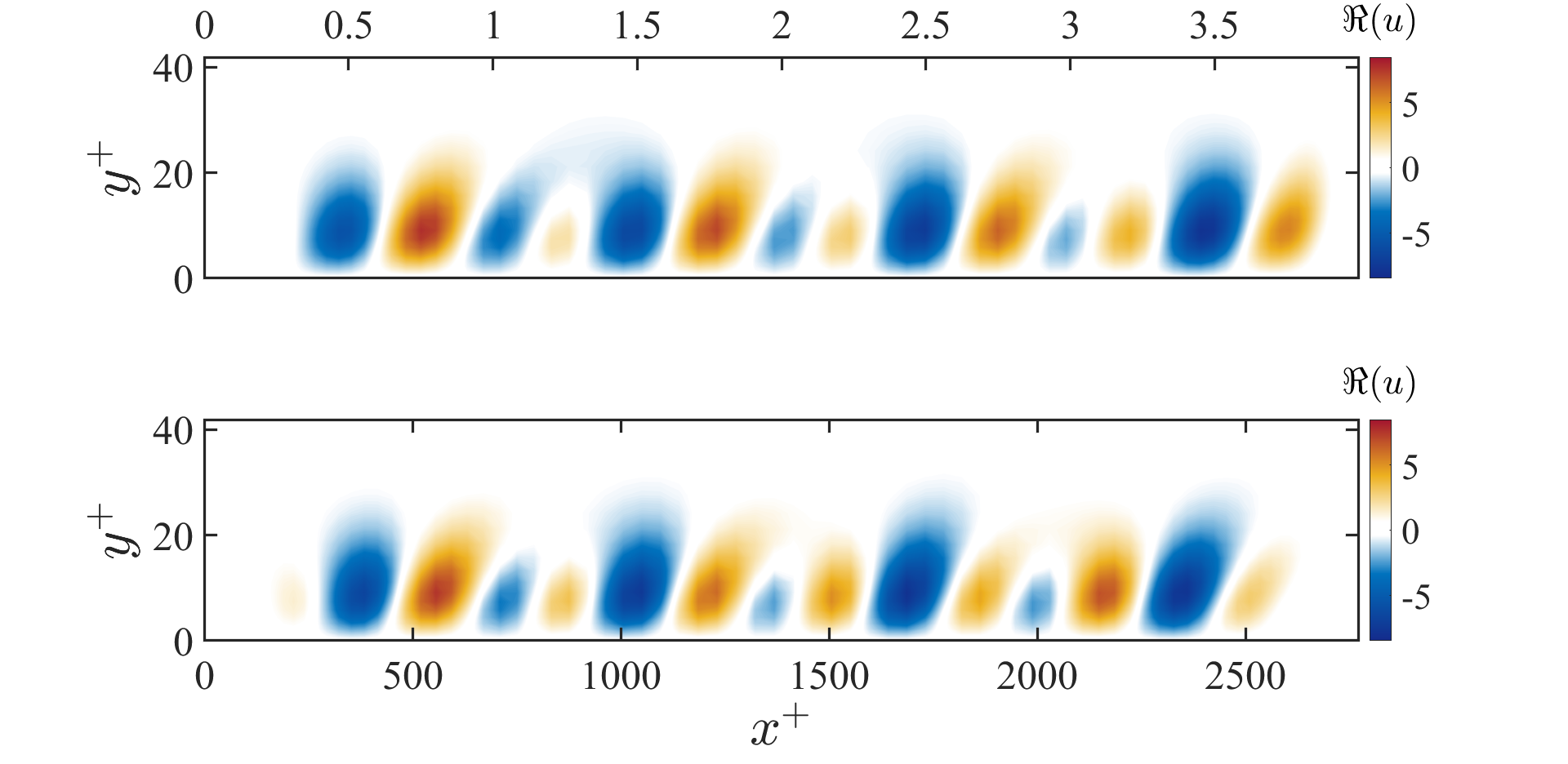}%
}
    \caption{First four resolvent response modes $(\boldsymbol{\psi}_j)$: real part of the streamwise component $u$. $j=1,2,3,4$ (a - d) $Re_{\tau} \approx 700$ and $[k_z,\omega] =  [183, 3.6]$. Top panels: true global modes, bottom panels: VRA model. Upper x-axis: represents outer units $x$, lower x-axis represents inner units $x^+$. Model basis parameters are: $N_{k_x} = 16$, $N_c = 3$, $N_{SVD} = 1$.  }
    \label{fig:BL_u_183}
\end{figure}

In Figures \ref{fig:BL_u_44} and  \ref{fig:BL_u_183} we compare the first four resolvent modes of the variational reconstruction and the modes computed directly through the classic resolvent analysis of the 2D resolvent for $[k_z,\omega] = [43.9,1.8]$ and $[k_z,\omega] = [183,3.6]$. The former's spanwise wavelength $\lambda_{z}^{+} \sim 100$ is representative of near wall streaks whereas the latter's spanwise wavelength $\lambda_{z}^{+} \sim 25$ is representative of smaller structure close to the wall \citep{kline_structure_1967}. In both cases we note that all modes display streamwise oscillations at wavelengths on the order of $\mathcal{\delta}_{99}$. Additionally, we also note the presence of a larger wavelength in the form of a modulating envelope with wavelength $L_{x}/j$ where $j$ is the rank of the mode. In both cases the characteristic streamwise wavelength and the modulating envelope of the modes are captured by the VRA model for both the optimal and the higher order modes. We note that this streamwise evolution in both shape and amplitude is not present in the VRA basis functions (see Appendix \ref{app:inputbasis}). Because of this streamwise scale separation, the VRA model requires basis functions with a large range of streamwise wavenumbers. Despite this the number of retained wave numbers $N_{k_x}$ is still significantly less than the required streamwise spatial discretization, $N_x$, of the full system.

For the wider (smaller $k_z$) modes plotted in Figure \ref{fig:BL_u_44}, we see that the VRA model predicts the mode shape and amplitude present in the SVD-based modes and replicates many of the general features. Especially in the interior of the domain the VRA modes capture the reference modes relatively accurately. However, near the streamwise boundaries there are some significant discrepancies. Here, the VRA modes have less support as compared to the reference modes. This difference is likely due to the basis functions not satisfying the same streamwise boundary conditions as the 2D modes. The basis has periodic boundary conditions while the 2D modes are treated with nonreflecting boundary conditions. The nonreflecting boundary conditions, through the sponge, cause the SVD modes to abruptly decay to $0$ near the inlet and outlet of the domain.

The narrower (larger $k_z$) modes plotted in Figure \ref{fig:BL_u_183} show relatively good agreement between the VRA prediction and the SVD-based modes throughout the domain. This is likely because in this case the shorter domain restricts the streamwise development of the mean flow, $(700 < Re_{\tau}<740)$, as opposed to the case of $k_z = 43.9$ where $(700 < Re_{\tau}<1040)$. Additionally, the narrower modes have less streamwise extent and are localized in the near wall region $y^{+} < 35$ where they are less susceptible to streamwise development of the wake \citep{ruan_direct_2021}. Since the mean flow is nearly parallel in this region, the fact that the 1D resolvent modes used in the VRA model are periodic in $x$ is less of an impediment. However, as seen for example in Figure \ref{fig:BL_u_183}c there is still some discrepancy between the suboptimal SVD and VRA based modes with the VRA mode being slightly shifted towards the inlet relative to the reference mode.

\begin{figure}
    \centering
    \subfloat[]{%
  \includegraphics[trim = 140 40 0 80,  scale =  0.17]{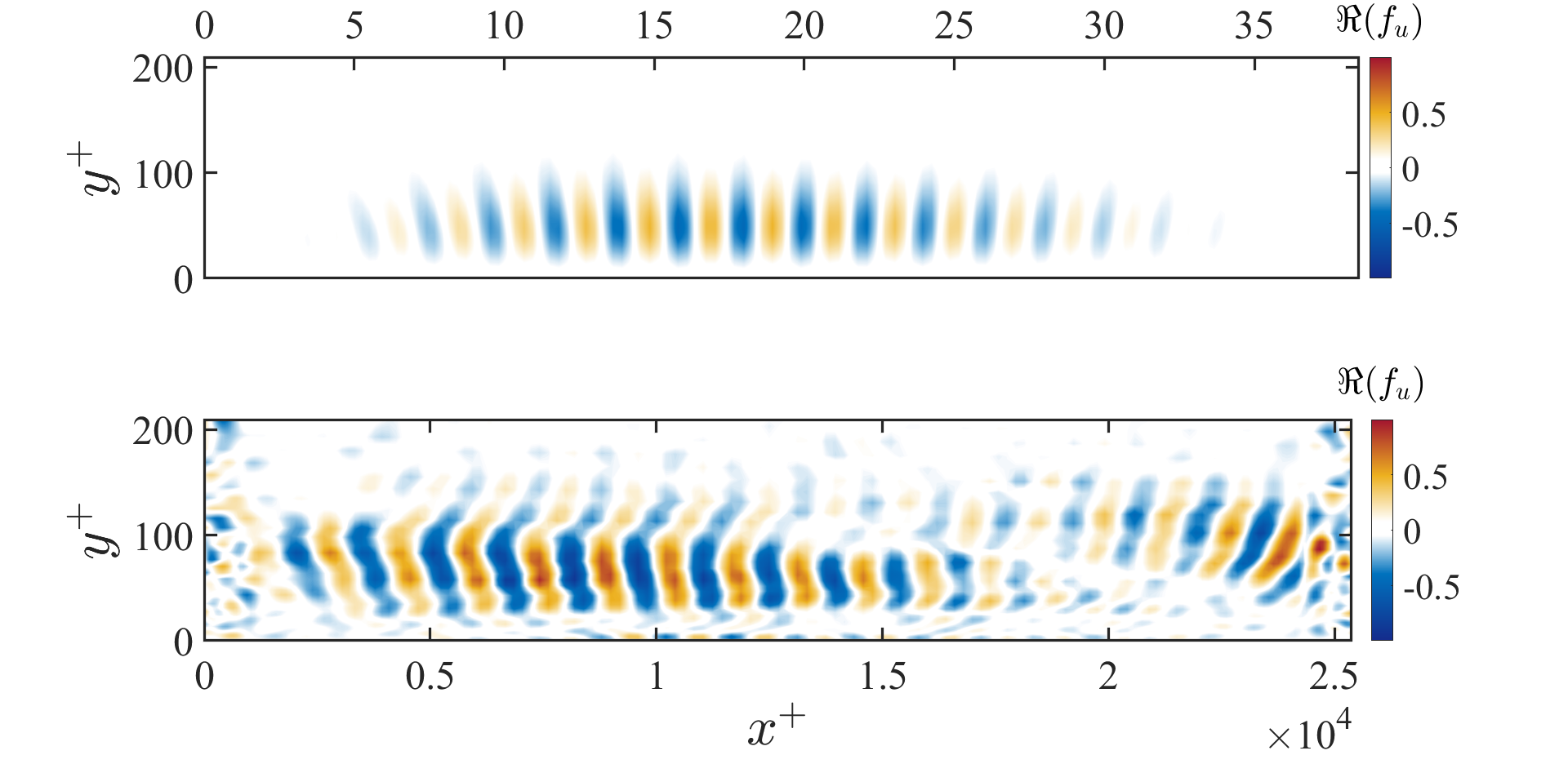}%
}
\subfloat[]{%
  \includegraphics[trim = 100 40 60 80,  scale =  0.17]{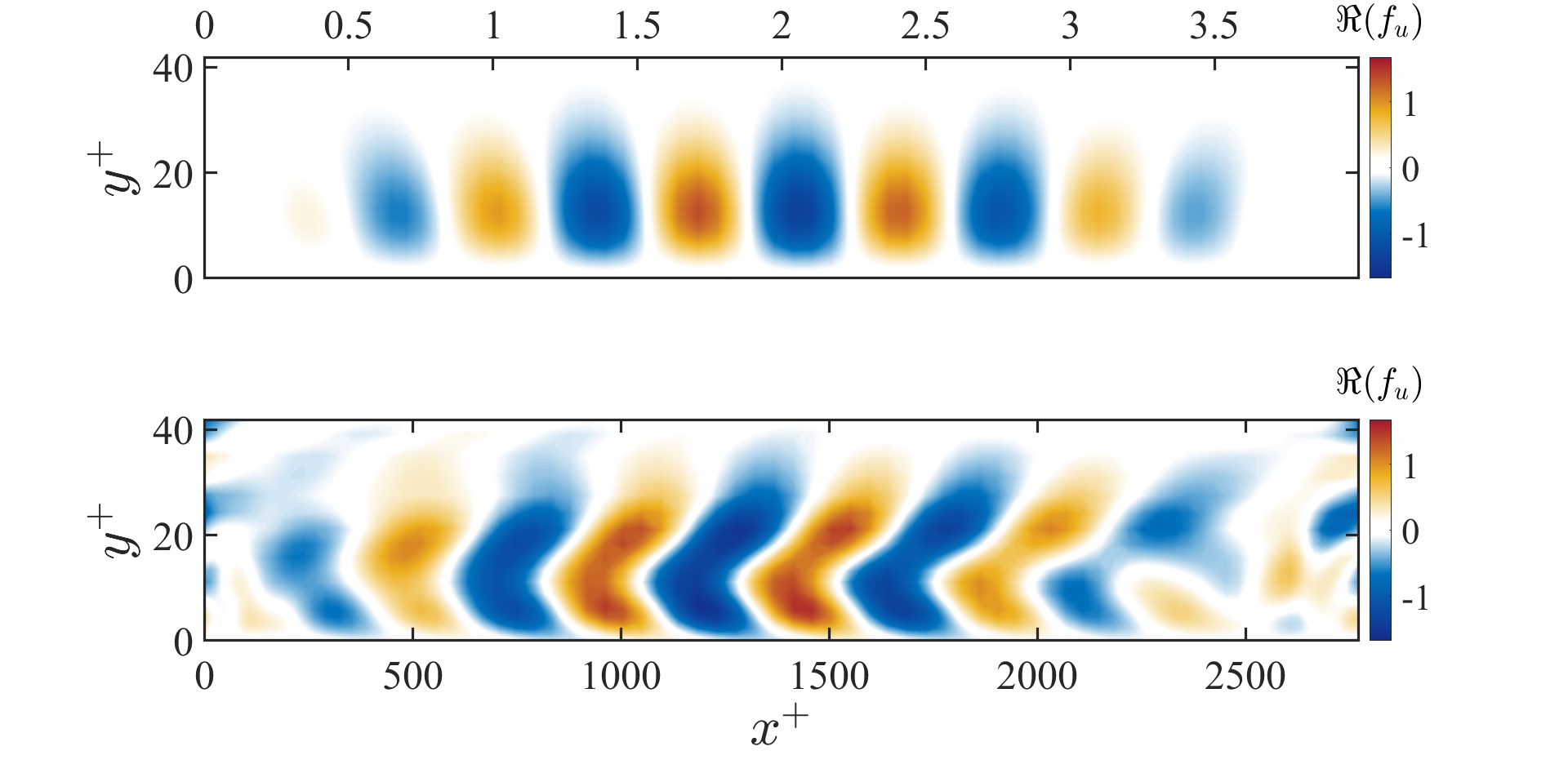}%
}

    \subfloat[]{%
  \includegraphics[trim = 140 40 0 20,  scale =  0.17]{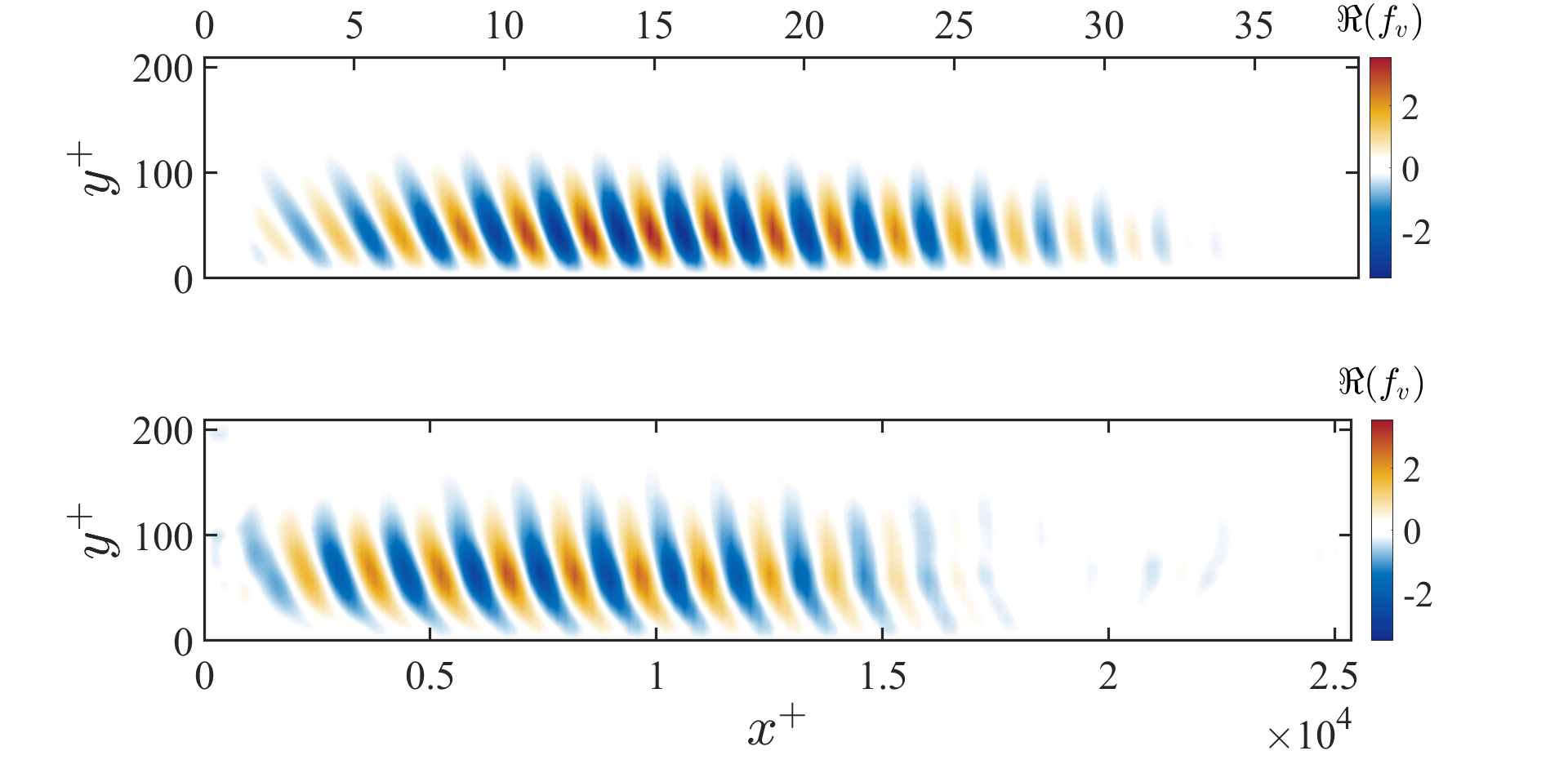}%
}
\subfloat[]{%
  \includegraphics[trim = 100 40 60 20,  scale =  0.17]{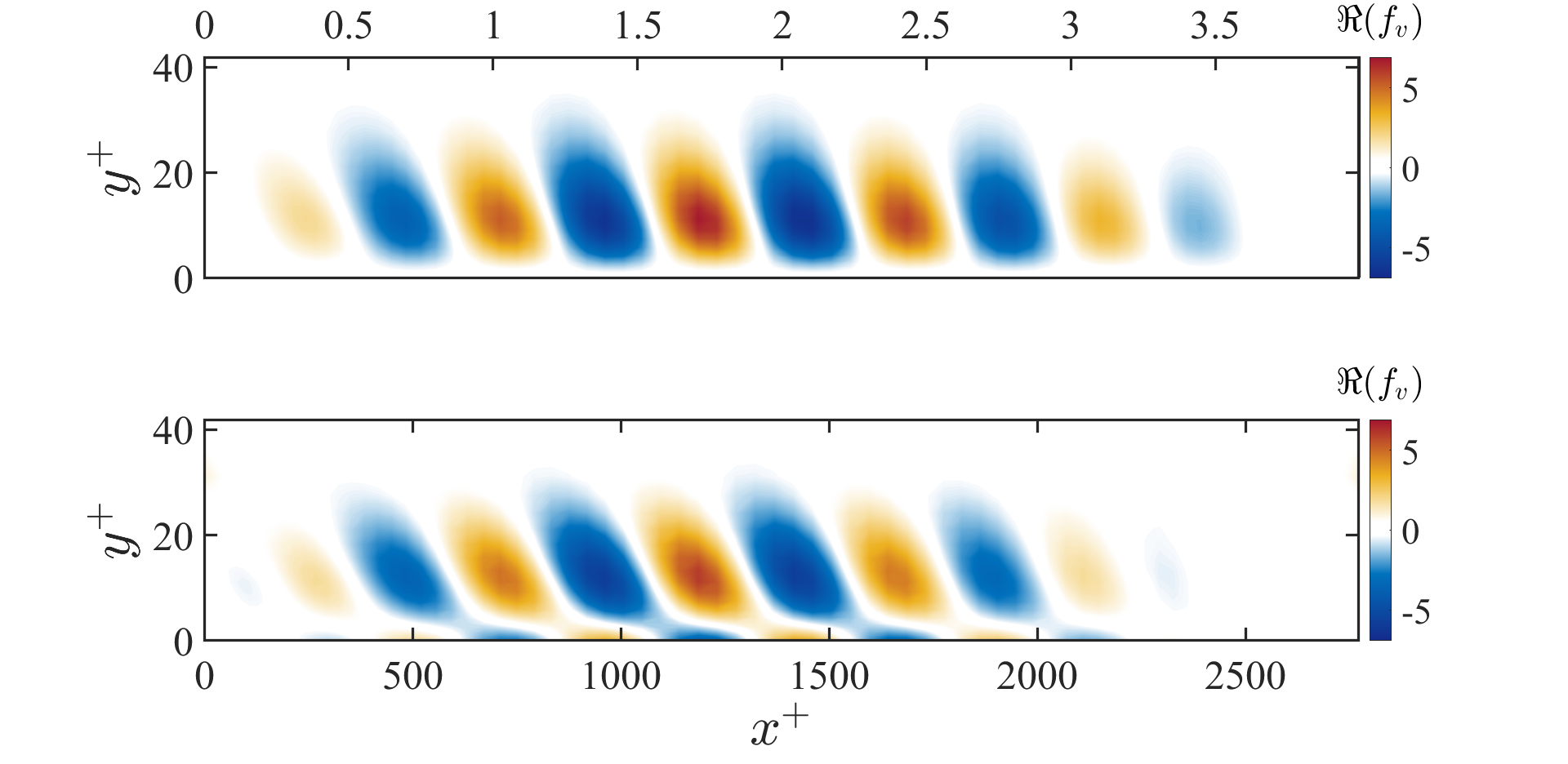}%
}

    \subfloat[]{%
  \includegraphics[trim = 140 40 0 20,  scale =  0.17]{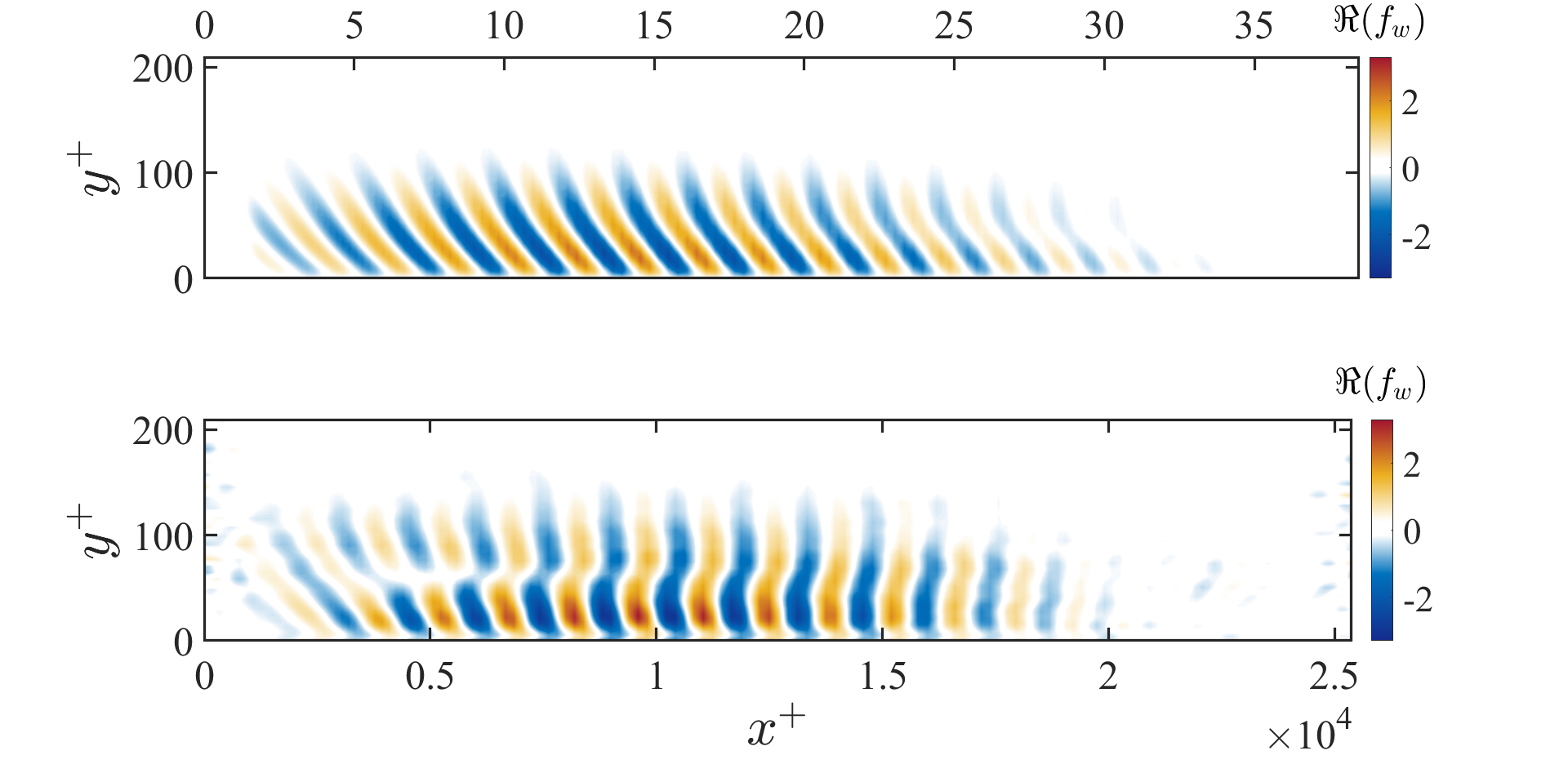}%
}
\subfloat[]{%
  \includegraphics[trim = 100 40 60 20,  scale =  0.17]{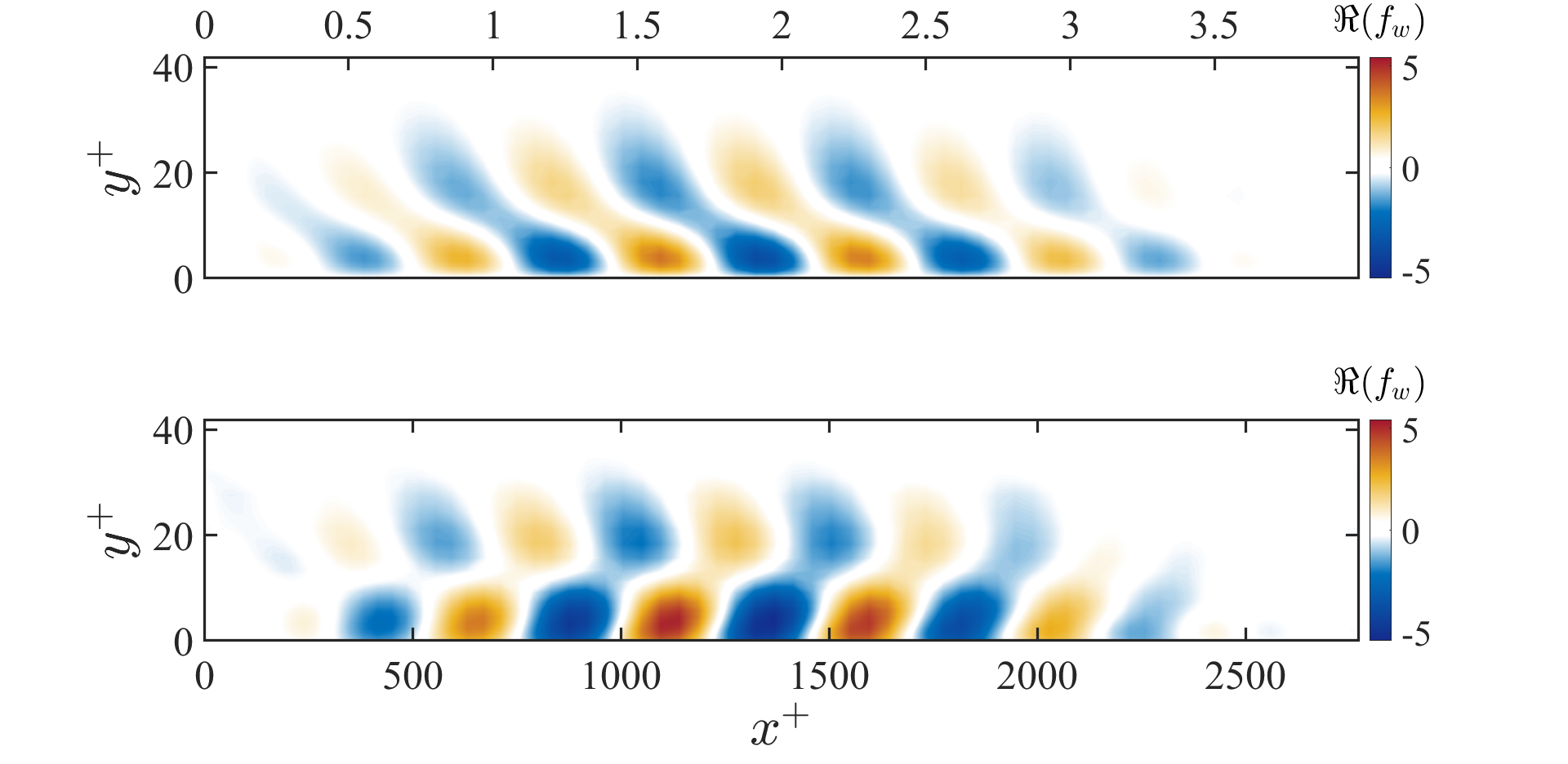}%
}

\caption{Leading resolvent forcing mode $(\boldsymbol{\phi}_1)$ for $Re_{\tau} \approx 700$. $[k_z,\omega] =  [43.9, 1.8]$, $f_u$(a), $f_v$(c), $f_w$(e). $[k_z,\omega] =  [183, 3.6]$, $f_u$(b), $f_v$(d), $f_w$(f). In each subplot, top panels: true global modes, bottom panels: VRA model.Upper x-axis: represents outer units $x$, lower x-axis represents inner units $x^+$. Model basis parameters are the same as in Table \ref{tab:BL}. }
\label{fig:BL_f}

\end{figure}

Figure \ref{fig:BL_f} shows all three components of the optimal forcing mode: $\boldsymbol{\phi}_1$ for both $k_z =43.9$ and $k_z = 183$. We plot all three components of the forcing modes in Figures \ref{fig:BL_u_44} and \ref{fig:BL_u_183} to illustrate the component-wise amplification present in non-normal operators. For the response modes the streamwise component accounts for $>95\%$ of the total norm of the leading modes investigated here, whereas for the leading forcing modes, the streamwise components account for less than $5\%$ of the total norm. In wall bounded flows this discrepancy in the amplification is associated with the lift up mechanism, where disturbances with large spanwise and wall normal components lead to flow responses with large streamwise components. Physically, this is related to the counter rotating vortices that lead to streamwise velocity streaks as recently reviewed by \citet{brandt_lift-up_2014}. In Figures \ref{fig:BL_sig44} and \ref{fig:BL_sig183}, we compare the exact singular values and the VRA prediction. Unlike the previous examples we have analyzed, we see that in both cases the VRA model significantly underpredicts the singular values. The error is greater for $k_z = 43.9$ with errors of approximately $33 \% $ in $\sigma_1$ compared to around $15 \%$ for $k_z = 183$.

In this example the VRA model largely fails to predict the shape of the forcing modes, most notably in the streamwise component of the forcing, and displays significant error in the prediction of the singular values. While the cross-stream components of the VRA approximations capture some of the features seen in the SVD-based forcing modes, the VRA modes exhibit a $\pi/2$ phase shift not seen in the SVD-based mode. Interestingly the phase shift seems to be centered at different wall normal locations for all three velocity components. We note that despite the differences in the shape, the VRA forcing modes still replicate the component amplitude trends of the SVD-based forcing modes. Again the significant difference in the VRA and SVD-based singular values and forcing modes, despite the similarity in the response modes, illustrates how $\mathbf{H}$ acts as a directional amplifier. The resolvent identifies the most amplified forcing mode, however $\mathbf{L}$ does not preferentially amplify the leading response. This is discussed in detail in \S\ref{sec:direction}.

\subsection{Outer Modes}\label{sec:BLouter}
\begin{figure}
    \centering
\subfloat[]{%
  \includegraphics[trim = 320 0 0 0,  scale =  0.4]{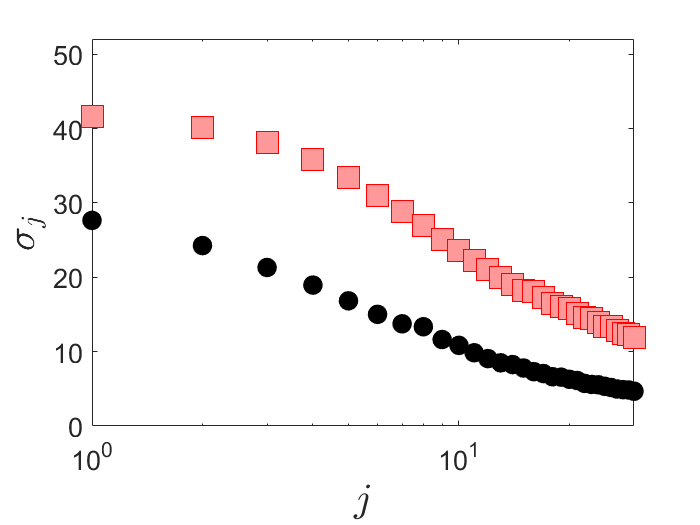}%
  \label{fig:BL_sig44}
}
\subfloat[]{%
  \includegraphics[trim = 20 0 260 0,  scale =  0.4]{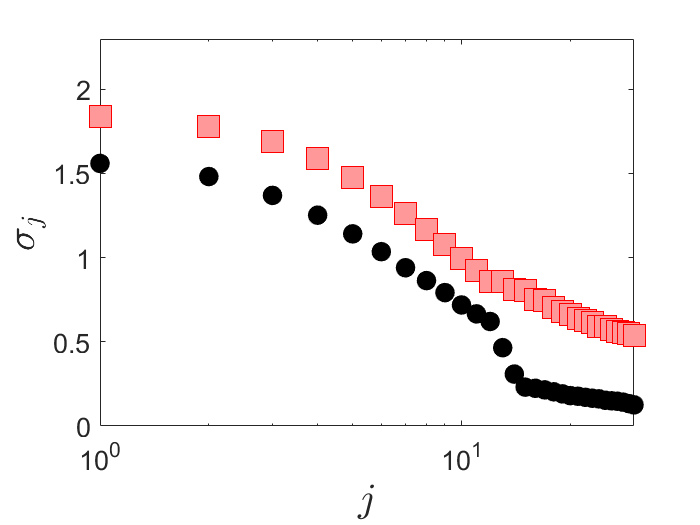}%
  \label{fig:BL_sig183}
}

\subfloat[]{%
  \includegraphics[trim = 140 0 0 0,  scale =  0.17]{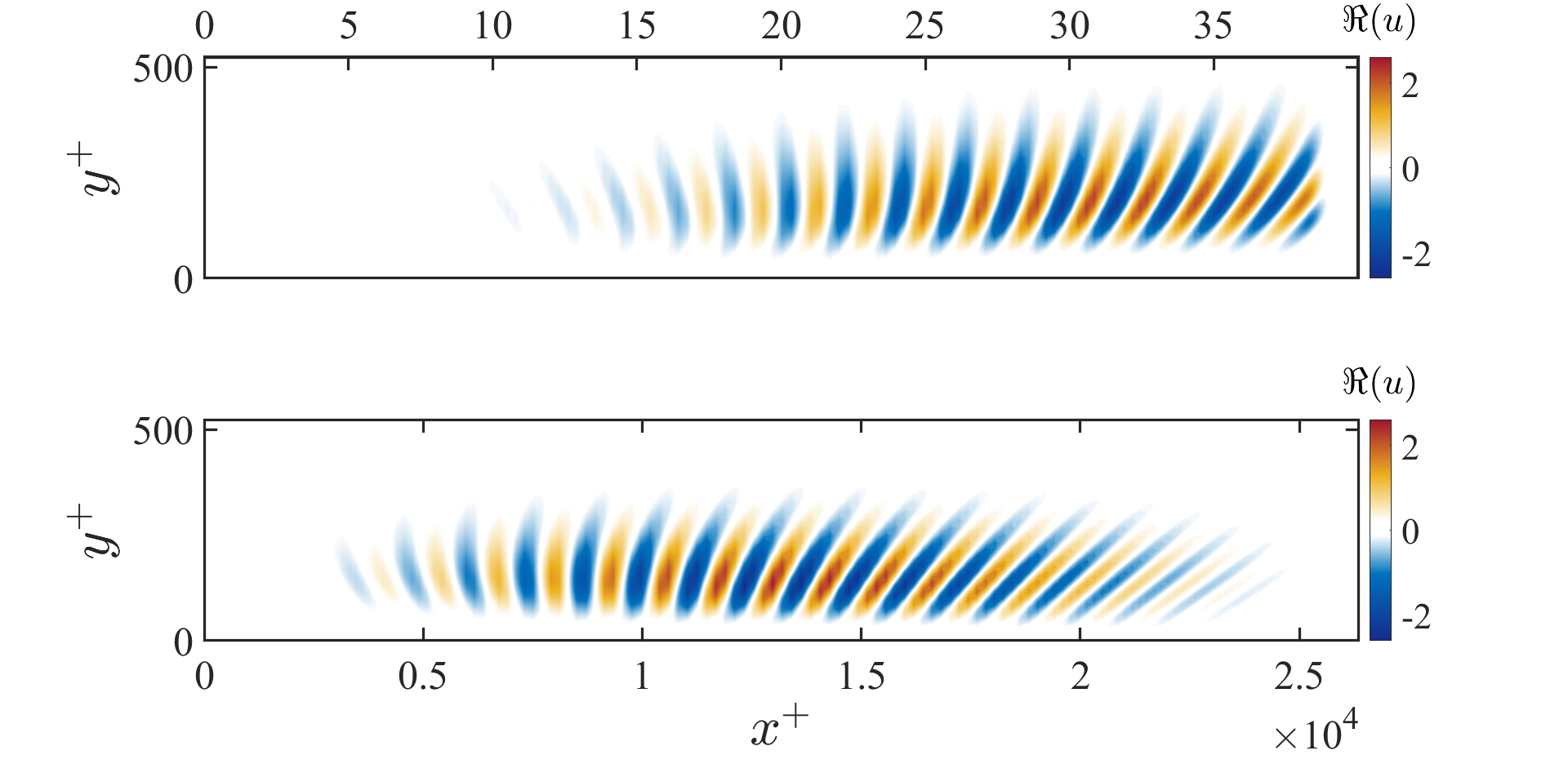}%
  \label{fig:BL_u_11}
}
\subfloat[]{%
  \includegraphics[trim = 100 0 60 0,  scale =  0.17]{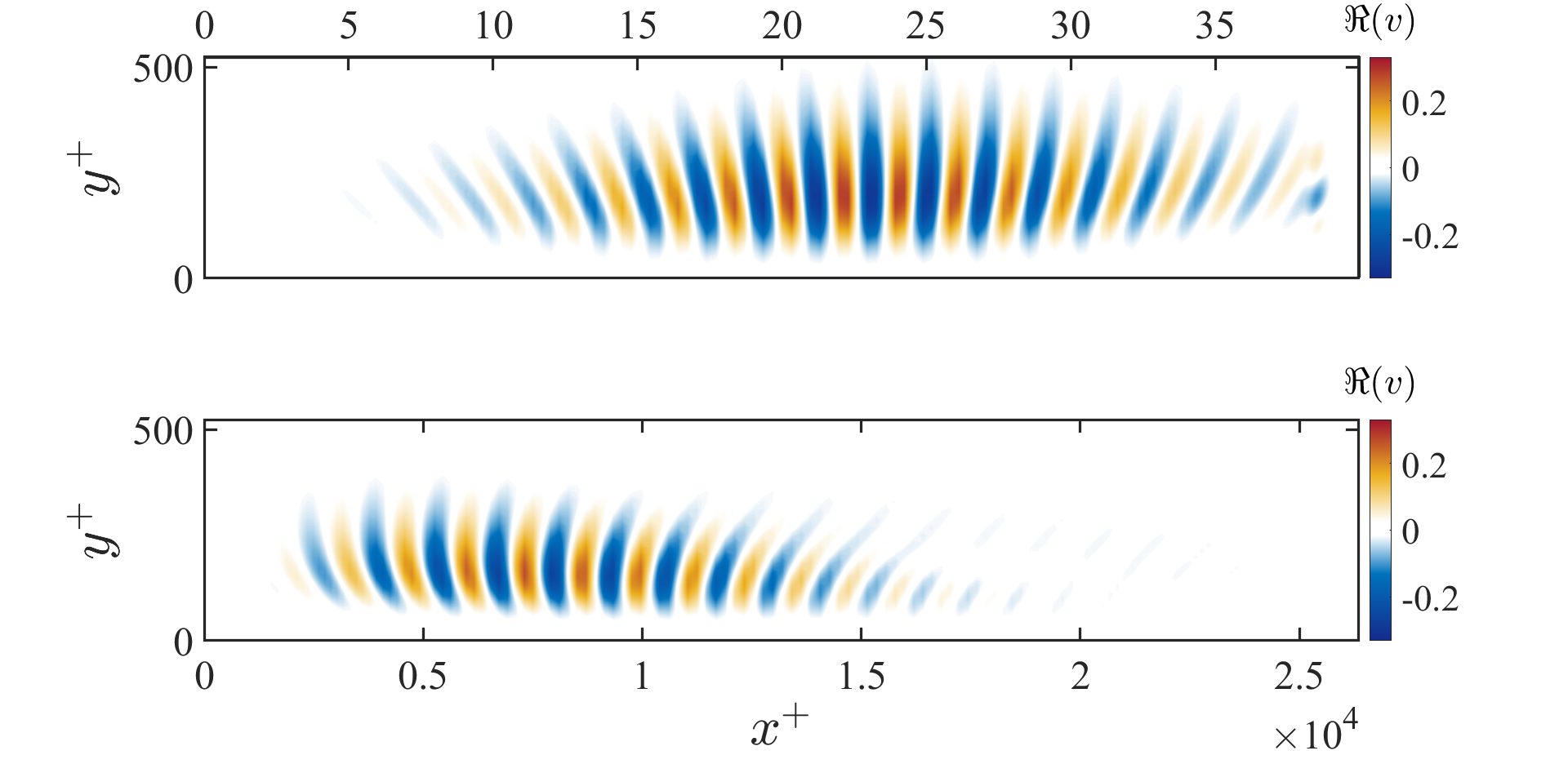}%
  \label{fig:BL_v_11}
}

    \caption{Singular values from the direct SVD (red squares) and variational reconstruction (black circles) at $Re_{\tau} \approx 700$. $[k_z,\omega] =  [43.9, 1.8]$ (a), $[k_z,\omega] =  [183, 3.6]$ (b). Real part of optimal resolvent response mode $(\boldsymbol{\psi}_1)$ for $[k_z,\omega] =  [11.0, 2.3]$: $u$ (c) and $v$ (d). Top panels: true global modes, bottom panels: VRA model. Upper x-axis: represents outer units $x$, lower x-axis represents inner units $x^+$.  }
    
\end{figure}

To illustrate the limits of our method we consider a wavenumber frequency combination for which the resolvent mode is localized in the wake region of the boundary layer: $[k_z,\omega] = [11.0,2.3]$. The model parameters, $N_{k_x}$, $N_c$, and $N_j$ (summarized in Table \ref{tab:BL}) were chosen such that further increasing the degrees of freedom no longer provided a meaningful speed up over the SVD of the original system. While the range of $Re_{\tau}$ is the same as for the mode with $k_z =44$, here the global resolvent mode has a much larger wall normal extent and is strongly affected by the streamwise development of the mean flow \citep{ruan_direct_2021}. Figures \ref{fig:BL_u_11} and  \ref{fig:BL_v_11} show the comparison of the VRA reconstruction of the resolvent response mode and the true reference response mode. As is clear from the figure, the VRA model completely fails to capture the broad support of the true mode in the outer wake region, and is instead much more localized closer to the wall and further upstream. Despite the lack of agreement between the VRA prediction and the RA mode, the VRA does reasonably predict the streamwise wavelength of the oscillations of this outer scaled mode and the relative amplitudes between $u$, $v$, and $w$ (not shown). This example illustrates that for strongly streamwise dependent flows, local, and thus streamwise periodic resolvent modes are inadequate for even qualitative reconstruction of the resolvent modes. More generally, if the boundary conditions of the modelling basis differ too much from those of the system being investigated the results of the VRA reconstruction may be inaccurate.
Better agreement could potentially be obtained by artificially altering the streamwise variation of the input basis to more closely match the desired result. However, such basis optimization is beyond the scope of this work.

\subsection{Computational Complexity}
Finally, in Table \ref{tab:computation} we compare the wall time and memory usage of the VRA model to the SVD of the original system for the modes in \S\ref{sec:BLinner}. We do not include the outer mode since this case the VRA method failed to even qualitatively replicate the true mode. The computations were all carried out on the Richardson computing cluster at Caltech using the same discretization and mode parameters as summarized in Table \ref{tab:BL}. The direct SVD computations include the inversion and SVD of $\mathbf{L}^{2D}$ using the LU decomposition and Arnoldi method described above. For the VRA model the computation includes the computation of the local resolvent mode basis as well as the construction and spectral decomposition of the variational matrices (\ref{EVP_basic}). Both methods require the construction of $\mathbf{L}^{2D}$ and thus we do not include it in this comparison. The construction of $\mathbf{L}^{2D}$ takes approximately 90 and 20 seconds for $k_z = 43.9$  and $k_z = 183$ respectively. For both cases we see a roughly $97\%$ reduction in wall time. The memory savings are significant but less drastic at $42\%$ and $76\%$ respectively. While the VRA model does not require any inversion it still requires knowledge of the full size $4N_{x}N_{y} \times 4N_{x}N_{y}$ matrix $\mathbf{L}^{2D}$ leading to these more modest gains in memory usage. We acknowledge that in this case the VRA method does not reproduce the the SVD modes exactly and so this comparison should be viewed in the context of a trade-off in cost and accuracy. However, considering that the VRA model replicates all the characteristic features of the SVD modes we believe our method alleviates a significant computational bottleneck in the computation of resolvent modes of non-periodic 2D systems such as the ZPGTBL considered here.

\begin{table}
    \centering
    \begin{tabular}{c|c|c|c|c|c|c|c|c|c}
       $k_z$ & $\omega$  & $N_x$ & $N_y$  & $N_{k_x}$  & $N_c$  & $N_{SVD}$ & $c_{min}$ & $c_{max}$ & $ \left(N_{k_x}\times N_c \times N_{SVD}\right)/\left(4 \times N_x\times Ny\right)$  \\
       \hline
        $43.9$  & $1.8 $  & $192$ & $81$  & 26  & 3  & 6 & $0.2U_{\infty}$ & $0.65U_{\infty}$ & $1/133$  \\
         $183$  & $3.6 $  & $96$ & $81$  & 16  & 3  & 1 & $0.05U_{\infty}$ & $0.25U_{\infty}$ & $1/648$  \\
        $11.0$  & $2.3$  & $192$ & $81$  & 32  & 6  & 10 & $0.6U_{\infty}$ & $0.99U_{\infty}$ & $1/33$  \\
    \end{tabular}
    \caption{Global parameters $(k_z, \omega)$, spatial discretization of the full system $(N_x,N_y)$, modeling parameters of the VRA model $(N_{k_x}, N_c, N_{SVD},c_{min}, c_{max})$ and model reduction from full system to VRA model.}
    \label{tab:BL}
\end{table}

\begin{table}
    \centering
    \begin{tabular}{c|c|c|}
        Method & Wall time & RAM used  \\
        \hline
        LU/Arnoldi SVD $(k_z = 43.9)$ & 72 min & 5.98 GB  \\
        VRA $(k_z = 43.9)$ & 2 min & 3.47 GB  \\
        LU/Arnoldi SVD $(k_z = 183)$ & 14 min & 5.34 GB  \\
        VRA $(k_z = 183)$ & $<$1 min & 1.26 GB  \\
    \end{tabular}
    \caption{Wall time and memory requirements for the LU/Arnoldi based SVD and the VRA model with the parameters in Table \ref{tab:BL}. The construction of linear operator $\mathbf{L}^{2D}$ is required for both methods and is thus not included in this comparison.}
    \label{tab:computation}
\end{table}

% BL PSI plots

% BL PHI plots

% BL Sigma and outer mode Plots

% Directional amplification / Sensitivity analysis section
\section{Sensitivity analysis: the influence of rank and condition number}\label{sec:direction}
In both \S\ref{sec:ECS} and \S\ref{sec:BL} we observed that even when the resolvent response modes, $\boldsymbol{\psi}_j$ were modeled accurately by the VRA method, the singular values, $\sigma_j$, and the forcing modes, $\boldsymbol{\phi}_j = \sigma_j \mathbf{L}\boldsymbol{\psi}_j$, may be susceptible to significant error. This is due to the directional amplification of the resolvent operator $\mathbf{H}$ which in the classical CT view of RA minimizes error in the response to errors in the forcing, but in this VRA framework amplifies errors in the predicted forcing due to errors in the response. This phenomenon can be demonstrated using (\ref{RAsvd}).We note that similar analysis has been performed by \citet{schmid_analysis_2014}, who considered the sensitivity of the eigenvalues and eigenvectors of the linearized NS operator to wide range of types of perturbations.  

Consider the action of $\mathbf{H}$ and $\mathbf{L}$ on arbitrary inputs $\tilde{\boldsymbol{\phi}}$ and $\tilde{\boldsymbol{\psi}}$, respectively:
\begin{equation} \label{RA_H_times_f}
    \mathbf{H}\tilde{\boldsymbol{\phi}} = \sum_{j}\sigma_{j}\boldsymbol{\psi}_{j}\left\langle \boldsymbol{\phi}_{j},\tilde{\boldsymbol{\phi}} \right\rangle 
\end{equation}
\begin{equation} \label{RA_L_times_q}
    \mathbf{L}\tilde{\boldsymbol{\psi}} = \sum_{j}\sigma_{j}^{-1}\boldsymbol{\phi}_{j}\left\langle \boldsymbol{\psi}_{j},\tilde{\boldsymbol{\psi}} \right\rangle.
\end{equation}
where $\tilde{\boldsymbol{\phi}}$ and $\tilde{\boldsymbol{\psi}}$ have unit norm. Suppose we chose $\tilde{\boldsymbol{\phi}} = a\boldsymbol{\phi}_{1} + b\mathbf{f}$ and $\tilde{\boldsymbol{\psi}} = a\boldsymbol{\psi}_{1} + b\mathbf{q}$ such that $\left\langle \boldsymbol{\phi}_{1},\mathbf{f}\right\rangle = 0$ and $\left\langle \boldsymbol{\psi}_{1},\mathbf{q}\right\rangle = 0$ as separate approximations for $\boldsymbol{\phi}_{1}$ and $\boldsymbol{\psi}_{1}$, respectively. Equation \ref{RA_H_times_f} demonstrates that the higher order response modes are weighted by $\sigma_{j}<\sigma_{1}$ for $j>1$, indicating that the component of $\tilde{\boldsymbol{\phi}_{1}}$ along $\boldsymbol{\phi}_{1}$ is weighed more heavily than the error $\mathbf{f}$ when approximating the leading response mode. On the contrary, (\ref{RA_L_times_q}) demonstrates that the output in the direction of $\boldsymbol{\phi}_{1}$ is weighted by the smallest singular value of $\mathbf{L}$, $\sigma_{1}^{-1}$, whereas the other components are weighted by the larger singular values, $\sigma_{j}^{-1}$ with $j>1$. When using (\ref{RA_definition_of_forcing}) to predict  $\boldsymbol{\phi}_{1}$ based on an approximation of $\boldsymbol{\psi}_{1}$, projection of the error onto higher order modes corrupts the prediction by weighing the output onto higher order forcing modes.

The differences between the error in approximating the gain in $\mathbf{H}$ and $\mathbf{L}$ can be quantified through a perturbation analysis of the singular values. The singular values are related to the resolvent response and forcing modes by 
\begin{equation}
    \sigma_j^{2} = \left(\mathbf{H}\boldsymbol{\phi}_j\right)^H\mathbf{Q}\left(\mathbf{H}\boldsymbol{\phi}_j\right) = \left(\left(\mathbf{L}\boldsymbol{\psi}_j\right)^H\mathbf{Q}\left(\mathbf{L}\boldsymbol{\psi}_j\right)\right)^{-1}.
\end{equation}
We consider the sensitivity of $\sigma_j$ to perturbation in either the resolvent forcing or response modes: $\boldsymbol{\psi}_{j,\epsilon} = \boldsymbol{\psi}_j + \epsilon \mathbf{r}$ and $\boldsymbol{\phi}_{j,\epsilon} = \boldsymbol{\phi}_j + \epsilon \mathbf{g}$, where $\epsilon \ll 1$ and $||\boldsymbol{\psi}_j|| =  ||\boldsymbol{\phi}_j||=||\mathbf{r}||=||\mathbf{g}||=1$. We define the perturbed singular value: $\sigma_{j,\epsilon,L} \equiv \sigma_j\left(\boldsymbol{\psi}_{j,\epsilon}\right)$ and $\sigma_{j,\epsilon,H} \equiv \sigma_j\left(\boldsymbol{\phi}_{j,\epsilon}\right)$. We may then derive the bounds on the error induced by the $\epsilon$ small perturbation in the singular modes:
\begin{equation}\label{sig_bound_L}
    \frac{|\sigma_{j,\epsilon,L} - \sigma_j|}{\sigma_j} \leq \epsilon \sigma_j ||\mathbf{L}||
\end{equation}
\begin{equation}\label{sig_bound_H}
    \frac{|\sigma_{j,\epsilon,H} - \sigma_j|}{\sigma_j} \leq \epsilon \frac{\sigma_1}{\sigma_j}.
\end{equation}
The details of the derivation are included in Appendix \ref{app:sigma_sensitivity}.
% error bounds in Phi
We can perform a similar analysis to investigate the sensitivity of the predicted forcing modes to perturbations in the response modes and vice versa.
\begin{equation}
    \boldsymbol{\phi}_{j,\epsilon,\psi} \equiv \sigma_{j,\epsilon,L} \mathbf{L} \left(\boldsymbol{\psi}_j + \epsilon \mathbf{r}\right)
\end{equation}
\begin{equation}
    \boldsymbol{\psi}_{j,\epsilon,\phi} \equiv \sigma_{j,\epsilon,H} \mathbf{H} \left(\boldsymbol{\phi}_j + \epsilon \mathbf{g}\right)
\end{equation}
Here the $\sigma_{j,\epsilon,L}$ and $\sigma_{j,\epsilon,H}$ are the same as defined above. The error in the resolvent modes may be bounded as follows
\begin{equation}\label{phi_bound}
    \|\boldsymbol{\phi}_{j,\epsilon,\psi}-\boldsymbol{\phi}_{j}\| \leq \epsilon \left(\sigma_j \|\mathbf{L}\| + 1\right)\sigma_j \|\mathbf{L}\|
\end{equation}
\begin{equation}\label{psi_bound}
    \|\boldsymbol{\psi}_{j,\epsilon,\phi}-\boldsymbol{\psi}_{j}\| \leq \epsilon \left( \frac{\sigma_1}{\sigma_j} + 1 \right)\frac{\sigma_1}{\sigma_j}
\end{equation}
where again we relegate the details to Appendix \ref{app:mode_sensitivity}. These results imply that as long as $\sigma_j/\sigma_1$ is not too large an $\mathcal{O}(\epsilon)$ perturbation to $\boldsymbol{\phi}$ leads to an error of $O(\epsilon)$ in $\sigma$ and $\boldsymbol{\phi}$; however, an $O(\epsilon)$ perturbation in $\boldsymbol{\psi}$ leads to an error in $\sigma$ and $\boldsymbol{\phi}$ that is expected to be larger by a factor of $\sigma_j \| \mathbf{L}\|$. To analyze how large the factor is expected to be we follow the analysis of \citet{symon_non-normality_2018} and consider the spectral decomposition of $\mathbf{L} = \mathbf{V}\boldsymbol{\Lambda}\mathbf{V}^{-1}$ which allows us to to rewrite (\ref{sig_bound_L}) as
\begin{equation}\label{sigma_error_bound_kappa}
    \frac{|\sigma_{j,\epsilon,L} - \sigma_j|}{\sigma_j} \leq \epsilon \kappa \frac{\sigma_j}{\sigma_{min}}   \leq \epsilon \kappa \frac{\sigma_1}{\sigma_{min}}   
\end{equation}
where 
\begin{equation}
    \sigma_{min} \equiv \min_{\sigma_j \in \boldsymbol{\Sigma(\mathbf{H})}}\sigma_j = \left(\max_{\lambda_j \in \boldsymbol{\Lambda}(\mathbf{L})} \left(\lambda_j\right)\right)^{-1}
\end{equation}
is the minimum singular value of the resolvent and $\kappa \equiv \|\mathbf{V}\|\|\mathbf{V}^{-1}\|$ is the condition number. The latter is always greater than one and quantifies the non-orthogonality of the eigenvectors, and thus the non-normality of the operator. The associated pseudo-resonance, where small perturbations to the operator leads to large perturbations to the eigenvalues \citep{trefethen_pseudospectra_2005-2}.  Thus there are two mechanisms which lead to an increased sensitivity of singular values and forcing modes to perturbations in the response modes. First, the relative resonant amplification of the mode quantified by  $\sigma_j/\sigma_{min}$, and second, the pseudo-resonant amplification of the linear dynamics quantified by $\kappa$.

% Toy Problem
\subsection{Perturbation analysis of a simplified example}
\begin{figure}
    \centering
    \includegraphics[trim = 80 0 0 0,  scale =  0.37]{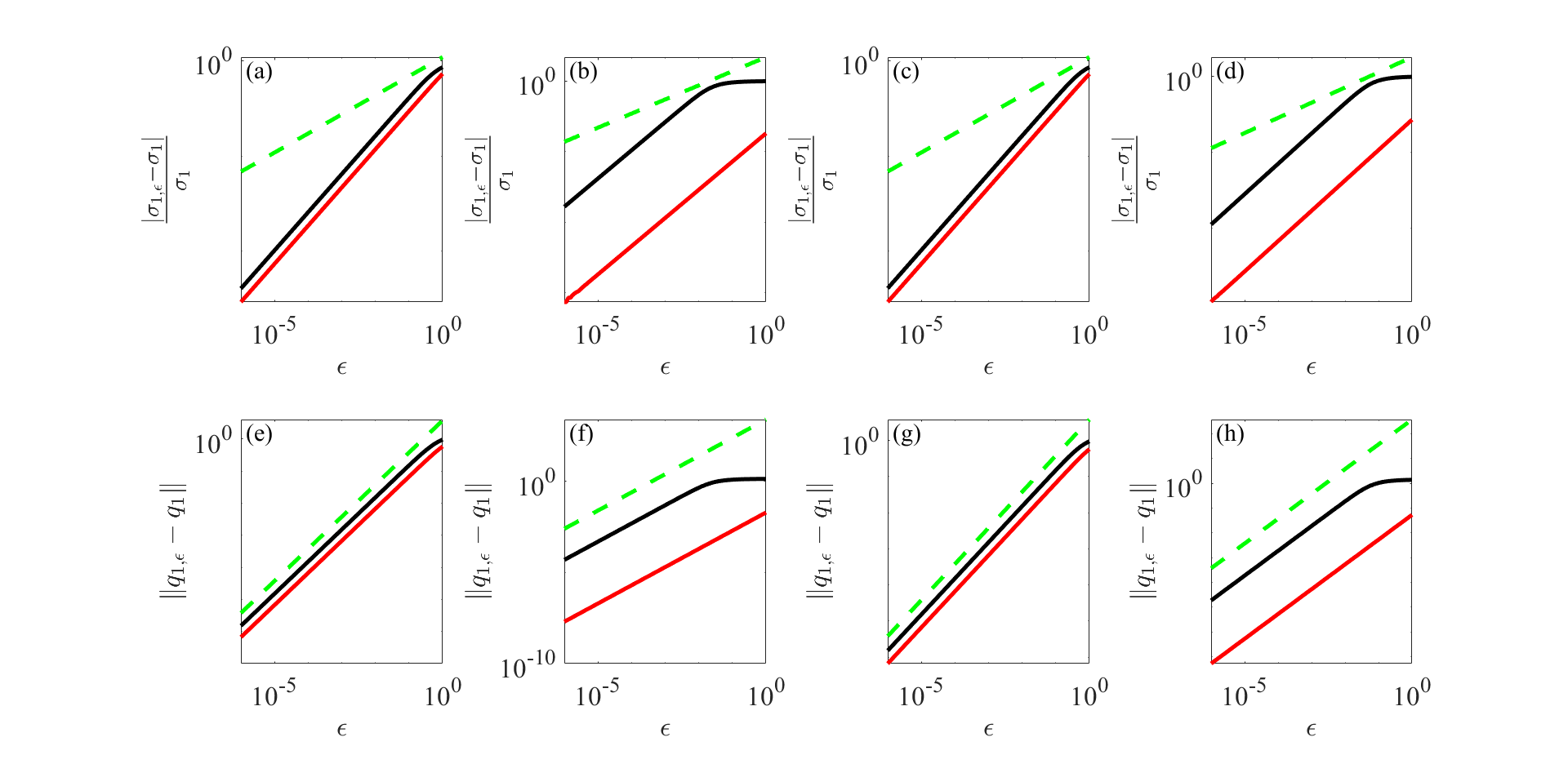}
    \caption{Top row: relative error in singular value, normal$(c=0)$: $b/a = 1.5$ (a), $b/a = 50$ (b), non-normal: $c/a = 0.1$ (c), $c/a = 5$ (d). Bottom row: relative error in singular modes, normal$(c=0)$: $b/a = 1.5$ (e), $b/a = 50$ (f), non-normal: $c/a = 0.1$ (g), $c/a = 5$ (h). Color code: error due to perturbation to  $\psi$ (solid black),  error due to perturbation to $\phi$ (solid red), and derived upper bound (dashed green). }
    \label{fig:toy_problem}
\end{figure}

To illustrate the effects of resonant and pseudo-resonant amplification on the error in singular values and singular modes we compute  $|\sigma_{1,\epsilon,L}  - \sigma_1|$, $|\sigma_{1,\epsilon,H} - \sigma_1|$, $|\phi_{1,\epsilon,L}  - \phi_1|$, and $|\psi_{1,\epsilon,H} - \psi_1|$,  for the model operator
\begin{equation}
    \mathbf{L} =
    \begin{bmatrix}
    a & c \\
    0 & b
    \end{bmatrix}
\end{equation}
for a range of $\epsilon$. To test the resonant amplification, we compare the error in singular values for normal operators $\mathbf{L}$ with the parameters set to $[a,b,c] = [1,1.5,0]$ and $[a,b,c] = [1,50,0]$. To test the pseudo-resonant effects, we introduce and vary the off-diagonal term $c$ that makes $\mathbf{L}$ non-normal. We compare $[a,b,c] = [1,1.5,0.1]$ and $[a,b,c] = [1,1.5,5]$. In each case we set the perturbation vectors $\mathbf{r}$, $\mathbf{g}$ to be orthogonal to $\psi_1$ and $\phi_1$ respectively. The error in singular values is plotted in the top row of Figure \ref{fig:toy_problem} and the error in the singular modes is plotted in the bottom row of Figure \ref{fig:toy_problem}. These plots reveal that as expected, the error grows with $\epsilon$, but when the resonant or pseudo-resonant effects are increased the error due to a perturbation in $\boldsymbol{\psi}$ is significantly greater by a factor of several orders of magnitude.  The error in singular modes is proportional to $\epsilon$ as predicted by the derived error bounds, while for small $\epsilon$ the error in singular values actually grows as $\epsilon^2$. This is due to the fact that for this toy problem the perturbation is chosen to be orthogonal to the singular vectors which causes the $\mathcal{O}(\epsilon^2)$ contribution to dominate.

\FloatBarrier

\subsection{Implications and limitations}
This analysis illustrates an inherent limitation of the VRA framework. The benefits of circumventing the inversion of the linear dynamics come at the cost of losing the directional amplification of the resolvent operator. While the VRA based estimation of the resolvent response modes is robust to the non-normality of the linear dynamics, since the matrices in (\ref{EVP_basic}) are normal, small errors in the resolvent response modes predicted by our method can lead to significant errors in the forcing modes and singular values if $\kappa$ is large and/or the resolvent operator is very low rank. This marks a difference between the success of VRA and RA algorithms that approximate the SVD with matrix sketching. The low rank nature of the resolvent operator assists the latter by improving the convergence but becomes a source of error in the former \citep{ribeiro_randomized_2020}.

Another factor leading to the increased error in the singular values and forcing modes we have observed is that the continuous linear differential operators being analyzed have unbounded spectra. Therefore the maximum eigenvalue of the discretized operator $\mathbf{L} \in \mathbb{C}^{n\times n}$ grows with its size $n$. In particular we expect the maximum eigenvalue of second order differential equations like the ones considered here to scale with $n^2$. However, depending on the numerical discretization used, the largest eigenvalues may be spurious, as in the case of Chebyshev differentiation matrices, where the largest eigenvalue scales with $n^{4}$ for these second order differential equations \citep{trefethen_spectral_2000}. This implies that the VRA reconstruction of the singular values and forcing modes becomes increasingly sensitive to errors in the response modes as the number of basis elements grows.

These are noteworthy limitations of our proposed method since the cost saving potential of the proposed method is greatest for larger systems and additionally in many flows of interest the resolvent operator is, in fact, low rank. Nonetheless, in most cases the aim of equation-driven modal analysis techniques such as resolvent analysis is to identify coherent structures or obtain an efficient modeling basis \citep{rosenberg_efficient_2019,nogueira_large-scale_2019,barthel_closing_2021}. In these cases the resolvent response modes, which our method can predict independent of condition number or singular value separation, are of primary interest. In the resolvent formulation of the nonlinear NSE, the forcing modes arise through their projection onto the nonlinear interaction of the response modes: $\langle \boldsymbol{\phi},\boldsymbol{\psi}\cdot\nabla\boldsymbol{\psi}\rangle$ \citep{mckeon_engine_2017,barthel_closing_2021}. As discussed in \S\ref{sec:direction}, the error in the forcing modes arises due to higher order $(j\gg1)$ response modes with very small $\sigma_j$ being amplified through the action of $\mathbf{L}$. However, since these higher order modes are not expected to be dynamically relevant \citep{morra_colour_2021}, especially if the resolvent is low rank, they generally will not have significant projection onto the actual nonlinear interaction of the response. This may in some cases ameliorate the practical implications of the error in forcing modes since even if there is significant error in $\boldsymbol{\phi}$, the error in the relevant metric: $\langle \boldsymbol{\phi},\boldsymbol{\psi}\cdot\nabla\boldsymbol{\psi}\rangle$ is expected to be small.

\FloatBarrier
%% Discussion Section
\section{Discussion}\label{sec:discussion}
\begin{figure}
    \centering
    \includegraphics[trim = 0 0 0 0,  scale =  0.95]{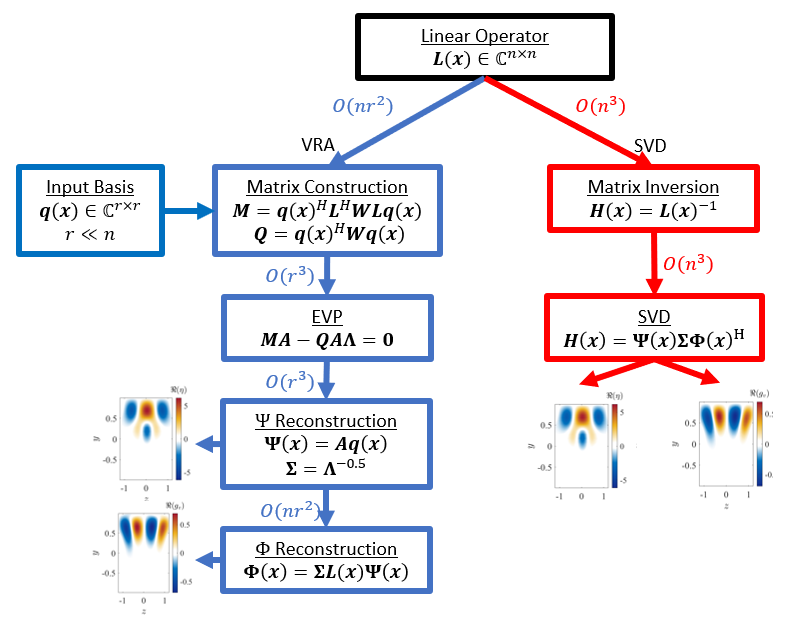}
    \caption{Comparison of the matrix operations and computational complexity involved in the VRA and SVD-based computations of resolvent modes. VRA operations are shown in blue, SVD operations are shown in red. }
    \label{fig:flowchart}
\end{figure}

The examples presented in this paper illustrate the avenues of progress enabled by the VRA formulation of resolvent analysis. First, circumventing the inversion of the linear operator in the definition of the resolvent modes allows for analytical manipulation. This facilitates the derivation of scaling laws and parametric dependencies as we have done in \S\ref{sec:analytical approximations}. Second, from a numerical point of view, the VRA method avoids the calculation of a matrix inverse and applying expensive linear algebra decompositions to the matrices. Figure \ref{fig:flowchart} outlines the matrix operations and computational complexity of the VRA method presented herein and the direct SVD. For a matrix of dimension $n \times n$, calculating the inverse, performing an LU decomposition, and applying an SVD are each $\mathcal{O}(n^{3})$ operations. The resolvent matrix, calculated as the inverse of a matrix, is in general, a dense matrix which leads to large memory costs in terms of storage. Even avoiding the inverse by applying the LU decomposition as explained in \S\ref{sec:BL} would require storage of large dense triangular matrices.  Typically when the LNS operator is discretized, the resulting matrix is sparse. Sparse matrices have the advantage that only their nonzero elements are stored and sparse matrix operations can be computed more efficiently. Even though the discretizations described herein use spectral methods, the discretized LNS operator described in \S\ref{sec:BL}, $\mathbf{L}^{2D}$, boasts sparsity of less than $1\%$. In the VRA method, the sparse discretized LNS operators are only used for matrix multiplication with the basis to create the $r\times r$ matrices $\mathbf{M}$ and $\mathbf{Q}$ for the eigenvalue problem in (\ref{EVP_basic}). Since the analytical form of the LNS operator is known, the matrix multiplications can be avoided altogether if the basis is defined with analytic functions, as demonstrated in \S\ref{sec:analytical approximations}.  Although the resulting matrices $\mathbf{M}$ and $\mathbf{Q}$ are dense, the eigenvalue problem can be solved almost trivially with standard methods as it scaled with $\mathcal{O}(r^{3})$ where $r \ll n$. Even if the number of basis elements, $r$, becomes large, the eigenvalue problem could be solved with approximate methods like the Arnoldi Algorithm with the Shift and Invert method. 

As discussed in \S\ref{sec:direction}, the VRA method is prone to error in predicting the singular values and forcing modes when there is strong non-normality or the operator is very low rank. In this sense the herein proposed VRA method provides a natural compliment to the recently developed randomized resolvent analysis method proposed by \citet{ribeiro_randomized_2020}, which is particularly effective when the resolvent is low rank. However, we reiterate that response modes can be modeled accurately regardless of these properties and at a fraction of the cost of a direct SVD. Furthermore, it is these response modes that are generally of primary interest. They have been shown to be an efficient basis for a variety of flows including turbulent jets \citep{schmidt_spectral_2018,pickering_optimal_2021}, boundary layers \citep{sipp_characterization_2013,rigas_nonlinear_2021}, exact coherent states, \citep{sharma_low-dimensional_2016,rosenberg_efficient_2019} and others. Notably, \citet{sharma_low-dimensional_2016} showed that using five response modes per Fourier mode for the N3L lower branch solution in a pipe, fluctuations were reconstructed retaining $98\%$ of the fluctuation energy. Using only one response mode per Fourier mode, they were able to reconstruct $95\%$ of the fluctuation energy. \citet{towne_spectral_2018} also studied the similarities between RA and SPOD. They found that the response modes and the data driven SPOD modes are equivalent when there is uncorrelated, white-noise forcing. This implies that in certain conditions RA could be used as a predictive tool to model near wall structures in the simulation of high Reynolds number wall bounded flows, where large numerical resolution is needed to resolve the near wall structures. Furthermore, since the proposed method is derived directly from the definition of the forced linear system, the method is not fundamentally limited to linear systems or a certain type of input basis. 

% Even if the underlying dynamics are highly non-normal the matrices in the eigenvalue problem (\ref{EVP_basic}) are self-adjoint (Hermitian). 

The primary limitation is that the spatial support of the input basis needs to overlap with the spatial support of the resolvent modes being estimated. In particular we saw in \S\ref{sec:BLouter} that a sufficiently strong mismatch between the boundary conditions of the input basis and the linear operator can lead to significant errors in the VRA reconstruction. In general, a critical layer mechanism (as in \S\ref{sec:ECS}) or scaling laws (as in \S\ref{sec:BL}) dictate the spatial localization and length scale of resolvent modes and thus one can reliably predict this region of support a priori. However, for flows where the general region of spatial support can not be predicted, a larger input basis with a broader range of wave numbers and spatial support may be necessary.  We found the most important parameter is the number of retained spatial wavenumbers, $N_{k_z}$ or $N_{k_x}$ and if the largest relevant wavenumber is not known a priori it may be necessary to progressively increase these parameters until convergence is obtained.  Additionally, unlike some recent equation-free methods such as \citet{herrmann_data-driven_2021} our method does rely on knowledge of the linearized dynamics of the system, which in some cases may not be known a priori. In this regard the primary challenge is generally lack of knowledge of the mean flow. However, recently several authors have developed methods to efficiently estimate the mean dynamics for a range of flows \citep{mantic-lugo_self-consistent_2014,mantic-lugo_self-consistent_2015,rosenberg_computing_2019}. Such techniques could be combined with the method presented in this work to efficiently compute resolvent modes in situations were the mean dynamics are unknown, or would be costly to compute directly, although this is beyond the scope of this  work.  

\FloatBarrier
\section{Conclusions}\label{sec:conclusion}
In this work we have suggested an alternative conceptual framework based on the calculus of variations from which to view resolvent analysis. In this variational framework the resolvent response modes are defined as the stationary points of an operator norm subject to a relevant norm constraint. To the best of the authors' knowledge this definition is novel, at least in the context of resolvent analysis. We proved that this variational formulation is equivalent to the standard SVD-based definition, and introduced a method to estimate the resolvent modes of complex systems as expansions in lower dimensional basis functions. The crucial advantage of this formulation and the method presented herein is the lack of reliance on the inversion of the linear operator, which from a theoretical point of view allows for easier analytical manipulation, and from a practical point of view enables drastic model reduction and leads to a significant reduction in computational complexity. 

The analytical advantages were illustrated on the example of streamwise constant structures in a turbulent channel flow, where we derived a closed form solution to the Euler-Lagrange equations governing the optimal resolvent mode. Resolvent modes have shown to encode physically relevant features of turbulence \citep{mckeon_engine_2017}, and therefore we believe the improved analytical tractability of the variational formulation will open the door to the further understanding and discovery of the underlying physics.  

The numerical advantages were illustrated first for both a 2D/3C equilibrium solution in plane Couette flow and a streamwise developing turbulent boundary layer. In the first example we showed that if the model basis satisfies the same boundary conditions as the full system, the VRA model converges to the true modes as more basis are included in the VRA model. In the second case we showed that even if the basis does not satisfy the correct boundary conditions, and the streamwise development of the mean flow is not too strong, the VRA model is able to reproduce the characteristic features of the SVD-based modes with a reduction of order of over two orders of magnitude resulting in an order of magnitude reduction in computation time and a $40-75\%$ reduction in RAM usage. As formulated here, the current method fails for flows with very strong streamwise development. In such cases more carefully chosen modeling basis, which already encode some of the anticipated streamwise development, is likely needed for our method to be viable.

Since the method estimates the response modes, we note that error between the true response mode and the VRA estimate is amplified when calculating the singular values and forcing modes. We note that this amplification in the error is due to the low rank nature and non-normality of the resolvent operator. Nonetheless, we believe that this conceptual approach to resolvent analysis can open the door for further analysis of the NSE and the discovery of new physics, as well as enabling the real time computation of resolvent modes in applications such as experiments and simulations where the cost of the standard SVD-based approach is prohibitive. 

% \newline
%% Acknowledgments and Appendixes
\begin{flushleft}
\textbf{Acknowledgments}
We thank Tony Leonard and Greg Chini for many inspiring discussions.
\newline

\textbf{Funding}
This work is supported by the Office of Naval Research under grants ONR N00014-17-1-2307 and N00014-17-1-3022.
\end{flushleft}

\appendix
% CR calculus
\section{Variation Over Complex fields}\label{app:CR}
The following derivation is an extension of the theory derived in \citep{wirtinger_zur_1927,brandwood_complex_1983}. Let $J=\langle F(q,q^*,\nabla q, \nabla q^*) \rangle \in \mathbb{R}$ where $q = a + i b\in \mathbb{C}^\infty$, with $a,b \in \mathbb{R}^\infty$. The functional $J$ can equivalently be written as $J=\langle F(a,b,\nabla a, \nabla b) \rangle$. The Euler-Lagrange equations defining stationary points of $J$ with respect to $a$ and $b$  are given by:
\begin{equation}
\frac{\delta F}{\delta a} \equiv \frac{\partial F}{\partial a} - \nabla \frac{\partial F}{\partial \nabla a} = 0
\end{equation}
\begin{equation}
\frac{\delta F}{\delta b} \equiv\frac{\partial F}{\partial b} - \nabla \frac{\partial F}{\partial \nabla b} = 0.
\end{equation}
Since $F \in \mathbb{R}$ a simple change of variables to $q$ and $q^*$ leads to

\begin{equation}
   \frac{\delta F}{\delta q} =\frac{1}{2} \left(\frac{\delta F}{\delta a} - i \frac{\delta F}{\delta b} \right)
\end{equation}

\begin{equation}
   \frac{\delta F}{\delta q^*} =\frac{1}{2} \left(\frac{\delta F}{\delta a} + i \frac{\delta F}{\delta b} \right)
\end{equation}
which implies that
\begin{equation}
\frac{\delta F}{\delta a}=\frac{\delta F}{\delta b} = 0 \Rightarrow \frac{\delta F}{\delta q} =  \frac{\delta F}{\delta q^*} = 0.
\end{equation}
Furthermore, since $F$, $a$, and $b$ are real functions it follows that 
\begin{equation}
\frac{\delta F}{\delta q} = 0 \Rightarrow \frac{\delta F}{\delta a}=\frac{\delta F}{\delta b} = 0
\end{equation}
\begin{equation}
\frac{\delta F}{\delta q^*} = 0 \Rightarrow \frac{\delta F}{\delta a}=\frac{\delta F}{\delta b} = 0 
\end{equation}
and therefore either of the above conditions is necessary and sufficient for stationarity.
%OS Eigenfunctions
\section{Orr-Sommerfeld Eigenfunctions}\label{app:LOS}
\begin{figure}
    \centering
    \includegraphics[trim = 50 0 0 0,  scale =  0.28]{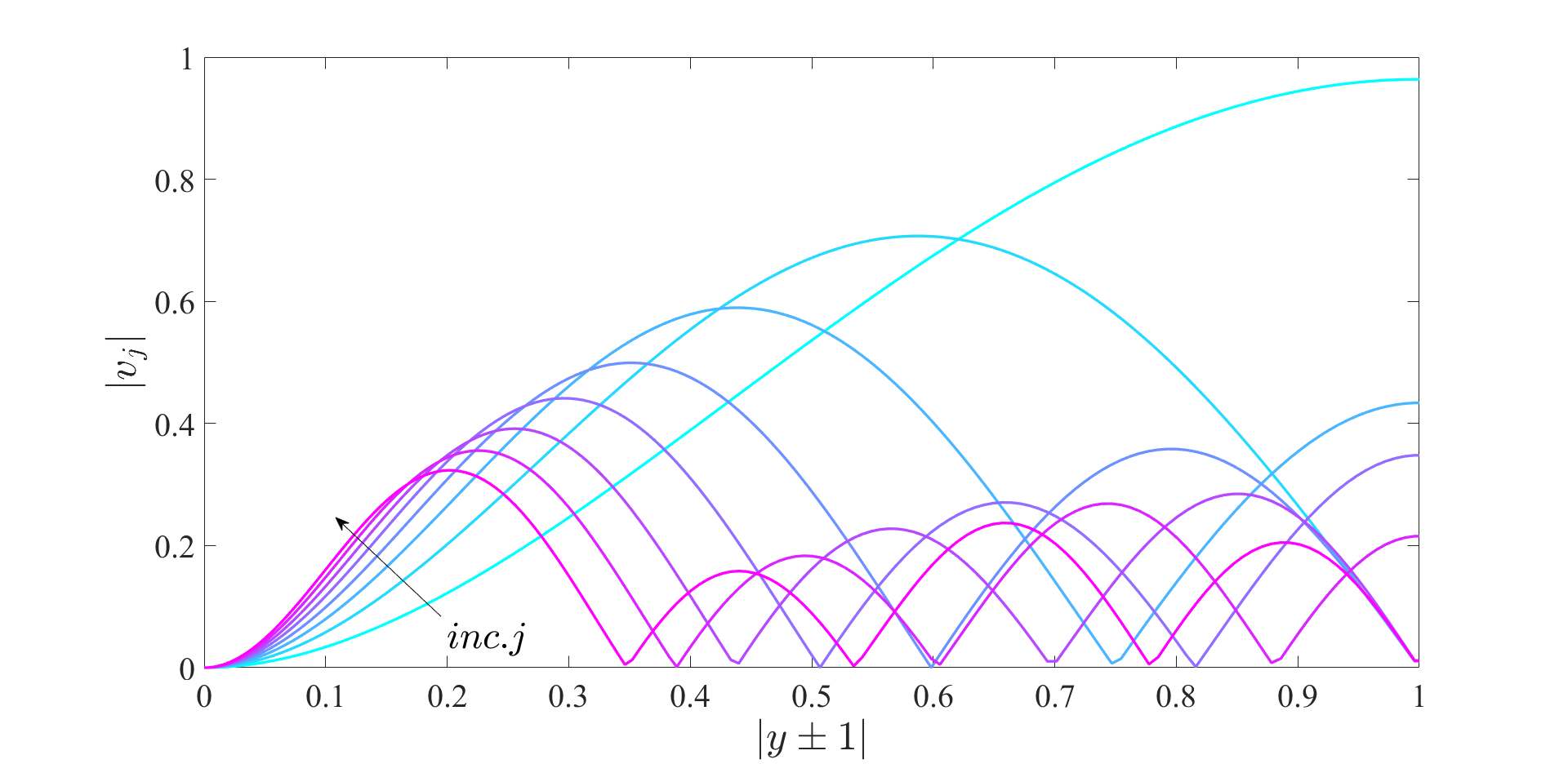}
    \caption{Absolute value of the Orr-Sommerfeld eigenfunctions $v_j$ $(j=1...8)$ for $k_x = 0$, $k_z = 6$, and $\omega = 0.1$.}
    \label{fig:inputbasis_kx0}
\end{figure}

\FloatBarrier
The Orr-Sommerfeld eigenvalue problem for $k_x = 0$ on the domain $y\in[-1,+1]$ is given by
\begin{equation}
-i\omega\nabla^2v_j - \frac{1}{R}\nabla^4v_j =\lambda^{OS}_j \nabla^2v_j 
\end{equation}
subject to the boundary condition $v(\pm1) = v_{y}(\pm1)  = 0$. This problem has been analyzed by several authors including \citet{dolph_application_1958,jovanovic_componentwise_2005} and the solutions are found to be:

\begin{equation}
\begin{aligned}
v_j(y;k_z)  =  A_j\left(\cos(\gamma_j(y+1))-\cosh(k_z(y+1))\right) + \\      B_j\left(\sin(\gamma_j(y+1))-\gamma_j k_z^{-1}\sinh(k_z(y+1))\right)
\end{aligned}
\end{equation}
\begin{equation}
\lambda^{OS}_j =\frac{1}{R}\left(\gamma_j^2 + k_z^2\right) - i\omega 
\end{equation}
where the $\gamma_j$ are defined as the roots of the following equation.
\begin{equation}
    \cos(2\gamma)\cosh(2k_z) - \left( \frac{k_z^2-\gamma^2}{2k_z\gamma}\right)\sin(2\gamma)\sinh(2k_z) - 1= 0
\end{equation}
The relative amplitudes $A_j$ and $B_j$ are defined for each $\gamma_j$ as the solutions of the following system.
\begin{equation}
     \begin{bmatrix}
\cos(2\gamma_j)-\cosh(2k_z) & \sin(2\gamma_j)-\left(\gamma_j k_z^-1\right)\sinh(2k_z)\\
-\gamma_j\sin(2\gamma_j)-k_z\sinh(2k_z &\gamma_j\left(\cos(2\gamma_n)-\cosh(2k_z)\right)
\end{bmatrix}
 \begin{bmatrix} 
A_j\\
B_j
\end{bmatrix}
=
 \begin{bmatrix} 
0\\
0
\end{bmatrix}
\end{equation}
In Figure \ref{fig:inputbasis_kx0} we plot the eigenfunctions $v_j$ for the same parameters plotted in \S\ref{sec:kx0}: $k_x=0$, $k_z = 6$, and $\omega = 0.1$. 

%Singular value scaling 
\section{Singular Value Scaling}\label{app:sigma}
The resolvent operator we consider in \S\ref{sec:kx0} is defined as 
\begin{equation}
\mathbf{H} = 
 \begin{bmatrix}
L_{OS} & 0\\
\bar{U}_y & L_{SQ}
\end{bmatrix}^{-1}
= 
 \begin{bmatrix}
L_{OS}^{-1} & 0\\
-L_{SQ}^{-1}\bar{U}_y L_{OS}^{-1} & L_{SQ}^{-1}
\end{bmatrix}.
\end{equation}
Noting the definitions (\ref{Lsq}) and (\ref{Los}), if $\omega = 0$, $\mathbf{H}$ may be written in the form
\begin{equation}
\mathbf{H} = 
 \begin{bmatrix}
R H_{vv} & 0\\
R^2 H_{uv} & R H_{uu}
\end{bmatrix}
\end{equation}
where $H_{vv}, H_{uv}, H_{uu} \neq f(R)$. This reveals that as $R\rightarrow \infty$, $\|H\| =\sigma_1 \rightarrow R^2\|H_{uv}\| \sim R^2$. If we further consider the limit $k_z\rightarrow \infty$ and rescale the wall normal coordinate $Y= k_z y$ we find
\begin{equation}
    \|H_{uv}\| = \nabla^{-2}\bar{U}_y\nabla^{-2} = k_z^{-3}\tilde{\nabla}^{-2}\bar{U}_y\tilde{\nabla}^{-2}
\end{equation}
where $\tilde{\nabla}^2= \frac{\partial^2}{\partial Y^2} -1$. Thus for $k_x = 0$ and as $\omega \rightarrow 0$, $R\rightarrow \infty$ and $k_z \rightarrow \infty$ we find that
\begin{equation}
    \sigma_1 \sim R^2 k_z^{-3}.
\end{equation}
A more in depth analysis can be found in \citep{jovanovic_componentwise_2005}.

%% Input basis Section
\section{Select Input Basis Elements}\label{app:inputbasis}
\begin{figure}
    \centering
    \includegraphics[trim = 100 0 0 0,  scale =  0.37]{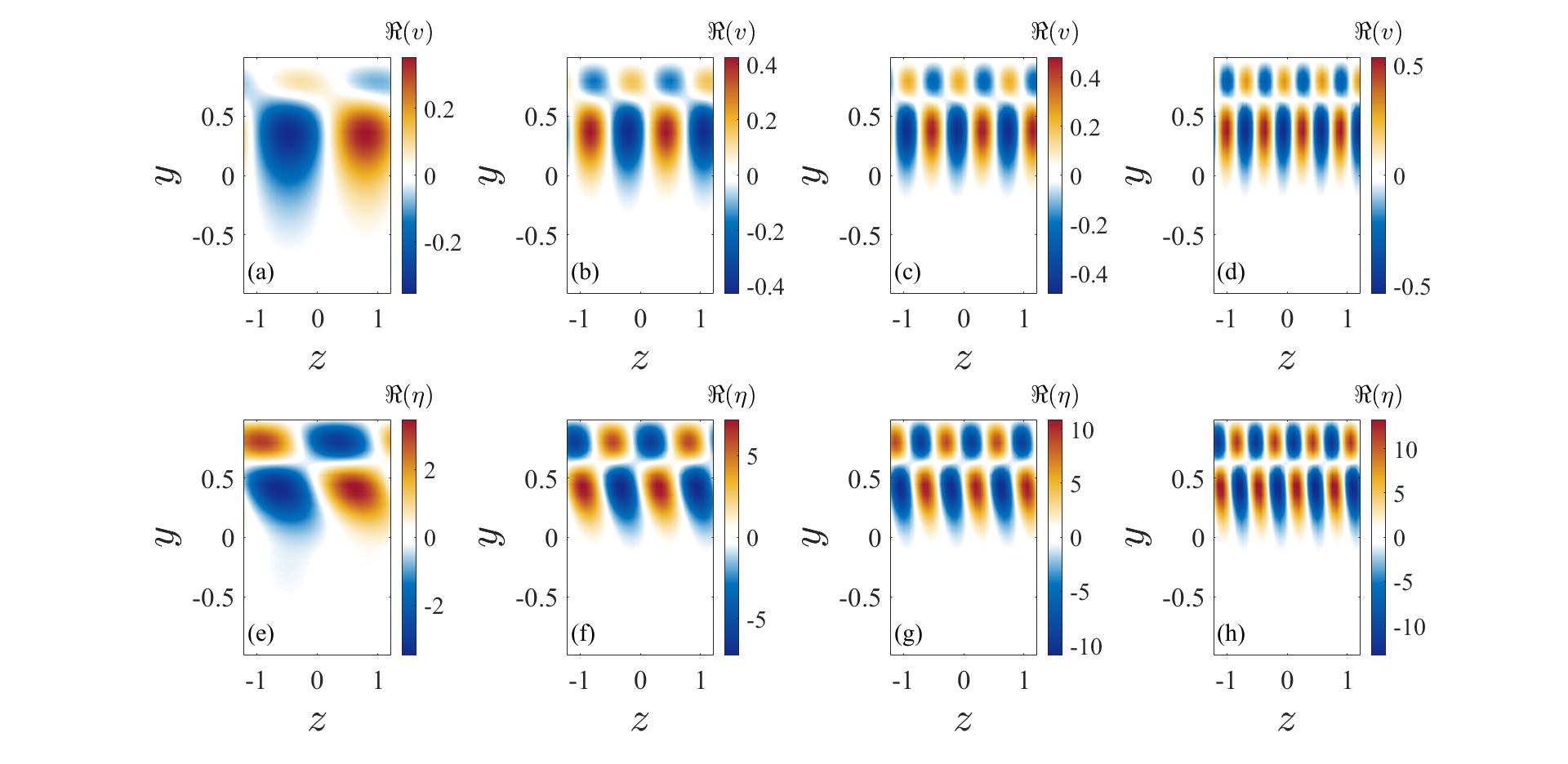}
    \caption{Select elements of input resolvent basis: $q(y,z) = \boldsymbol{\psi}_{k_x,k_z,\omega,j}(y) e^{i k_z z}$ for $k_x = 0.5$, $\omega = 0.375$, $j=1$ and $L_z k_z /2\pi =$ 1 (a,e), 2 (b,f), 3 (c,g), and 4 (d,h). Top row: $v$, bottom row $\eta$.}
    \label{fig:inputbasis_ECS}
\end{figure}

\begin{figure}
    \centering
    \subfloat[]{
    \includegraphics[trim = 140 20 0 20,  scale =  0.17]{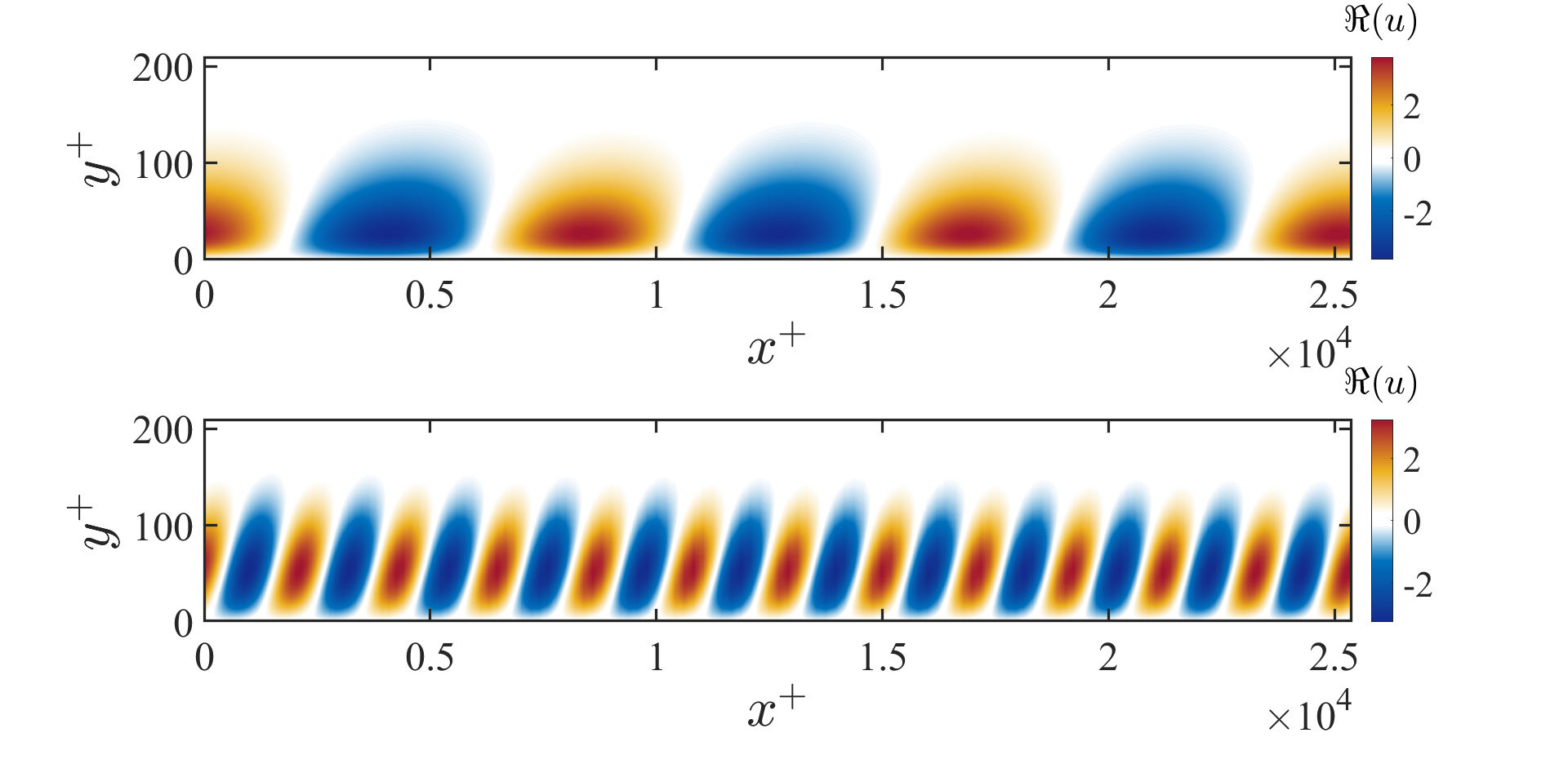}
}
\subfloat[]{
    \includegraphics[trim = 100 20 60 20,  scale =  0.17]{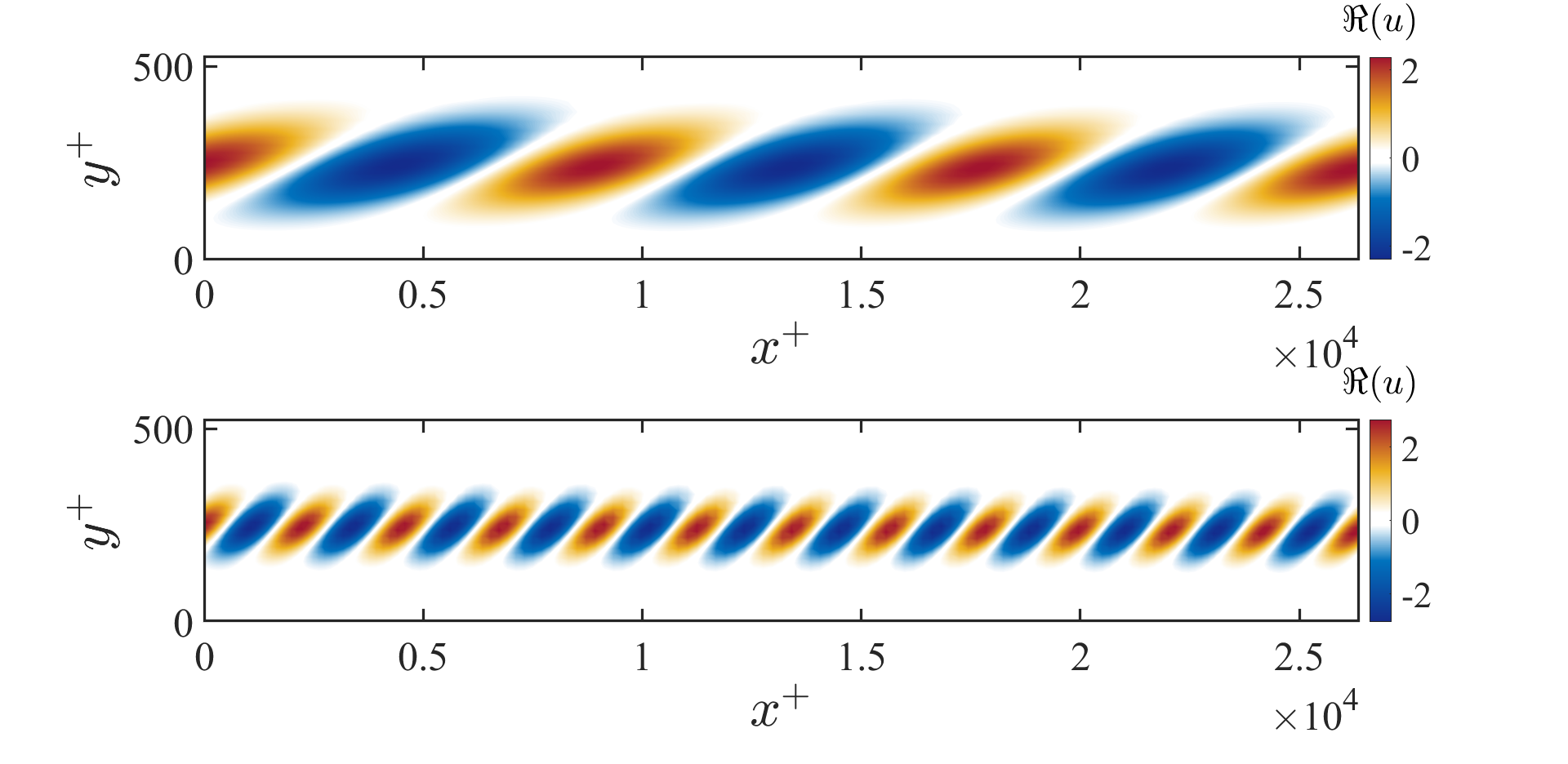}
    }
    \caption{Select elements of input resolvent basis: $q(x,y) = \boldsymbol{\psi}_{k_x,k_z,\omega,j}(y) e^{i k_x x}$ for $[k_z,\omega] = [44.0, 0.65 U_{\infty}/k_x]$ (a) and $[k_z,\omega] = [11.0, 0.8 U_{\infty}/k_x]$ (b). In both cases $ L_x k_x / 2 \pi = 3$ (top panel) and $12$ (lower panel), and in all cases $j=1$. }
    \label{fig:inputbasis_BL}
\end{figure}

In this section we plot a selection of representative input basis elements used in the 2D examples presented in this work. Figure \ref{fig:inputbasis_ECS} shows four of the local resolvent modes used in the VRA reconstruction of the 2D resolvent modes computed about EQ1 in in \S\ref{sec:ECS}. Figure \ref{fig:inputbasis_BL} show two of the local resolvent modes used in the reconstruction of the 2D resolvent modes computed about the ZPGTBL in \S\ref{sec:BL}. Figure \ref{fig:inputbasis_BL}a represents a ``wall-attached" mode used in the reconstruction of the global resolvent inner-mode with $[k_z,\omega] = [44.0,1.8]$. Figure \ref{fig:inputbasis_BL}b represents a ``wall-detached" mode used for the outer-mode with $[k_z,\omega] = [11.0,2.3]$. 

\FloatBarrier

\FloatBarrier
% Sensitivity analysis
\section{Singular Value Sensitivity}\label{app:sigma_sensitivity}
The true singular value/mode pairs satisfy
\begin{equation}
    \sigma_j^{2} = \left(\mathbf{H}\boldsymbol{\phi}_j\right)^H\mathbf{Q}\left(\mathbf{H}\boldsymbol{\phi}_j\right) = \left(\left(\mathbf{L}\boldsymbol{\psi}_j\right)^H\mathbf{Q}\left(\mathbf{L}\boldsymbol{\psi}_j\right)\right)^{-1}.
\end{equation}
Consider a perturbation to either $\boldsymbol{\psi}_j$ or $\boldsymbol{\phi}_j$:
\begin{equation}
\begin{split}
    \boldsymbol{\psi}_{j,\epsilon} = & \boldsymbol{\psi}_j + \epsilon \mathbf{r}\\
    \boldsymbol{\phi}_{j,\epsilon} = & \boldsymbol{\phi}_j + \epsilon \mathbf{g}
\end{split}
\end{equation}
where $||\boldsymbol{\psi}_j|| = ||\boldsymbol{\phi}_j|| = ||\mathbf{r}|| = ||\mathbf{g}|| = 1$ and $\epsilon \ll 1$. The error in the singular value due to a perturbation in $\boldsymbol{\psi}$ may be bounded as follows.
\begin{equation}
    \sigma^{-2}_{j,\epsilon,L} =\left(\mathbf{L}\left(\boldsymbol{\psi} + \epsilon \mathbf{r}\right)\right)^H\mathbf{Q}\left(\mathbf{L}\left(\boldsymbol{\psi} + \epsilon \mathbf{r}\right)\right)= \sigma_j^{-2} + 2\epsilon \Re\{\left(\mathbf{L}\boldsymbol{\psi}\right)^H\mathbf{Q}\left(\mathbf{L}\mathbf{r}\right)\} + \mathcal{O}(\epsilon^2)
\end{equation}
Using the definition $\mathbf{L}\boldsymbol{\psi}_j = \sigma^{-1}_j\boldsymbol{\phi}$ and rearranging slightly we find
\begin{equation} \label{append_error_sigma_L}
    \sigma_{j,\epsilon,L} = \sigma_j\left( 1 + 2\epsilon \sigma_j \Re\{\left(\boldsymbol{\phi}\right)^H\mathbf{Q}\left(\mathbf{L}\mathbf{r}\right)\} \right)^{-0.5}.
\end{equation}
Taylor expanding for small $\epsilon$ gives
\begin{equation}
    \sigma_{j,\epsilon,L} = \sigma_j\  - \epsilon \sigma^2_j \Re\{\left(\boldsymbol{\phi}\right)^H\mathbf{Q}\left(\mathbf{L}\mathbf{r}\right)\} + \mathcal{O}(\epsilon^2)
\end{equation}
further applying the Cauchy-Schwartz inequality, and noting that $\|\boldsymbol{\phi}_j\| = \|\mathbf{r}\| = 1$ leads to 
\begin{equation}\label{D_sigma_psi}
   \frac{ |\sigma_{j,\epsilon,L} - \sigma_j\|}{\sigma_j}  \leq \epsilon \sigma_j \|\mathbf{L}\|.
\end{equation}

% Error in sigma due to changes in phi
Conversely, the error in the singular value due to a perturbation in $\boldsymbol{\phi}$ may be bounded as follows.
\begin{equation}
    \sigma^{2}_{j,\epsilon,H} = \left(\mathbf{H}\left(\boldsymbol{\phi} + \epsilon \mathbf{g}\right)\right)^H\mathbf{Q}\left(\mathbf{H}\left(\boldsymbol{\phi} + \epsilon \mathbf{g}\right)\right)= \sigma_j^{2} + 2\epsilon \Re\{\left(\mathbf{H}\boldsymbol{\psi}\right)^H\mathbf{Q}\left(\mathbf{H}\mathbf{g}\right)\} + \mathcal{O}(\epsilon^2)
\end{equation}
Using the definition $\mathbf{H}\boldsymbol{\phi}_j = \sigma_j\boldsymbol{\psi}$ and rearranging slightly we find
\begin{equation}
    \sigma_{j,\epsilon,H} = \sigma_j\left( 1 + 2\epsilon \sigma^{-1}_j \Re\{\left(\boldsymbol{\psi}\right)^H\mathbf{Q}\left(\mathbf{H}\mathbf{g}\right)\} \right)^{0.5}.
\end{equation}
Taylor expanding for small $\epsilon$ gives
\begin{equation} \label{append_error_sigma_H}
    \sigma_{j,\epsilon,H} = \sigma_j\  + \epsilon  \Re\{\boldsymbol{\psi}^H\mathbf{Q}\left(\mathbf{H}\mathbf{g}\right)\} + \mathcal{O}(\epsilon^2)
\end{equation}
again applying the Cauchy - Schwartz inequality and noting that given that $\|\boldsymbol{\psi}_j\| = \|\mathbf{g}\| = 1$ and $\|\mathbf{H}\| = \sigma_1$ leads to
\begin{equation}\label{D_sigma_phi}
   \frac{ |\sigma_{j,\epsilon,H} - \sigma_j\|}{\sigma_j}  \leq \epsilon \frac{\sigma_1}{\sigma_j}.
\end{equation}

% Error in Phi due to changes in psi
\section{Singular Mode Sensitivity}\label{app:mode_sensitivity}
Here we derive bounds on the sensitivity of $\boldsymbol{\phi}_j$.
\begin{equation}
    \boldsymbol{\phi}_{j,\epsilon,\psi} = \sigma_{j,\epsilon,L} \mathbf{L} \left(\boldsymbol{\psi}_j + \epsilon \mathbf{r}\right)= \sigma_{j,\epsilon,L} \mathbf{L}\boldsymbol{\psi}_j +  \epsilon \sigma_{j,\epsilon,L} \mathbf{L} \mathbf{r}
\end{equation}
Again we assume $\|\boldsymbol{\psi}_j\| = \|\mathbf{r}\| = 1$ and $\epsilon \ll 1$.
Subtracting $\boldsymbol{\phi}_j = \sigma_j\mathbf{L}\boldsymbol{\psi}_j$ from both sides, and rearranging the right hand side slightly results in
\begin{equation}
    \boldsymbol{\phi}_{j,\epsilon,\psi} - \boldsymbol{\phi}_{j} = \left(\sigma_{j,\epsilon,L} - \sigma_j\right) \mathbf{L}\boldsymbol{\psi}_j +  \epsilon \left(\sigma_{j,\epsilon,L} - \sigma_j\right) \mathbf{L} \mathbf{r} + \epsilon \sigma_j \mathbf{L} \mathbf{r} .
\end{equation}
We note from the results of appendix \ref{app:sigma_sensitivity} that $\left(\sigma_{j,\epsilon,L} - \sigma_j\right) \sim \epsilon$ which allows us to write
\begin{equation}
    \boldsymbol{\phi}_{j,\epsilon,\psi} - \boldsymbol{\phi}_{j} = \left(\sigma_{j,\epsilon,L} - \sigma_j\right) \mathbf{L}\boldsymbol{\psi}_j  + \epsilon \sigma_j \mathbf{L} \mathbf{r} + \mathcal{O}(\epsilon^2).
\end{equation}
Next we analyze the norm of both the left and right hand side, which upon application of the triangle inequality and finally, using the Cauchy-Schwartz inequality as well as (\ref{D_sigma_psi}) results in
\begin{equation}
    \|\boldsymbol{\phi}_{j,\epsilon,\psi}-\boldsymbol{\phi}_{j}\| \leq \epsilon \left(\sigma_j \|\mathbf{L}\| + 1\right)\sigma_j \|\mathbf{L}\|.
\end{equation}
% error in psi due to error in phi
The same analysis may be applied to derive bounds on the sensitivity of $\boldsymbol{\psi}_j$:
\begin{equation}
    \boldsymbol{\psi}_{j,\epsilon,\phi} \equiv \sigma_{j,\epsilon,H}^{-1} \mathbf{H} \left(\boldsymbol{\phi}_j + \epsilon \mathbf{g}\right)  = \sigma_{j,\epsilon,H}^{-1} \mathbf{H}\boldsymbol{\phi}_j +  \epsilon \sigma_{j,\epsilon,H}^{-1} \mathbf{H} \mathbf{g}
\end{equation}
where again we assume $\|\boldsymbol{\phi}_j\| = \|\mathbf{g}\| = 1$ and $\epsilon \ll 1$.
Subtracting $\boldsymbol{\psi}_j = \sigma^{-1}_j\mathbf{H}\boldsymbol{\phi}_j$ from both sides, and rearranging the right hand side slightly results in
\begin{equation}
    \boldsymbol{\psi}_{j,\epsilon,\phi} - \boldsymbol{\psi}_{j} = 
    \frac{\sigma_j -\sigma_{j,\epsilon,H}}{\sigma_j \sigma_{j,\epsilon,H}} \mathbf{H}\boldsymbol{\phi}_j +
    \epsilon \frac{\sigma_j -\sigma_{j,\epsilon,H}}{\sigma_j \sigma_{j,\epsilon,H}} \mathbf{H} \mathbf{g} +
    \epsilon \sigma_j^{-1} \mathbf{H} \mathbf{g} .
\end{equation}
Taylor expanding about $\epsilon = 0$ and noting that $\left(\sigma_{j,\epsilon,H} - \sigma_j\right) \sim \epsilon$ results in
\begin{equation}
    \boldsymbol{\psi}_{j,\epsilon,\phi} - \boldsymbol{\psi}_{j} = 
    \frac{\sigma_j -\sigma_{j,\epsilon,H}}{\sigma_j^2} \mathbf{H}\boldsymbol{\phi}_j +
    \epsilon \sigma_j^{-1} \mathbf{H} \mathbf{g} + \mathcal{O}(\epsilon^2).
\end{equation}
Next we analyze the norm of both the left and right hand side, which upon application of the triangle inequality and finally, using the Cauchy-Schwartz inequality as well as (\ref{D_sigma_phi}) results in
\begin{equation}
    \|\boldsymbol{\psi}_{j,\epsilon,\phi}-\boldsymbol{\psi}_{j}\| \leq \epsilon \left( \frac{\sigma_1}{\sigma_j} + 1 \right)\frac{\sigma_1}{\sigma_j}.
\end{equation}

\FloatBarrier
\bibliography{references2_vra.bib}

\end{document}